\PassOptionsToPackage{dvipsnames}{xcolor}
\documentclass[preprint]{aastex631}
\usepackage{placeins}
\usepackage{amsmath}
\usepackage{bm}
\usepackage{capt-of}   
\usepackage{svg}
\usepackage{empheq}
\usepackage{enumitem}
\usepackage{hyperref}
\usepackage{booktabs}
\usepackage{hhline}
\usepackage{tabularx}
\usepackage{longtable}
\usepackage{numprint}
\usepackage{listings}
\usepackage{xspace}
\usepackage{graphicx}
\usepackage{float} 
\usepackage{mwe}
\usepackage{MnSymbol}
\usepackage{url}
\usepackage{pythonhighlight}
\lstnewenvironment{mypython}[1][]{\lstset{style=mypython,#1}}{}

\newcommand{\ang}{\ensuremath{\mathrm{\AA}}\xspace}

\newcommand{\rf}{\textcolor{black}}

\newcommand{\projname}{DESI Strong Lens Foundry\xspace}

\newcommand{\ho}{\ensuremath{H_0}\xspace}

\newcommand{\desionefivefour}{DESI~J154.6972-01.3590\xspace}
\newcommand{\desionesixfive}{DESI~J165.6876-06.0423\xspace}
\newcommand{\desizeroninefour}{DESI~J094.5639+50.3059\xspace}
\newcommand{\desitwothreefour}{DESI~J234.4783+14.7232\xspace}
\newcommand{\desitwofiveseven}{DESI~J257.4348+31.9046\xspace}
\newcommand{\desitwothreeeight}{DESI~J238.5690+04.7276\xspace}
\newcommand{\desitwofoursix}{DESI~J246.0062+01.4836\xspace}

\newcommand{\gigal}{\texttt{GIGA-Lens}\xspace}
\newcommand{\thetaEfit}{\ensuremath{2.646\pm0.002\twopr}\xspace}
\newcommand{\gammafit}{\ensuremath{1.37\pm0.02}\xspace}
\newcommand{\lamlam}{\ensuremath{\lambda\lambda}\xspace}

\newcommand{\rhat}{\ensuremath{\hat{R}}\xspace}

\newcommand{\HST}{\emph{Hubble Space Telescope}\xspace}
\newcommand{\euc}{\emph{Euclid}\xspace}
\newcommand{\hst}{\emph{HST}\xspace}

\newcommand{\jwst}{\emph{JWST}\xspace}

\newcommand{\tf}{\texttt{TensorFlow}\xspace}
\newcommand{\coverage}{20,000\xspace}

\newcommand{\ser}{S{\'e}rsic\xspace}

\newcommand{\twopr}{\ensuremath{^{\prime \prime}}\xspace}
\newcommand{\tE}{\ensuremath{\theta_E}\xspace}
\newcommand{\gma}{\ensuremath{\gamma}\xspace}
\newcommand{\zd}{\ensuremath{z_d\xspace}}
\newcommand{\zs}{\ensuremath{z_s\xspace}}

\submitjournal{ApJ}

\shorttitle{\projname V}
\shortauthors{Huang, \'{A}lvarez, \'{U}beda, Bhamre, Xu et al.}

\begin{document}

\title{\projname V:\\
A Sample of \hst-Observed Strong Lenses Modeled with \gigal}


\correspondingauthor{Xiaosheng Huang}
\email{xhuang22@usfca.edu}

\author[0000-0001-8156-0330]{Xiaosheng~Huang}
\affiliation{Department of Physics \& Astronomy, University of San Francisco, San Francisco, CA 94117, USA}
\affiliation{Physics Division, Lawrence Berkeley National Laboratory, 1 Cyclotron Road, Berkeley, CA 94720, USA}

\author[0009-0006-0284-3863]{David~\'{A}lvarez-Garc\'{i}a}
\affiliation{Physics Division, Lawrence Berkeley National Laboratory, 1 Cyclotron Road, Berkeley, CA 94720, USA}
\affiliation{Department of Physics, 
Complutense University of Madrid, 28040 Madrid, Spain}

\author[0009-0005-4355-0293]{M\'{o}nica~\'{U}beda}
\affiliation{Physics Division, Lawrence Berkeley National Laboratory, 1 Cyclotron Road, Berkeley, CA 94720, USA}
\affiliation{Department of Physics, 
Complutense University of Madrid, 28040 Madrid, Spain}

\author[0009-0007-3732-5211]{Vikram~Bhamre}
\affiliation{Department of Physics, University of California, Berkeley, Berkeley, CA 94720, USA}

\author[0009-0006-4623-3629]{Sean~Xu}
\affiliation{Department of Physics, University of California, Berkeley, Berkeley, CA 94720, USA}
\affiliation{Physics Division, Lawrence Berkeley National Laboratory, 1 Cyclotron Road, Berkeley, CA 94720, USA}

\author[0009-0003-4697-7079]{S.~Baltasar}
\affiliation{Physics Division, Lawrence Berkeley National Laboratory, 1 Cyclotron Road, Berkeley, CA 94720, USA}
\affiliation{Department of Physics, 
Complutense University of Madrid, 28040 Madrid, Spain}

\author[0009-0009-8206-0325]{N.~Ratier-Werbin}
\affiliation{Physics Division, Lawrence Berkeley National Laboratory, 1 Cyclotron Road, Berkeley, CA 94720, USA}
\affiliation{Department of Physics, 
Complutense University of Madrid, 28040 Madrid, Spain}

\author[0009-0009-0407-2419]{F.~Urcelay}
\affiliation{Institute of Astrophysics, Pontificia Universidad Católica de Chile, Santiago, Chile}

\author[0000-0002-2350-4610]{S.~Agarwal}
\affiliation{University of Chicago, Department of Astronomy, Chicago, IL 60615, USA}



\author[0000-0001-7101-9831]{A.~Cikota}
\affiliation{Gemini Observatory / NSF's NOIRLab, Casilla 603, La Serena, Chile}



\author[0000-0002-6876-8492]{Yuan-Ming~Hsu}
\affiliation{Department of Physics, National Taiwan University, No. 1, Section 4, Roosevelt Road, Taipei 106319, Taiwan}

\author[0009-0006-5989-4899]{E.~Lin}
\affiliation{Department of Physics, University of California, Berkeley, Berkeley, CA 94720, USA}



\author[0000-0002-5042-5088]{D.~J.~Schlegel}
\affiliation{Physics Division, Lawrence Berkeley National Laboratory, 1 Cyclotron Road, Berkeley, CA 94720, USA}

\author[0000-0002-1804-3960]{E.~Silver}
\affiliation{Department of Physics, Harvard University, Cambridge, MA 02138, USA}

\author[0000-0002-0385-0014]{C.~J.~Storfer}
\affiliation{Institute for Astronomy, University of Hawai'i, Honolulu, HI 96822-1897, USA}


\author[0009-0008-0518-8045]{M.~Tamargo-Arizmendi}
\affiliation{Physics Division, Lawrence Berkeley National Laboratory, 1 Cyclotron Road, Berkeley, CA 94720, USA}

\begin{abstract}
We present six galaxy-scale strong lenses with \emph{HST} imaging modeled using \texttt{GIGA-Lens}. This is Paper~V of the DESI Strong Lens Foundry series. These systems were discovered in the DESI Legacy Imaging Surveys using ML/AI methods and confirmed with DESI, Keck/NIRES, and VLT/MUSE spectroscopy. They span $z_d = 0.39\, \text{ --}\, 1.1$ and $z_s = 1.4\, \text{--}\, 3.3$.
This is the first \emph{HST} strong lens sample modeled with full forward modeling---all lens and source parameters sampled simultaneously in a single inference---with explicit convergence validation using both $\hat{R}$ and effective sample size (ESS) for each system. 
All inferred parameters satisfy $\hat{R} < 1.1$ and ${\rm ESS} \gtrsim 10{,}000$, demonstrating that \texttt{GIGA-Lens} achieves statistically robust inference even for some of the most complex galaxy-scale lenses known. 
These results pave the way for scaling to much larger, high-resolution strong lens samples from \emph{HST}, \emph{Euclid}, \emph{JWST}, and \emph{Roman}. Convergence-validated modeling will be critical for key science goals, including constraining the mass-density profile of galaxies, detecting low-mass dark matter (sub)halos, and delivering precise and accurate cosmological constraints.

\end{abstract}

\section{Introduction}\label{sec:intro}
Strong gravitational lensing systems 
are a powerful tool for astrophysics and cosmology.
They have been used to study how dark matter is distributed in galaxies and \rf{galaxy} clusters 
\citep[e.g.,][]{kochanek1991a, blandford1992a, broadhurst2000a,  koopmans2002a, 
bolton2006a,  koopmans2006a, bradac2008a, vegetti2009a, huang2009a, jullo2010a,
tessore2016a,  jauzac2018a, 
shajib2019a, meneghetti2020a, odonell2025}.
Furthermore, carefully modeling the mass profiles of galaxy-scale strong lenses for a large number of lensing systems 
over a wide range of redshifts makes it possible to  study the structural evolution 
of massive elliptical galaxies \citep[e.g., ][]{bolton2012b, filipp2023a}.

They are uniquely suited to probe 
dark matter substructure beyond the local universe and line-of-sight low-mass halos
\citep[e.g.,][]{vegetti2014a, hezaveh2016a, vegetti2018a, ritondale2019a, diazrivero2020a, cagan-sengul2022a, nierenberg2023a}, 
to test the predictions of the cold dark matter (CDM) model.

Recent measurements of the Hubble constant \ho span a range of $\sim$10\% \citep[e.g.,][]{abbott2017a, abbott2018b, riess2019a, wong2019a, freedman2019a, freedman2020a, planck2020a, khetan2020a, philcox2020a, choi2020a, dhawan2023a}, 
and significant tension between predictions for \ho based on early-universe observables 
and direct late-universe measurements remain \citep[e.g.,][]{verde2019a}.
Multiply-lensed supernovae (SNe) and quasars have been used for measuring time delays and \ho \citep{refsdal1964a, treu2010a, oguri2010a, wong2019a, suyu2023a, kelly2023a, pascale2025a}.

Furthermore time delay \ho measurements are a powerful complement to other independent measurements of the dark energy equation of state
\citep[e.g.,][]{linder2011a, treu2016a, pierel2021a}.
In addition, multi-source-plane systems can constrain cosmology by providing angular diameter distance ratios along a single line of sight \citep[e.g.,][]{collett2014a, sharma2022a, sharma2023a}.
Finally, for nearby strong lensing galaxies, 
extra-galactic tests of General Relativity can be performed 
by combining lens modeling with 
spatially resolved 
stellar kinematic observations \citep{collett2018a}.

The DESI Legacy Surveys\footnote{\url{http://www.legacysurvey.org/}}
\citep[][Schlegel et al. in prep]{dey2019a},
for which at least $z$ band is observed with a 4-m telescope,
covers $>$\coverage~deg$^2$,
four times the size of the Dark Energy Survey \citep{des2005a} footprint.
We identified $\sim 4000$ new strong lenses in the Legacy Surveys Data Release~7, 8, 9, and 10 \citep[][respectively]{huang2020a, huang2021a, storfer2024a, inchausti2025} by using residual neural networks (and EfficientNet for the DR10 search). 
We have also found 436 lensed quasars \citep{dawes2023a} using autocorrelation for a lensed quasar candidate sample.
Very recently, we have found over 6000 lens candidates in the DESI spectroscopic data \citep[][]{hsu2025, karp2025}.
The entire catalog of these lens candidates can be found on our project website.\footnote{\url{https://sites.google.com/usfca.edu/neuralens/}}

With so many lens discoveries,
a fast and robust modeling pipeline is needed.
We therefore developed \gigal \citep{gu2022a}. 
A GPU-accelerated, fully forward-modeling Bayesian pipeline that can speed up the lens modeling time by two orders of magnitude. 
In \citet{cikota2023a}, we applied \gigal to a observed lensing system, DESI~J253.2534+26.8843.
It was then applied to a group-scaled strong lens (DES~J0248-3955), also using ground-based imaging data \citep{Urcelay2025}.
Most recently we applied it to a system with \hst data, DESI~J165.4754-06.0423 
\citep[][Paper~I in this series]{huang2025a}.

This is paper V of the \projname paper series. 
In this paper, we present six lensing systems that have been fully confirmed with both spectroscopy and \hst imaging.
These systems were first identified as lens candidates in \citep[][henceforth, H20, H21]{huang2020a, huang2021a}.
As mentioned in \citet{huang2025b}, Paper~II of this series, for systems with candidate status, we follow the naming convention of ``DESI-'' followed by the RA and Dec (with a sign prefix) in decimal format to the fourth decimal place (with rounding) for both coordinates. Such a naming scheme is easy to parse and precise to $< 1''$.
With full confirmation, which is the case for the systems in this work, we follow the official DESI naming convention of  ``DESI J'' (note the space) followed by RA and Dec (with a sign prefix) in decimal format to the fourth decimal place (with truncation) for both coordinates. 
Simply put, the presence of ``J" indicates a system has now been fully confirmed. 
This paper is organized as follows.  
The six systems from our \HST SNAP program to be modeled in this work are presented in   
\S~\ref{sec:ls-hst}, followed by a brief description of the spectroscopic observations in \S~\ref{sec:spect}.
In \S~\ref{sec:lens-model}, we highlight two representative systems modeled with \gigal, 
while the results for the remaining systems are provided in Appendix~\ref{sec:lens-models-app} (with the effective PSF generation described in Appendix~\ref{sec:psf_section}).
We discuss our results in \S~\ref{sec:discussion} and conclude in \S~\ref{sec:conclusion}.


\FloatBarrier
\section{\HST SNAP Program}\label{sec:ls-hst}

We presented our \emph{Hubble} SNAP program GO-15867 (PI: X.~Huang) in Paper~I in this series.
This program followed up on a subset of the lens candidates found in H20 and H21, using the DESI Legacy Imaging Surveys data. 
All targets have $3 \times$399.23~sec exposure, 
for a total of 1197.7~sec, 
on WFC3 using the F140W filter in the NIR channel.
The targets were approximately centered in the WFC3 aperture, with no CR split, because we wanted to keep read noise down. 
Of the 112 systems submitted to \hst, 
a total of 53 were targeted. 
Two of them were not successfully observed due to the loss of the guide star.
The native pixel size is $0.13\twopr$, and each image is drizzled to $0.065''$.
The six systems that will be modeled in the paper are presented in Figure~\ref{fig:hs-6}.

\begin{minipage}[H]{\linewidth}
\makebox[\linewidth]{  \includegraphics[keepaspectratio=true,scale=.75]{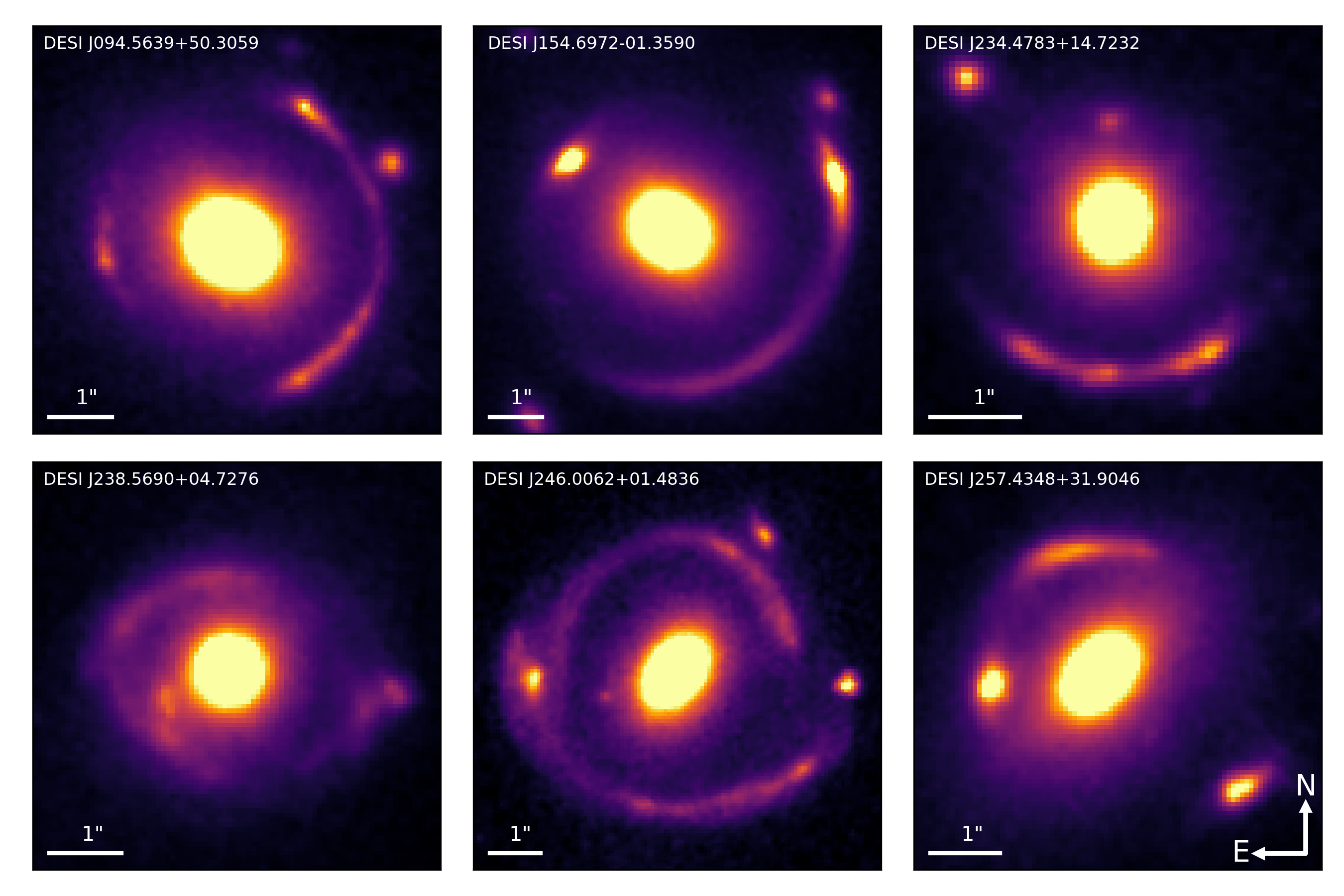}}
\captionof{figure}{
Six systems observed by the \hst SNAP program GO-15867 modeled in this paper. 
The naming convention is RA and Dec in decimal format.
North is up, and East to the left. 
The systems are arranged in ascending RA.
}
\label{fig:hs-6}
\end{minipage}

\section{Spectroscopic Observations}\label{sec:spect}
The DESI Strong Lens Secondary Target program (introduced in Paper~II in this series) has observed a large fraction of our lens candidates.
The success rate of obtaining redshifts for lensing galaxies by DESI is high.
These are nearly always bright elliptical galaxies with a number of identifiable features in the optical range,
typically absorption lines
and the 4000 \ang break. 
For lensed sources, 
some of the spectra have no clear features,  
likely due to the fact that they are too faint and/or their redshifts place key spectral features outside the optical range. 
Lensed sources are typically star forming galaxies for which the [\ion{O}{2}] doublet emission feature is often used to anchor redshift fits. 
This makes redshift measurements in the optical range for $z_s \gtrsim 1.6$ challenging.\footnote{It is true that for $z_s \gtrsim 2.0$, 
Ly-$\alpha$ will 
be in the optical range, 
though
this feature is not always present.}
For these systems, we determine their redshifts from our on-going Keck NIRES program \citep[][Paper~III in this series; and Storfer, Tamargo et al., Paper~VI, in prep.]{agarwal2025}.
Finally some of our systems are being observed by a VLT MUSE program (PI: A.~Cikota). Most of these are only observable from the southern hemisphere. 
But our MUSE program does target some systems that are in the DESI strong lens program but not yet observed by DESI, because they are of high priority \citep[e.g.,][]{cikota2023a}.
The MUSE results are presented in Paper~IV of this series \citep{lin2025}.

\section{Lens Modeling}\label{sec:lens-model}
In Paper~I, as a demonstration, we used the open-source \gigal\footnote{\url{https://github.com/giga-lens/gigalens}} framework \citep[][Gu22]{gu2022a} to model one lens system with \hst data, \href{https://www.legacysurvey.org/viewer/?ra=165.4753&dec=-6.0424&layer=ls-dr9&pixscale=0.262&zoom=16}{\desionesixfive}.
In the following, we present lens models for this and six additional systems in Table~\ref{tab:model-sys}, arranged in increasing lens redshift.
The effective PSF generation is described in Appendix~\S~\ref{sec:psf_section}.
In this section, we present the lens modeling results for two systems. For \desitwothreefour\ (\S~\ref{sec:desi234}), we fix five light profile parameters for a field galaxy based on preliminary modeling, and then optimize and sample the remaining 36 parameters \emph{simultaneously} in the main modeling process. Such a field galaxy has little to no impact on the lensing of the background source. To emphasize this, we refer to these field galaxies as “environmental” galaxies throughout this work.
For \desitwothreeeight\ (\S~\ref{sec:desi238}), we do not even fix the parameters of an environmental galaxy: all 38 are modeled and sampled simultaneously.
The remaining systems are presented in Appendix~\ref{sec:lens-models-app}.

\gigal is a fully forward-modeling Bayesian lens modeling pipeline.
Briefly, it consists of three stages: finding the maximum a posteriori (MAP) for the lensing parameters via multi-start gradient descent, determining a surrogate multidimensional Gaussian covariance matrix for these parameters using stochastic variational inference (SVI), and finally sampling with Hamiltonian Monte Carlo (HMC).
All three stages use gradient descent with automatic differentiation and take advantage of GPU acceleration.  

Our mass model consists of an elliptical power law (EPL) for the lens mass profile and external shear. 
The Einstein radius is denoted $\theta_E$ (arcsec), and $\gamma$ is the EPL mass-slope parameter.
$(x, y)$ denote the center coordinates, specified independently for the lens mass, lens light, and source light. 
($\gamma_{1, \text{ext}}$, $\gamma_{2, \text{ext}}$) are the external shear components.
The parameters $\epsilon_1$ and $\epsilon_2$ specify the eccentricities.
We model lens light with one or more \ser profiles.
For source light, we use \ser profiles and/or shapelets \citep{birrer2015a}.
For the lens and source surface brightness scaling, we either model the light intensities as parameters or solve for them by linear inversion.
Both are mathematically justified. 
A systematic, side-by-side evaluation of these approaches---particularly their impact on uncertainties---will be explored in a future study.

We aim for 1) full forward modeling (the modeling of all lens and source parameters at the same time, instead of taking a staged approach) and 2) statistical convergence.
Both goals have been achieved for all seven systems (including \desionesixfive from Paper~I). 


We apply the two widely used diagnostics for convergence.
1) the potential scale reduction factor (PSRF, or \rhat), also known as the Gelman-Rubin statistic, which compares the inter- and intra-chain estimates for model parameters.
2) The effective sample size (ESS).
To achieve good convergence \rhat needs to be below 1.1 \citep{gelman1992a, gelman2014a}.
This indicates that the between- and within-chain estimates are in good agreement, or the chains have ``mixed well''.
In addition, the ESS generally needs to be higher than 1000 \citep[e.g.,][]{mandel2022a}. 
For all seven systems, we achieve $\rhat < 1.1$ and ESS $\gtrsim 10,000$, satisfying the convergence criteria. We will present each system in turn.




\begin{center}
\begin{deluxetable}{lcccccccccc}[H]
\tabletypesize{\scriptsize}
\tablecaption{Seven Modeled Systems \label{tab:model-sys}}
\tablehead{
\colhead{System} & 
\colhead{\zd} &  
\colhead{\zs} &  
\colhead{\tE $('')$} &
\colhead{\gma} & 
\colhead{Modelers} &
\colhead{$\rhat_\mathrm{ceiling}$} &
\colhead{Sec} &
\colhead{Source Model} &
\colhead{Params (tot/fix)}\\[-6pt]
\colhead{[1]} & \colhead{[2]} & \colhead{[3]} & \colhead{[4]} & \colhead{[5]} &
\colhead{[6]} & \colhead{[7]} & \colhead{[8]} & \colhead{[9]} & \colhead{[10]}
}
\startdata
\href{https://www.legacysurvey.org/viewer/?ra=154.6972&dec=-01.3590&layer=ls-dr10-grz&pixscale=0.262&zoom=16}
{\desionefivefour} 
& 0.3884$^\S$ & 1.4304*  & 2.8982$^{0.0010}_{0.0011}$ & 1.411$\pm0.016$ &   MU & $ 1.005$& \ref{sec:desi154} & 4 \ser's & 44/6 \\
\href{https://www.legacysurvey.org/viewer/?ra=165.6876&dec=-06.0423&layer=ls-dr10-grz&pixscale=0.262&zoom=16}
{\desionesixfive} 
& 0.4834* & 1.6748 & $2.6263^{+0.0017}_{-0.0016}$ &  $1.372^{+0.023}_{-0.022}$ & 
SB, NRW & $ 1.06$ & \ref{sec:desi165} & S\'{e}rsic+shapelets  & 35/0\\
& & & & & & & & & ($n_{max} = 6$) & \\
\href{https://www..legacysurvey.org/viewer/?ra=094.5639&dec=+50.3059&layer=ls-dr10-grz&pixscale=0.262&zoom=16}
{\desizeroninefour} 
& 0.552*$^\ddag$ & 3.3332   & 2.295$\pm0.001$ & 2.548$^{+0.043}_{-0.042}$    & VB  & 1.09 &  \ref{sec:desi094} & 2 \ser's & 44/6 \\
\href{https://www.legacysurvey.org/viewer/?ra=234.4783&dec=+14.7232&layer=ls-dr10-grz&pixscale=0.262&zoom=16}
{\desitwothreefour} 
& 0.7313* & 2.4780   & $1.548_{-0.003}^{+0.004}$& 
$2.057_{-0.032}^{+0.028}$&	DA & $ 1.009 $  & \ref{sec:desi234} & 2 \ser's & 41/5\\
\href{https://www.legacysurvey.org/viewer/?ra=257.4348&dec=31.9046&layer=ls-dr10-grz&zoom=16}
{\desitwofiveseven} 
& 0.7464* & 2.1200  & 1.9899$\pm$0.0024 & 2.066$\pm0.023$  &  DA & $  1.001$  & \ref{sec:desi257} & shapelets & 29/0\\
& & & & & & & & & ($n_{max} = 3$) & \\
\href{https://www.legacysurvey.org/viewer/?ra=238.5690&dec=+04.7276&layer=ls-dr10-grz&pixscale=0.262&zoom=16}
{\desitwothreeeight} 
& 0.7768* & 1.7210  & 1.475$^{+0.003}_{-0.002}$  & 1.937$^{+0.019}_{-0.020}$   & SX & $  1.08$ & \ref{sec:desi238}  & 2 \ser's & 38/0\\
\href{https://www.legacysurvey.org/viewer/?ra=246.0062&dec=+01.4836&layer=ls-dr10-grz&pixscale=0.262&zoom=16}
{\desitwofoursix} 
& 1.092$^\diamond$ & 2.3685   & 2.698$\pm0.001$ & 2.616$\pm0.017$ &  DA, SX & $ 1.037$  & \ref{sec:desi246} & 5 \ser's & 78/28\\
\enddata
\tablecomments{Column [1]: System names arranged in ascending order of lens redshifts. [2] and [3] show  lens redshifts (\zd) and source redshifts (\zs), respectively. 
The redshifts without a superscript are obtained from Keck NIRES, with the
facilities used to obtain the rest shown at the bottom of these notes.
[4]: Best-fit Einstein radii.
[5]: Best-fit mass density slope.
[6]: Modelers, with XH for every system.
[7]: The ceiling of \rhat for each system, i.e., the \rhat for all parameters are lower than the value shown for that system. 
[8]: The section that describes the modeling of each system. The model for \desionesixfive was presented in Paper~I and was summarized in this work.
[9]: Source light model. 
For the cases where we use shapelets, $n_{max}$ is given.
[10]: 
\emph{We emphasize that only the light profile parameters of environmental galaxies in the cutout image are fixed based on preliminary modeling, and only for four of the systems.} In each of these cases, 5 - 6 parameters are fixed, except for \desitwofoursix, which has an especially complex environment. For \desionesixfive and \desitwothreeeight, even the environmental galaxies are included in the full forward modeling.  \desitwofiveseven does not have environmental galaxies bright enough to require modeling.
Finally, the effective sample size (ESS) for HMC is $\gtrsim 10,000$ for all systems.\\
* DESI; $^\S$ SDSS; 
$^\ddag$ Lick~Kast; and
$^\diamond$ VLT~MUSE. \\
}
\end{deluxetable}
\end{center}

\vspace{-1.25cm}
\subsection{\desitwothreefour}\label{sec:desi234}
\href{https://www.legacysurvey.org/viewer/?ra=234.4783&dec=14.7231&layer=ls-dr10-grz&zoom=16}{\desitwothreefour}\footnote{The first mention of each system in the text is hyperlinked to the Legacy Surveys’ \href{https://www.legacysurvey.org/viewer}{skyviewer web portal}.} was discovered in H21, with a ResNet probability of 0.98, and a human score of 3.0 (out of 4.0), corresponding to a B grade.
The lens redshift, \zd = 0.7313 is from DESI (based on the 4000 \ang break, Ca~H\&K, and [\ion{O}{2}] emission).
The source redshift, \zs = 2.477 is from Keck NIRES (Storfer, Tamargo et al., in prep; Paper~VI).
This system has an Einstein radius of \(1.55''\).
We use a cutout with $64\times64$ pixels and a $33\times33$ pixel empirical PSF, generated using the procedure outlined in Section~\ref{sec:psf_section}.

\begin{figure}[H]
    \centering
    \includegraphics[trim={0.45cm 0.45cm 0.45cm 0.52cm}, clip,width=0.48\linewidth]{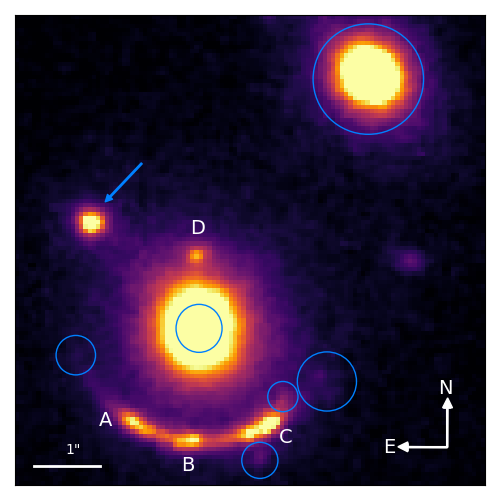}
    \captionof{figure}{
    \desitwothreefour.
    The four lensed images are labeled counterclockwise as A, B, C and D. We mask out the light from five faint objects (blue circles). We model the light of the environmental galaxy (blue arrow) to the upper left of the lensing galaxy and then fix its parameters in the main modeling process. }
    \label{fig:234mask}
\end{figure}

\desitwothreefour is shown in Figure \ref{fig:234mask}. We model the lens light with two elliptical \ser profiles. 
We mask a region around its center, 
with a brightness threshold of 1.0 (100 pixels, or 0.4225~arcsec$^2$). We model the source using two elliptical \ser profiles.
In addition, we mask four small and faint objects, indicated in Figure~\ref{fig:234mask} by small circles: two to the northwest of image C, one to the south of image C, and one to the northeast of image A. We also mask a very bright object near the top right corner. 
We model the light of the galaxy (blue arrow in Figure~\ref{fig:234mask}) to the upper left of the lensing galaxy with a circular \ser profile. 

Throughout this work, the first figure for each system uses blue arrows to mark environmental galaxies whose light is modeled (but whose mass is not), while blue circles mark objects that are masked and not modeled. For four of the systems, including this one (see Table~\ref{tab:model-sys}), these light parameters are then held fixed in the main inference. For two others, these parameters are sampled together with the rest of the model parameters. Finally, for \desitwofiveseven, there are no environmental galaxies bright enough to be modeled.

\begin{figure}[H]
    \centering
    \includegraphics[trim={0.4cm 0.65cm 0.4cm 0.25cm}, clip,scale=0.62]{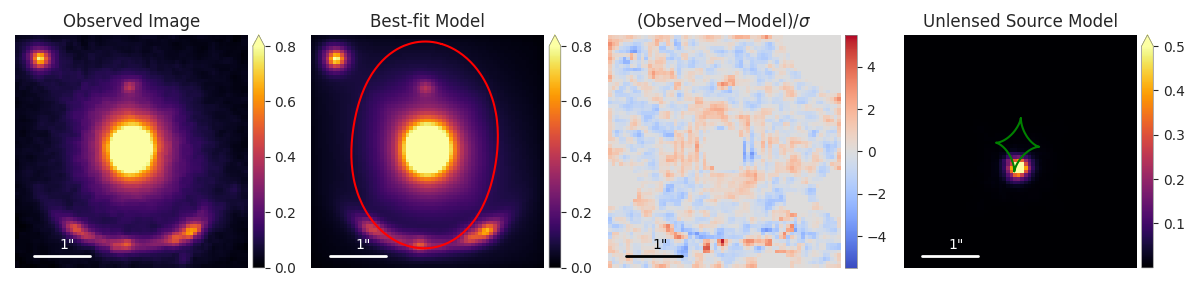}
    \caption{Best-fit Model \desitwothreefour. From left to right, we show: the observed Hubble image, our best-fit model with the critical curve, the reduced residual, and the unlensed source with the caustic.}
    \label{fig:bestfitmodeldavid}  
\end{figure}

\begin{figure}[H]
    \centering
   \includegraphics[width=0.85\linewidth]{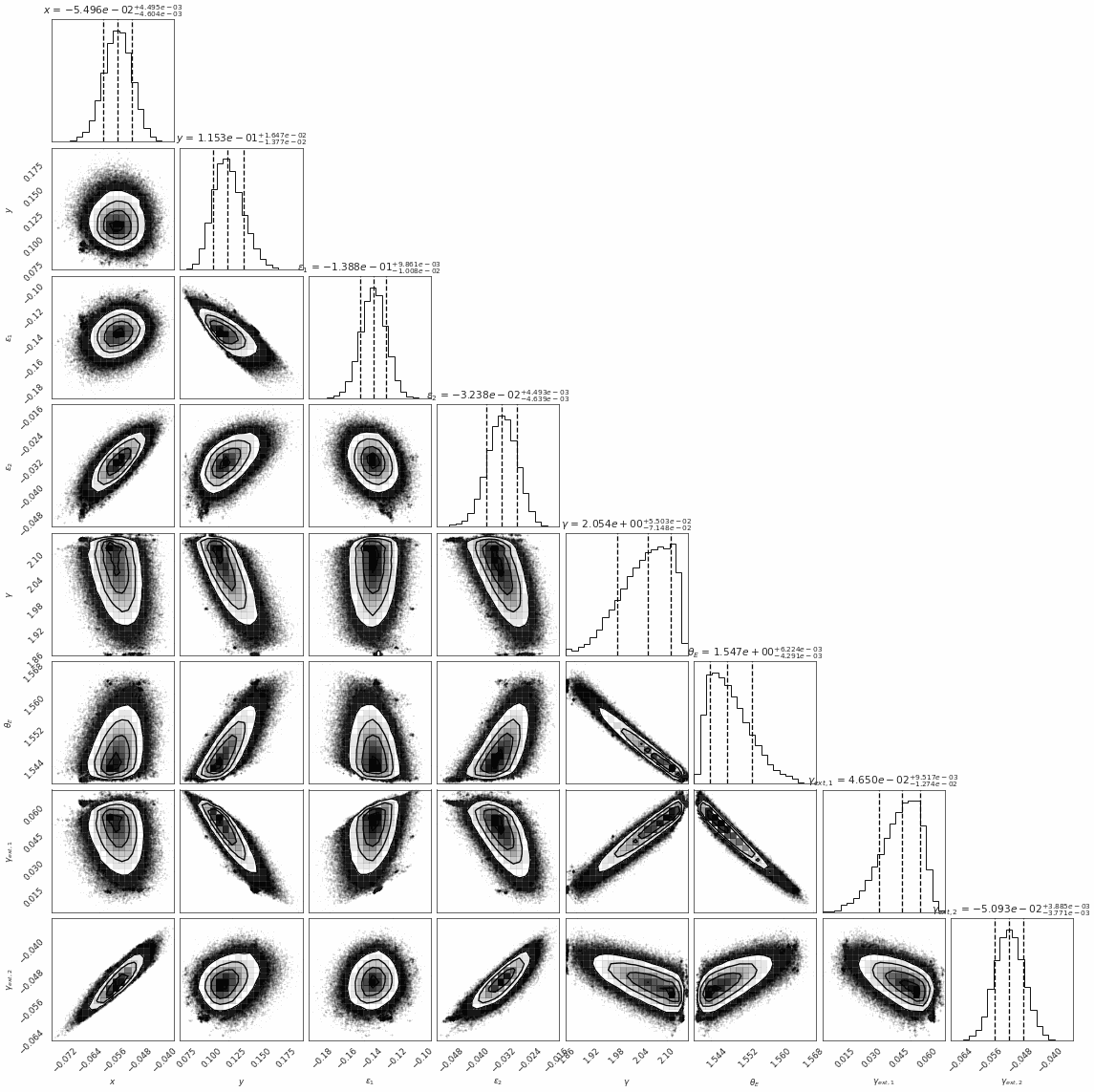}
    \caption{Corner plot of the eight mass parameters for \desitwothreefour}
    \label{fig:234cornerplotmass}
\end{figure}

We fit for the light intensities, $I_e$ (instead of using linear inversion).
We achieved a reduced $\chi^2~=~0.8789$.
The residuals are excellent (Figure \ref{fig:bestfitmodeldavid}).
The best-fit lens and external shear parameters are presented in Table \ref{tab:234best-fit-mass-params} and their sampling results are shown in Figure \ref{fig:234cornerplotmass}. 
The best-fit light parameters are presented in Table~\ref{tab:234best-fit-light-params}.

\vspace{-.3cm}

\begin{deluxetable}{lccccccc}[H]
\tabletypesize{\scriptsize}
\tablecaption{Best-fit mass parameters for \desitwothreefour.
\label{tab:234best-fit-mass-params}}
\renewcommand{\arraystretch}{1.4}
\tablehead{
    \colhead{$\theta_E$} &
    \colhead{$\gamma$} &
    \colhead{$\epsilon_1$} &
    \colhead{$\epsilon_2$} &
    \colhead{$x$} &
    \colhead{$y$} &
    \colhead{$\gamma_{ext, 1}$} &
    \colhead{$\gamma_{ext, 2}$} 
}
\startdata
$1.5466_{-0.0043}^{+0.0062}$& 
$2.054_{-0.071}^{+0.055}$&	
$-0.139\pm0.010$&
$-0.0324_{-0.0046}^{+0.0045}$& 
$-0.0550_{-0.0046}^{+0.0045}$&
$0.115_{-0.014}^{+0.017}$&	
$0.046_{-0.013}^{+0.010}$&	
$-0.0509_{-0.0038}^{+0.0039}$ \\
\enddata
\end{deluxetable}

\begin{deluxetable}{l|rr|rr}[H]
\tablecaption{Best-fit light parameters for \desitwothreefour.}
\label{tab:234best-fit-light-params}
\renewcommand{\arraystretch}{1.4}
\setlength{\tabcolsep}{8pt}
\tablehead{
    \colhead{Parameter} & \multicolumn{2}{c}{Lens light} & \multicolumn{2}{c}{Source light} \\
    \colhead{} & \colhead{Comp 1} & \colhead{Comp 2} & \colhead{Comp 1} & \colhead{Comp 2}
}\startdata
$\bm{R_e}$ &
$3.87_{-0.28}^{+0.33}$&$0.422_{-0.035}^{+0.039}$&$0.0341^{+0.0027}_{-0.0026}$&$0.0475_{-0.0044}^{+0.0048}$\\[4pt]
$\bm{n}$ &               
$2.22_{-0.17}^{+0.19}$&$5.80_{-0.26}^{+0.28}$&$5.71_{-0.42}^{+0.22}$&
$7.82_{-0.30}^{+0.14}$\\[4pt]
$\bm{\epsilon_{1}}$ &    
$0.047_{-0.012}^{+0.013}$&$-0.0969_{-0.0035}^{+0.0034}$&$0.196\pm0.022$&
$0.255^{+0.029}_{-0.039}$\\[4pt]
$\bm{\epsilon_{2}}$ &    
$-0.259_{-0.021}^{+0.020}$&$0.0217_{-0.0047}^{+0.0049}$&$0.216_{-0.027}^{+0.026}$& 
$0.2962^{+0.0029}_{-0.0060}$\\[4pt]
$\bm{x}$ &
$0.136_{-0.019}^{+0.022}$&$-0.0020\pm0.0005$&$-0.0385_{-0.0047}^{+0.0045}$& 
$-0.0939_{-0.0074}^{+0.0071}$\\[4pt]
$\bm{y}$ &
$0.0706_{-0.0087}^{+0.0091}$&$0.0340\pm0.0006$&$-0.298_{-0.024}^{+0.033}$&
$-0.281_{-0.024}^{+0.032}$\\[4pt]     
$\bm{I_e}$ &
$2.05_{-0.25}^{+0.28}$ & $111_{-15}^{+17}$  & $198_{-28}^{+34}$ & $14.7_{-2.2}^{+2.6}$  \\
\enddata
\end{deluxetable}

\begin{figure}[H]
    \centering
    \includegraphics[trim={0.45cm 0.45cm 0.45cm 0.4cm}, clip,width=0.7\linewidth]{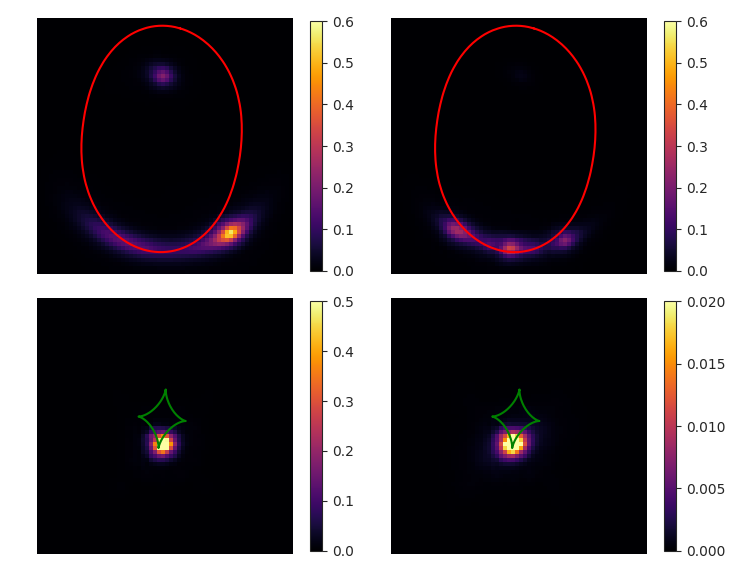}
    \caption{Source light \ser components in the lensed and unlensed plane for \desitwothreefour from left to right,  component 1 and 2, respectively.}
    \label{fig:234components}
\end{figure}

Figure \ref{fig:234components} shows the two components of the source, with the first one (left column) being considerably brighter than the second (right column).

For MAP, we used 10,000 samples ($n_\mathrm{MAP}$ in Gu22) and 2000 steps.
It ran for 2 minutes and 33 seconds. 
SVI was configured with 4000 samples ($n_\mathrm{VI}$ in Gu22) and 4000 steps, using a quadratic transition through the first 3500 steps from an initial learning rate of $0$ to a $-3 \times 10^{-3}$ final rate (for details, see Gu22). It ran for 2 minutes and 41 seconds. 
For HMC, we used 8 chains, 1500 burn-in steps, and 30,000 total results, with 20 steps between samples. The target acceptance probability was set at 0.75, the initial step size at $\epsilon = 0.03$, $L = 3$ initial leapfrog steps, and a maximum of 30 leapfrog steps. It ran for 11 hours and 43 minutes.
\emph{All \rhat values are below 1.01.} We achieve an average $\hat{R}$ of 1.0020, with the maximum value being 1.0087.
We note that for this system we tested how low we could drive $\hat{R}$. If we were satisfied with merely “presentable’’ corner plots, or even simply requiring $\hat{R} < 1.1$, the total execution time would have been an order of magnitude lower.

 We determine the Einstein radius of the lens to be $\theta_E = 1.5466''$$^{+0.0062}_{-0.0043}$ and the slope of the power law mass profile to be $\gamma = 2.0540^{+0.0553}_{-0.0714}$. The total mass within the critical curve is $8.312^{+0.108}_{-0.062} \times 10^{11} M_{\odot}$. 
 The magnification 
results using the three methods in the Appendix of Paper~I are shown in Table~\ref{tab:magnifications_234}.
The uncertainties for the magnification values of image B are the largest due to its proximity to the critical curve.

\begin{deluxetable}{lccccc}[H]
    \tabletypesize{\footnotesize}
    \tablecaption{Magnification estimates for \desitwothreefour using three methods. Values are shown for images A through D. \label{tab:magnifications_234}}
    \tablewidth{0pt}
    \tablehead{
        \colhead{Method} & \colhead{$A$} & \colhead{$B$} & \colhead{$C$} & \colhead{$D$} & \colhead{Total}
    }
    \startdata
    $1^{\text{st}}$ & $19.8\pm2.2$  & $38.5\pm8.2$ & $7.13\pm0.22$ & $0.49\pm0.22$ & $67.6\pm8.5$ \\
    $2^{\text{nd}}$ & $17.3 \pm1.3$ & $28.7\pm5.2$ & $8.14\pm0.91$ & $0.23\pm0.15$ & $54.6\pm5.4$ \\
    $3^{\text{rd}}$ & $18.8 \pm2.0$ & $34.7\pm5.3$ & $7.17\pm0.49$ & $0.49\pm0.09$ & $62.3\pm5.7$ \\
    \enddata
\end{deluxetable}


\vspace{-1.cm}
\subsection{\desitwothreeeight}\label{sec:desi238}
\href{https://www.legacysurvey.org/viewer/?ra=238.5690&dec=+04.7276&layer=ls-dr10-grz&pixscale=0.262&zoom=16}{\desitwothreeeight} was discovered in H21, with a ResNet probability of 1.00, and a human score of 2.5 (out of 4.0), corresponding to a C grade. As stated in H21,  all candidates with a human inspection score $\geq 2.5$ are likely lensing systems. The lens redshift, \zd = 0.7768 is from DESI DR1 (based on the 4000 \ang break, Ca~H\&K absorption). The source redshift, \zs = 1.72 is from Keck NIRES (Paper~VI, in prep.).

\begin{figure}[H]
    \centering
    \includegraphics[trim={0cm 0cm 0cm 0.4cm}, clip, width=0.637\linewidth]{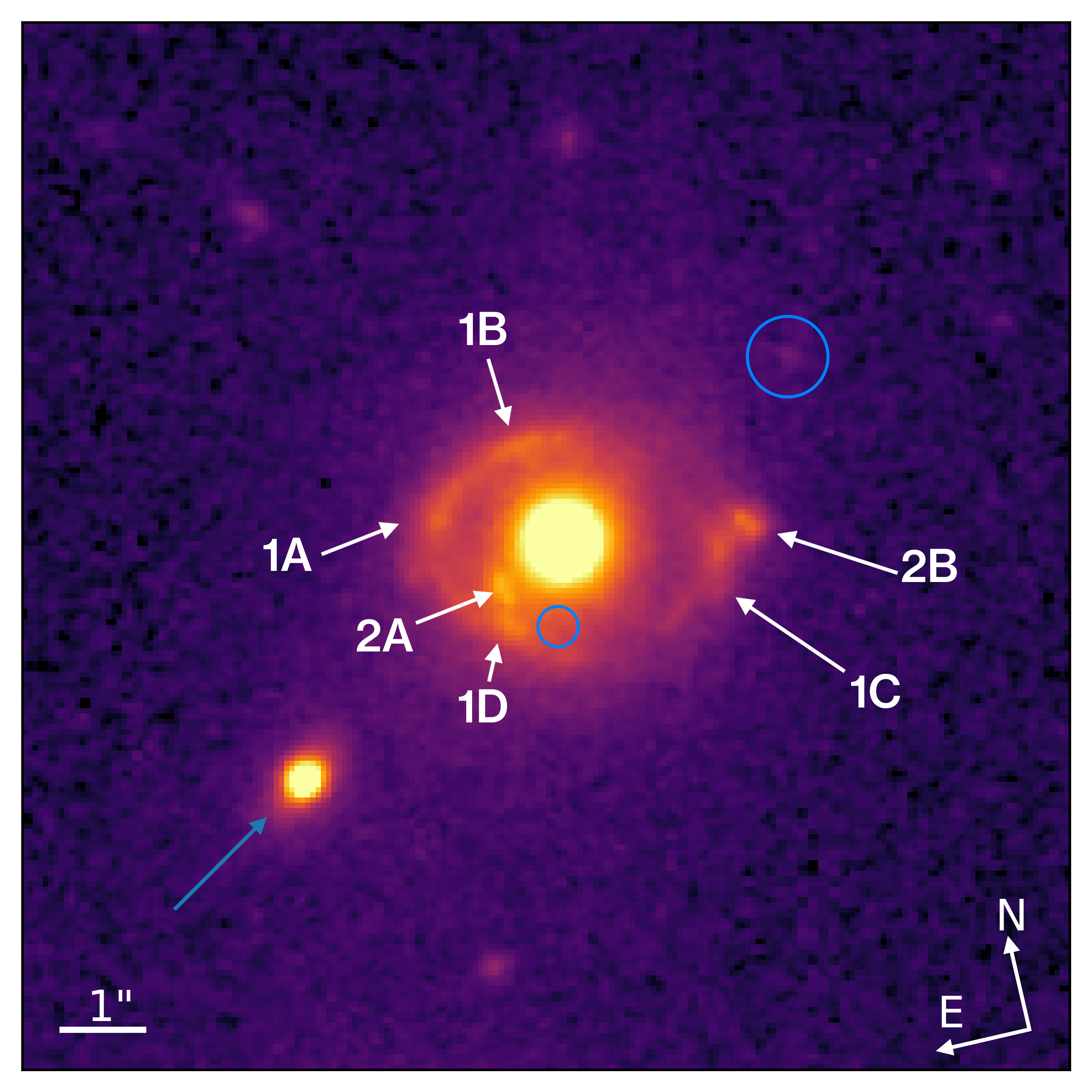}
    \caption{\desitwothreeeight. We mask out pixels in the two circles, and model the light of the environmental galaxy in the lower left (blue arrow). The two sources are lensed into a quad (images labeled 1A, 1B, 1C, and 1D) and a double (2A and 2B).} 
    \label{fig:238cutout}
\end{figure}

This system has an Einstein radius of $\theta_E=1.5''$, slightly larger than the \tE values of the simulated systems in the original \gigal paper ($1.1''$). Due to its relatively small size in the set of systems modeled in this work, uncluttered environment 
(Figure~\ref{fig:238cutout}), it is relatively straightforward to model. We use a cutout with $120\times 120$ pixels and a $27\times27$ pixel empirical PSF, generated using the procedure outlined in Section~\ref{sec:psf_section}.

We model the lens light with two elliptical \ser profiles. 
We mask out two small objects, one to the northwest (upper right) and another near a lensed arc (circles in Figure~\ref{fig:238cutout}). We model the source using two elliptical \ser profiles. In addition, we model the light but not mass of an environmental galaxy to the southeast with an elliptical \ser profile (arrow in Figure~\ref{fig:238cutout}), for a total of 38 parameters. 

\begin{figure}[H]
    \centering
    \includegraphics[trim={6.2cm 10.4cm 4.5cm 0.9cm}, clip, width=1.02\linewidth]{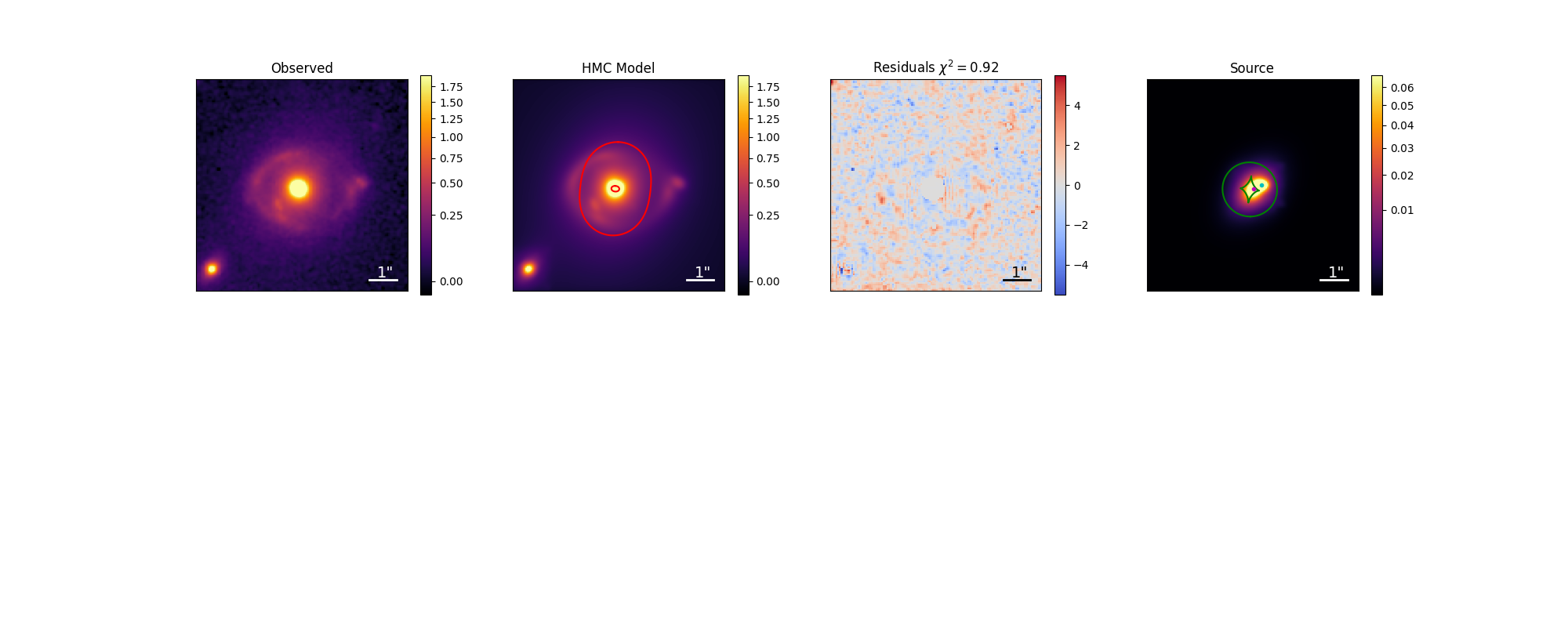}
    \caption{Best-fit Model \desitwothreeeight. From left to right, we show: the observed Hubble image, our best-fit model with the critical curve, the reduced residuals, and the unlensed source with the caustic. For calculating $\chi^2$ we mask a circular region with a width of 12~pixels ($0.8''$) at the center of the lensing galaxy. This area is not masked during modeling. The unmasked $\chi^2$ is 1.004. Only the two circles in Figure~\ref{fig:238cutout} are masked during modeling.}
    \label{fig:238model}
\end{figure}

\begin{figure}[H]
    \centering
   \includegraphics[width=0.85\linewidth]{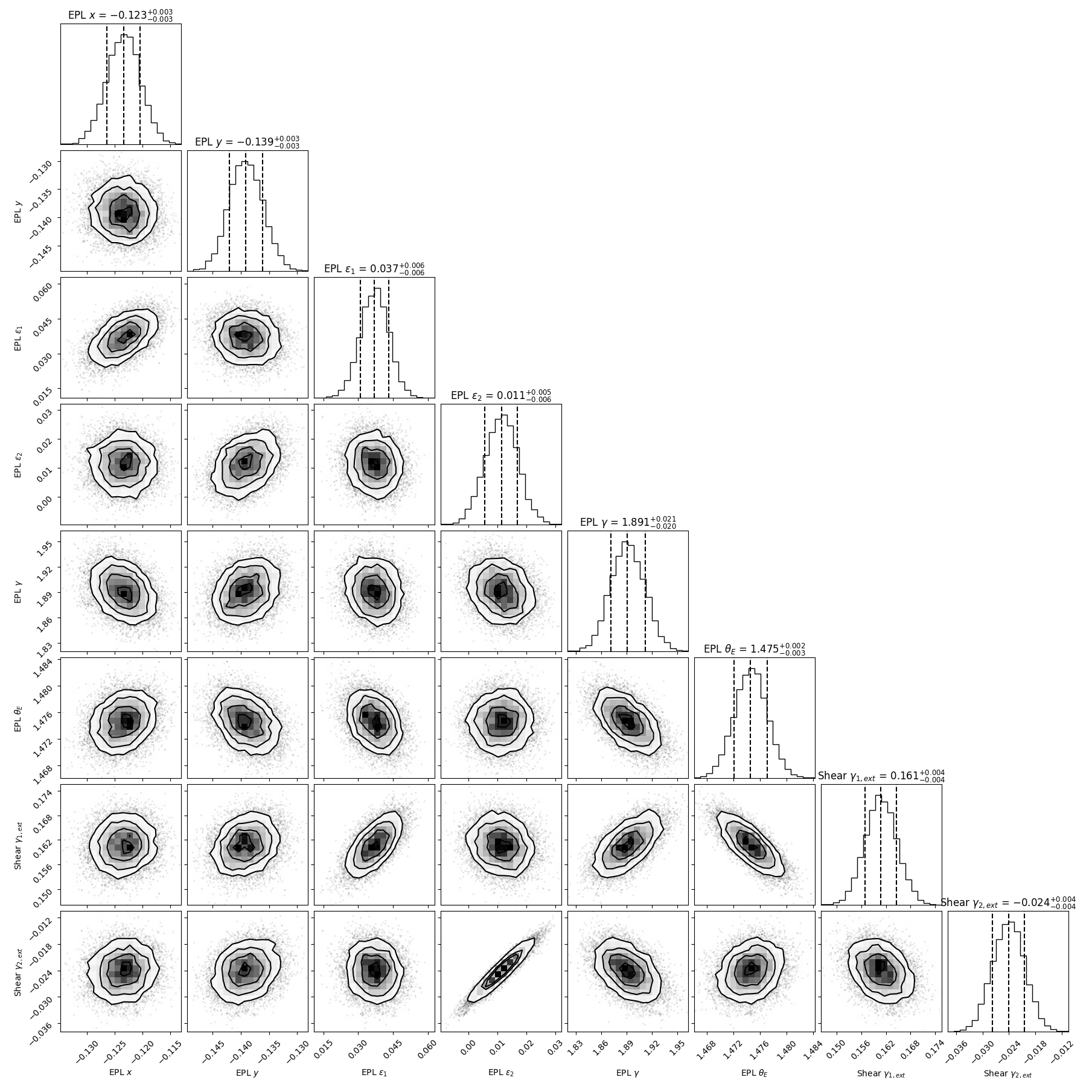}
    \caption{Corner plot of the eight mass parameters for \desitwothreeeight.}
    \label{fig:238masscorner}
\end{figure}

This system is modeled with linear inversion to solve for the light intensities. 
We achieved a reduced $\chi^2=0.95$ with superb residuals (Figure \ref{fig:238model}).  
The best-fit lens and external shear parameters are presented in Table \ref{tab:238mass} and their sampling results are shown in Figure \ref{fig:238masscorner}. 
The best-fit light parameters are presented in Table~\ref{tab:238light}.

\begin{deluxetable}{lccccccc}[H]
\tabletypesize{\scriptsize}
\tablecaption{Best-fit mass parameters for \desitwothreeeight.
\label{tab:238mass}}
\renewcommand{\arraystretch}{1.4}
\tablehead{
    \colhead{$\theta_E$} &
    \colhead{$\gamma$} &
    \colhead{$\epsilon_1$} &
    \colhead{$\epsilon_2$} &
    \colhead{$x$} &
    \colhead{$y$} &
    \colhead{$\gamma_{ext, 1}$} &
    \colhead{$\gamma_{ext, 2}$} 
}
\startdata
$1.475_{-0.003}^{+0.002}$&      
$1.89\pm0.02$&	
$0.037\pm0.006$&  
$0.011_{-0.006}^{+0.005}$&  
$-0.123\pm0.003$&  
$-0.139\pm0.003$&	
$-0.161\pm0.004$&	
$-0.024\pm0.004$\\ 
\enddata
\end{deluxetable}

\begin{deluxetable}{l|rr|rr|c}[H]
\tablecaption{Best-fit light parameters for \desitwothreeeight}
\label{tab:238light}
\renewcommand{\arraystretch}{1.4}
\setlength{\tabcolsep}{8pt}
\tablehead{
    \colhead{Parameter} & 
    \multicolumn{2}{c}{Lens} & 
    \multicolumn{2}{c}{Source} & 
    \colhead{Environmental galaxy} \\
    \colhead{} & 
    \colhead{Comp 1} & \colhead{Comp 2} & 
    \colhead{Comp 1} & \colhead{Comp 2} & 
    \colhead{}
}
\startdata
$\bm{R_e}$ & $0.452_{-0.006}^{+0.007}$ & $0.63\pm{0.01}$ & $0.072\pm{0.002}$ & $0.247\pm{0.007}$ & $0.058\pm{0.002}$ \\[2pt]
$\bm{n}$ & $2.375_{-0.06}^{+0.09}$ & $3.884_{-0.06}^{+0.05}$ & $0.31_{-0.05}^{+0.07}$ & $1.18\pm{0.04}$ & $9.95_{-0.08}^{+0.04}$ \\[2pt]
$\bm{\epsilon_{1}}$ & $-0.122\pm{0.006}$ & $-0.095\pm{0.003}$ & $0.30\pm{0.02}$ & $-0.053\pm{0.009}$ & $-0.133\pm{0.006}$ \\[2pt]
$\bm{\epsilon_{2}}$ & $0.147\pm{0.005}$ & $0.103_{-0.002}^{+0.003}$ & $-0.06\pm{0.02}$ & $0.207\pm{0.008}$ & $0.289_{-0.006}^{+0.005}$ \\[2pt]
$\bm{x}$ & $-0.0975\pm{0.0003}$ & $-0.108\pm{0.001}$ & $0.33\pm{0.01}$ & $0.042\pm{0.003}$ & $-3.3155\pm{0.0004}$ \\[2pt]
$\bm{y}$ & $-0.0330_{-0.0010}^{+0.0004}$ & $-0.067\pm0.002$ & $0.013\pm0.005$ & $-0.151\pm0.003$ & $-3.0873\pm0.0006$ \\[2pt]
\enddata
\end{deluxetable}

\begin{figure}[H]
    \centering
    \includegraphics[width=0.5\linewidth]{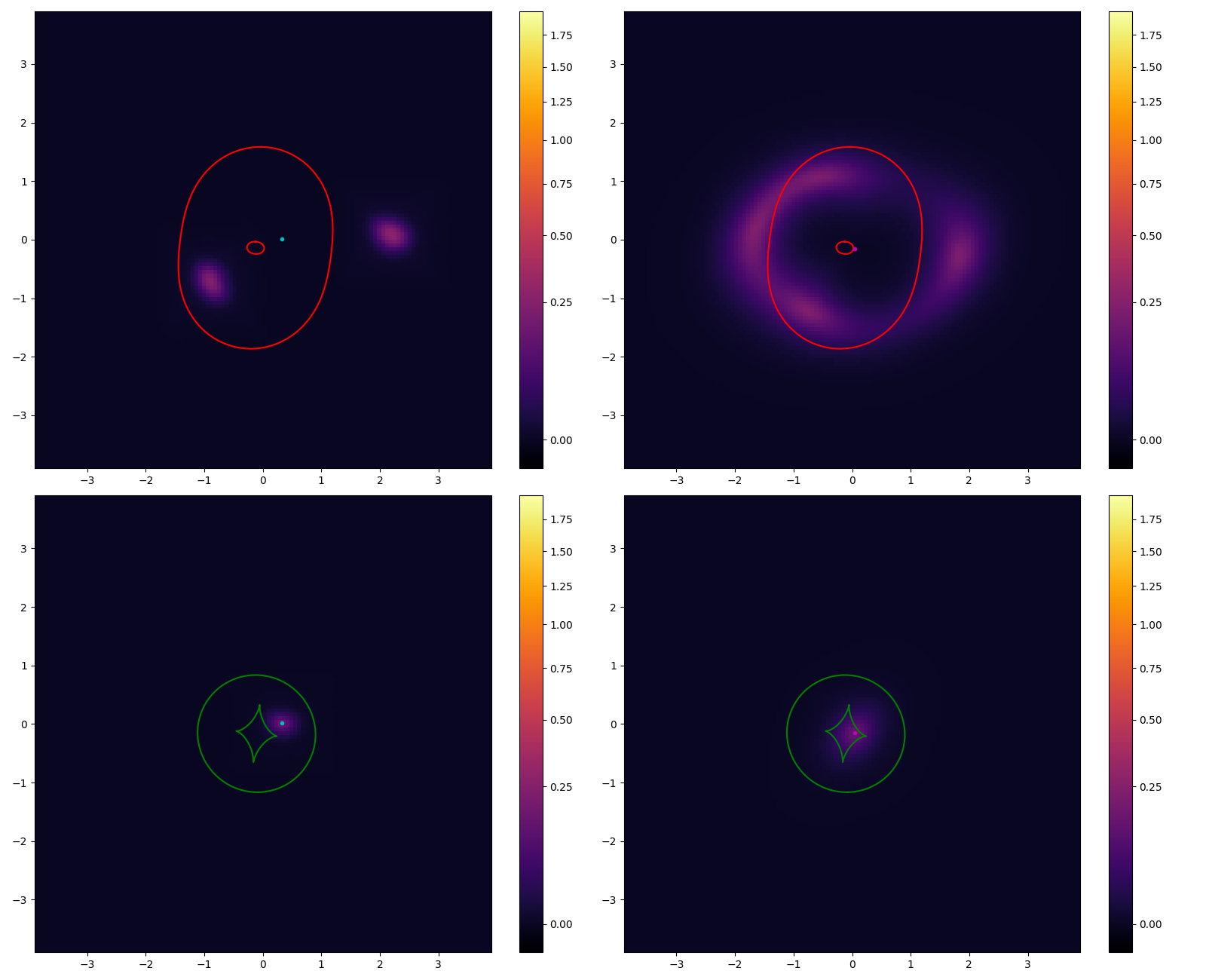}
    \caption{From left to right, source component 1 and 2, respectively, in the lensed and unlensed plane for \desitwothreeeight.}
    \label{fig:238-comps}
\end{figure}

We show the two \ser components for the source in (Figure~\ref{fig:238-comps}). One source (component 1) is lensed into a double, while the other (component 2) is lensed into an Einstein cross.
The redshift for component 2 is $\zs = 1.7210$. Component~1 was targeted by DESI and Keck~NIRES; however, so far we have not detected any clear spectral features. Given the proximity of the unlensed positions of these two components, they possibly indicate a source with two components or a pair of merging galaxies.

For MAP, we used $n_\mathrm{MAP}$ = 4000 samples over 1000 steps, with square root transition over all 1000 iterations from an initial learning rate of $-1\times10^{-2}$ to a final value of $-1\times10^{-4}$. It ran for five minutes and 55 seconds. 
For SVI, we used $n_\mathrm{VI}=500$  samples over 6000 iterations, with a square root transition 
from an initial learning rate of $1\times10^{-6}$ to a final value of $1\times10^{-3}$. This stage lasted 10 minutes. 
Finally, for HMC we employed 16 chains over 160,000 steps, sampling every 100 steps for a total of 1,200 samples per chain after the 400 sample burn-in. We also discard the first half of the samples and only use the latter 600/chain for reporting the posterior distribution. The target acceptance probability was set at 0.75, with an initial step size of \( \epsilon = 0.03, L = 3 \) initial leapfrog steps, and a maximum of 30 leapfrog steps. HMC ran for 193 minutes and 36 seconds. The total execution time for all three modeling steps was 209 minutes and 31 seconds.
We have successfully sampled this system with $\hat R$ values below 1.1 for all parameters.

The best-fit Einstein radius $\theta_E=1.475''$$_{-0.003}^{+0.002}$ and the mass profile slope, $\gamma=1.89\pm0.02$. The total projected mass within the critical curve is $9.91\pm0.04\times10^{11}M_{\odot}$.  For magnifications, we use three different methods as described in the Appendix of Paper~I. The results are shown in Table~\ref{tab:238magnification}.

\begin{deluxetable}{lccccc||ccc}[H]
\label{tab:238magnification}
\tablecaption{Magnifications for \desitwothreeeight using three methods. We drew 250 random samples from the HMC chains for the calculation.}
\tabletypesize{\footnotesize}
\renewcommand{\arraystretch}{1.4}
\tablehead{
    \colhead{Method} &
    \colhead{$1A$} &
    \colhead{$1B$} &
    \colhead{$1C$} &
    \colhead{$1D$} & 
    \colhead{Total 1} &
    \colhead{$2A$} & 
    \colhead{$2B$} & 
    \colhead{Total 2}
}
\startdata
$1^{\text{st}}$ & $8.4\pm0.4$ & $-5.9\pm0.4$ & $3.7\pm0.1$ & $-3.9\pm0.2$ & $22.0\pm1.1$ & $-2.4\pm0.1$ & $2.9\pm0.1$ & $5.3\pm0.2$\\
$2^{\text{nd}}$ & $8.5\pm0.4$ & $-6.0\pm0.4$ & $3.8\pm0.2$ & $-4.1\pm0.2$ & $22.5\pm1.1$ & $-2.5\pm0.1$ & $2.9\pm0.1$ & $5.4\pm0.2$ \\
$3^{\text{rd}}$ & $8.1\pm0.4$ & $-5.8\pm0.3$ & $3.7\pm0.1$ & $-3.9\pm0.2$ & $21.5\pm1.0$ & $-2.4\pm0.1$ & $2.9\pm0.1$ & $5.3\pm0.2$ \\
\enddata
\end{deluxetable}

\begin{figure}[H]
    \centering
   \includegraphics[width=\linewidth]{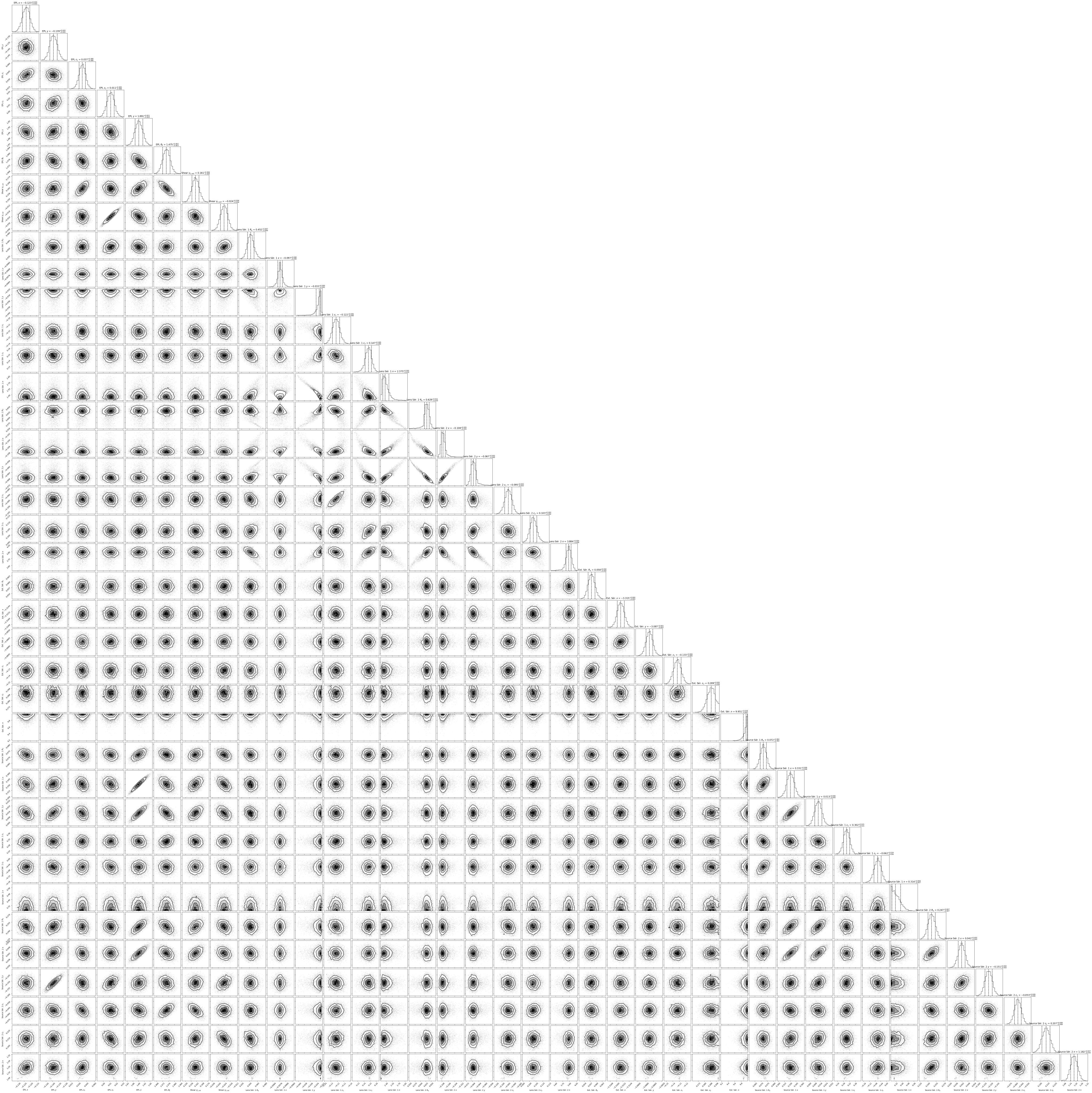}
    \caption{Corner plot of all 38 model parameters for \desitwothreeeight}
    \label{fig:238full-corner}
\end{figure}

\newpage
Finally, for this system, we even include the environmental galaxy in the fully forward modeling, with all 38 parameters sampled simultaneously, with no fixed parameters. The complete posterior distributions are shown in Figure~\ref{fig:238full-corner}.

\newpage
\section{Discussion}\label{sec:discussion}
\subsection{Full Forward Modeling and Convergence Metrics}
In this work, we have achieved two main goals: 1) full forward modeling for all six systems---all parameters describing the lens and the source are sampled simultaneously in a single inference. 
2) All six systems meet explicit, quantitative criteria for statistical convergence based on both the Gelman--Rubin statistic ($\hat{R}$) and effective sample size (ESS). 
This is despite the fact that the systems modeled in this work span a wide range of image morphology, Einstein radii, and environmental complexity (Table~\ref{tab:model-sys}). 
Together with the system in Paper~I, this brings the total to seven. 

This demonstrates that fully Bayesian modeling of complex, bright, galaxy-scale lenses with high-resolution space-based imaging can yield robust inferences, even for highly nonlinear strong-lensing models with $\gtrsim40$ parameters determined \emph{simultaneously}. These results underscore the steadily increasing maturity of strong gravitational lens modeling and provide a foundation for exploiting the thousands of high-quality lenses with high resolution soon to arrive from \hst, \euc, \emph{Roman}, and \jwst \citep[e.g.,][]{nightingale2025, silver2025}.

This approach stands in contrast to ``staged'' modeling strategies (e.g., optimize lens light first, then fix those parameters before optimizing mass, and so on). While computationally convenient, such sequential procedures have two major drawbacks: 1) they effectively impose more restrictive priors and risk converging to a local rather than global optimum, where the posterior probability mass is the greatest; and
2) they can artificially suppress the posterior variance of key quantities such as the mass density slope, \gma. By jointly modeling and sampling \emph{all} parameters pertaining to the lens mass, lens and source light, and external shear, our posterior space is not constrained by the order in which individual components are optimized.

\subsection{Incorporation of Environmental Galaxies in Full Forward Modeling}

We have, in fact, gone beyond simultaneously modeling the lens and source parameters in this work. For two systems in this sample---\desionesixfive and \desitwothreeeight---we also include the light of the environmental galaxies directly in the full forward modeling. \desitwofiveseven does not have an environmental galaxy bright enough to be modeled. For each of the remaining four systems, we fix only five or six light profile parameters associated with a single environmental galaxy, except in the most complex case, \desitwofoursix (\S~\ref{sec:desi246}), which has 78 parameters in total. For this system, 28 light profile parameters for three environmental galaxies are fixed for sampling tractability, while the 50 parameters pertaining to the lens and the source are sampled simultaneously---an unprecedented scale for a galaxy-scale lens using \hst-quality imaging.

Our next step is to incorporate the environmental galaxies in these four systems into full forward modeling, then extend this approach to all 26 galaxy-scale lenses in our \hst\ program (see Paper~I), and ultimately to group- and cluster-scale lenses \citep[e.g.,][]{sheu2024}.

\subsection{Mass Slope Evolution}
The predicted evolution of $\gma$ with redshift is subtle, and robust inference requires reliable characterization of the intrinsic scatter in these measurements. Published $\gma$--$z$ distributions, including the recent AGEL results of \citet{sahu2024}, reveal a tight locus of slopes near the isothermal value at $z_d \lesssim 0.3$, but a wider distribution at $0.5 \lesssim z_d \lesssim 0.8$. Our results reinforce this broader picture: lenses in this redshift range span $\gma = 1.4$--$2.5$, comparable to the full range seen in the literature.
We also report the mass density slope for an elliptical galaxy at $z_d > 1$: $\gma = 2.616 \pm 0.17$ for DESI~J246.0062+01.4836 at $\zd = 1.092$.

\subsection{Lens Modeling Time}
Finally, we note that the modeling times quoted in this work reflect the time required to achieve convergence for all parameters sampled according to the $\hat{R}$ and ESS criteria. If only ``presentable'' corner plots and reasonable residuals were desired,
execution times would be an order of magnitude or more shorter. The timing benchmarks presented here should therefore be interpreted as practical estimates for rigorous, convergence-validated modeling.

Together, these advances establish a framework that can support a new generation of precision strong lensing studies. With validated posteriors, the data presented here---and especially larger samples to come---can confidently address high-impact questions in galaxy evolution and cosmology.


\section{Conclusion}\label{sec:conclusion}

In this Paper V of the DESI Strong Lens Foundry series, 
we present the first sample of strong lenses with \hst modeled with full forward modeling and quantitative convergence validation ($\hat{R}$ and ESS).
We use the open-source \gigal.
For each of the seven systems in this sample, we achieve $\hat{R} < 1.1$ and ESS $> 10{,}000$ for every lens and source parameter and establish a methodological foundation for large-sample studies of galaxy-scale strong lenses.
Reporting quantitative sampling-convergence diagnostics such as $\hat{R}$ and ESS has become standard in cosmological inference, and strong-lensing analyses are now positioned to benefit from the same transparent practices.

In this sample, we observe increased scatter in $\gamma$ at intermediate redshifts ($z \sim 0.5\, \text{--}\, 0.8$), consistent with earlier reports. 
Our sample also extends mass density slope inference for elliptical galaxies beyond $z =1$, with a convergence-validated measurement of $\gma = 2.62 \pm 0.17$ for DESI~J246.0062+01.4836 at $\zd = 1.092$.
In our view, a more definitive conclusion on the mass slope evolution awaits a uniform, convergence-tested, and full forward modeling effort across a large sample of systems---including the known lenses---so that statistical uncertainties are robustly characterized and intrinsic redshift evolution can then be distinguished from modeling systematics and galaxy-to-galaxy diversity.

Modeling real systems with high-resolution data is inherently more challenging than fitting simulated systems (Gu22), as we previously demonstrated in our application of \gigal\ to \hst\ observations \citep[][Paper~I]{huang2025a}. The present analysis further shows that large \tE, complex environments, and source light structure
can significantly increase the cutout size and number of model parameters, making efficient sampling more challenging.
We nevertheless confirm that with \gigal, as in Paper~I, convergence-validated, full forward modeling is scalable to the dimensionality required by real systems. Continued improvements in GPU-parallel inference and adaptive covariance learning (N.~Ratier-Werbin et al. in prep.)
will further expand feasibility to much larger lens samples.
With continued hardware and software advances in GPU-accelerated modeling and flexible parameterizations, strong lensing is poised to deliver precision constraints on galaxy evolution, dark matter physics, and cosmology in the era of thousands, if not tens of thousands, of high-quality strong lenses.




\newpage
\section*{Acknowledgment}\label{sec:acknowledgement}
We thank Peter Harrington, Nestor Demeure, Steve Farrell, and Rollin Thomas at the National Energy Scientific Computing Center (NERSC) for their consultation and advice.
This research used resources of the National Energy Research Scientific Computing Center (NERSC), a U.S. Department of Energy Office of Science User Facility operated under Contract No. DE-AC02-05CH11231 and the Computational HEP program in The Department of Energy's Science Office of High Energy Physics provided resources through the ``Cosmology Data Repository" project (Grant \#KA2401022).
X.H. acknowledges the University of San Francisco Faculty Development Fund. 
A.D.'s research is supported by National Science Foundation's National Optical-Infrared Astronomy Research Laboratory, 
which is operated by the Association of Universities for Research in Astronomy (AURA) under cooperative agreement with the National Science Foundation.

This paper is based on observations at Cerro Tololo Inter-American Observatory, National Optical
Astronomy Observatory (NOAO Prop. ID: 2014B-0404; co-PIs: D. J. Schlegel and A. Dey), which is operated by the Association of
Universities for Research in Astronomy (AURA) under a cooperative agreement with the
National Science Foundation.

This project used data obtained with the Dark Energy Camera (DECam),
which was constructed by the Dark Energy Survey (DES) collaboration.
Funding for the DES Projects has been provided by 
the U.S. Department of Energy, 
the U.S. National Science Foundation, 
the Ministry of Science and Education of Spain, 
the Science and Technology Facilities Council of the United Kingdom, 
the Higher Education Funding Council for England, 
the National Center for Supercomputing Applications at the University of Illinois at Urbana-Champaign, 
the Kavli Institute of Cosmological Physics at the University of Chicago, 
the Center for Cosmology and Astro-Particle Physics at the Ohio State University, 
the Mitchell Institute for Fundamental Physics and Astronomy at Texas A\&M University, 
Financiadora de Estudos e Projetos, Funda{\c c}{\~a}o Carlos Chagas Filho de Amparo {\`a} Pesquisa do Estado do Rio de Janeiro, 
Conselho Nacional de Desenvolvimento Cient{\'i}fico e Tecnol{\'o}gico and the Minist{\'e}rio da Ci{\^e}ncia, Tecnologia e Inovac{\~a}o, 
the Deutsche Forschungsgemeinschaft, 
and the Collaborating Institutions in the Dark Energy Survey.
The Collaborating Institutions are 
Argonne National Laboratory, 
the University of California at Santa Cruz, 
the University of Cambridge, 
Centro de Investigaciones En{\'e}rgeticas, Medioambientales y Tecnol{\'o}gicas-Madrid, 
the University of Chicago, 
University College London, 
the DES-Brazil Consortium, 
the University of Edinburgh, 
the Eidgen{\"o}ssische Technische Hoch\-schule (ETH) Z{\"u}rich, 
Fermi National Accelerator Laboratory, 
the University of Illinois at Urbana-Champaign, 
the Institut de Ci{\`e}ncies de l'Espai (IEEC/CSIC), 
the Institut de F{\'i}sica d'Altes Energies, 
Lawrence Berkeley National Laboratory, 
the Ludwig-Maximilians Universit{\"a}t M{\"u}nchen and the associated Excellence Cluster Universe, 
the University of Michigan, 
{the} National Optical Astronomy Observatory, 
the University of Nottingham, 
the Ohio State University, 
the OzDES Membership Consortium
the University of Pennsylvania, 
the University of Portsmouth, 
SLAC National Accelerator Laboratory, 
Stanford University, 
the University of Sussex, 
and Texas A\&M University.


\software{
    TensorFlow \citep{TensorFlow},
    TensorFlow Probability \citep{dillon2017a}, 
    JAX \citep{bradbury2018a}, 
    Optax \citep{optax2020},
    lenstronomy \citep{birrer2018a},
    Matplotlib \citep{hunter2007a},
    photutils \citep{bradley2023a}
    seaborn \citep{waskom2021a},
    corner.py \citep{foreman2016a},
    NumPy \citep{harris2020a}
} 

\bibliographystyle{aasjournal}

\bibliography{DLF-papre-V}

@ARTICLE{odonell2025,
       author = {{O'Donnell}, Jackson H. and {Jeltema}, Tesla E. and {Roberts}, M. Grant and {Nightingale}, James and {Flowers}, Abigail and {Aldas}, Dhruv},
        title = "{A Constraint on Dark Matter Self-Interaction from Combined Strong Lensing and Stellar Kinematics in MACS J0138-2155}",
      journal = {arXiv e-prints},
         year = 2025,
        month = aug,
          eid = {arXiv:2508.20179},
        pages = {arXiv:2508.20179},
          doi = {10.48550/arXiv.2508.20179},
archivePrefix = {arXiv},
       eprint = {2508.20179},
 primaryClass = {astro-ph.CO},
       adsurl = {https://ui.adsabs.harvard.edu/abs/2025arXiv250820179O}
}

@ARTICLE{nightingale2025,
       author = {{Nightingale}, James W. and {Mahler}, Guillaume and {McCleary}, Jacqueline and {He}, Qiuhan and {Hogg}, Natalie B. and {Amvrosiadis}, Aristeidis and {Gozaliasl}, Ghassem and {Mercier}, Wilfried and {Scognamiglio}, Diana and {Berman}, Edward and {Leroy}, Gavin and {Liu}, Daizhong and {Massey}, Richard J. and {Shuntov}, Marko and {von Wietersheim-Kramsta}, Maximilian and {Franco}, Maximilien and {Paquereau}, Louise and {Ilbert}, Olivier and {Allen}, Natalie and {Toft}, Sune and {Akins}, Hollis B. and {Casey}, Caitlin M. and {Kartaltepe}, Jeyhan S. and {Koekemoer}, Anton M. and {McCracken}, Henry Joy and {Rhodes}, Jason D. and {Robertson}, Brant E. and {Drakos}, Nicole E. and {Faisst}, Andreas L. and {Jin}, Shuowen},
        title = "{The COSMOS-Web Lens Survey (COWLS) I: discovery of >100 high redshift strong lenses in contiguous JWST imaging}",
      journal = {\mnras},
         year = 2025,
        month = oct,
       volume = {543},
       number = {1},
        pages = {203-222},
          doi = {10.1093/mnras/staf1253},
archivePrefix = {arXiv},
       eprint = {2503.08777},
 primaryClass = {astro-ph.GA},
       adsurl = {https://ui.adsabs.harvard.edu/abs/2025MNRAS.543..203N}
}

@ARTICLE{karp2025,
       author = {{Karp}, Juliana S.~M. and {Schlegel}, David J. and {Huang}, Xiaosheng and {Padmanabhan}, Nikhil and {Bolton}, Adam S. and {Storfer}, Christopher J. and {Aguilar}, J. and {Ahlen}, S. and {Bailey}, S. and {Bianchi}, D. and {Brooks}, D. and {Castander}, F.~J. and {Claybaugh}, T. and {Cuceu}, A. and {de la Macorra}, A. and {Della Costa}, J. and {Doel}, P. and {Font-Ribera}, A. and {Forero-Romero}, J.~E. and {Gazta{\~n}aga}, E. and {Gontcho}, S. Gontcho A and {Gutierrez}, G. and {Honscheid}, K. and {Ishak}, M. and {Jimenez}, J. and {Joyce}, R. and {Juneau}, S. and {Kirkby}, D. and {Kremin}, A. and {Lamman}, C. and {Landriau}, M. and {Le Guillou}, L. and {Manera}, M. and {Martini}, P. and {Meisner}, A. and {Miquel}, R. and {Moustakas}, J. and {Nadathur}, S. and {Percival}, W.~J. and {Poppett}, C. and {Prada}, F. and {P{\'e}rez-R{\`a}fols}, I. and {Rossi}, G. and {Sanchez}, E. and {Schubnell}, M. and {Sprayberry}, D. and {Tarl{\'e}}, G. and {Weaver}, B.~A. and {Zhou}, R. and {the DESI Collaboration}},
        title = "{The DESI Single Fiber Lens Search. I. Four Thousand Spectroscopically Selected Galaxy-Galaxy Gravitational Lens Candidates}",
      journal = {arXiv e-prints},
         year = 2025,
        month = dec,
          eid = {arXiv:2512.04275},
        pages = {arXiv:2512.04275},
          doi = {10.48550/arXiv.2512.04275},
archivePrefix = {arXiv},
       eprint = {2512.04275},
 primaryClass = {astro-ph.GA},
       adsurl = {https://ui.adsabs.harvard.edu/abs/2025arXiv251204275K}
}

@ARTICLE{hsu2025,
       author = {{Hsu}, Yuan-Ming and {Huang}, Xiaosheng and {Storfer}, Christopher J. and {Inchausti}, Jose Carlos and {Schlegel}, David and {Moustakas}, John and {Aguilar}, J. and {Ahlen}, S. and {Anand}, A. and {Bailey}, S. and {Bianchi}, D. and {Brooks}, D. and {Castander}, F.~J. and {Claybaugh}, T. and {Cuceu}, A. and {de la Macorra}, A. and {Della Costa}, J. and {Dey}, Arjun and {Dey}, Biprateep and {Doel}, P. and {Forero-Romero}, J.~E. and {Gazta{\~n}aga}, E. and {Gontcho}, S. Gontcho A and {Gutierrez}, G. and {Huterer}, D. and {Joyce}, R. and {Kehoe}, R. and {Kirkby}, D. and {Kisner}, T. and {Kremin}, A. and {Lahav}, O. and {Landriau}, M. and {Le Guillou}, L. and {Manera}, M. and {Meisner}, A. and {Miquel}, R. and {Nadathur}, S. and {Palanque-Delabrouille}, N. and {Percival}, W.~J. and {Prada}, F. and {P{\'e}rez-R{\`a}fols}, I. and {Rossi}, G. and {Sanchez}, E. and {Schubnell}, M. and {Silber}, J. and {Sprayberry}, D. and {Tarl{\'e}}, G. and {Weaver}, B.~A. and {Zhou}, R. and {Zou}, H.},
        title = "{A New Way to Discover Strong Gravitational Lenses: Pair-wise Spectroscopic Search from DESI DR1}",
      journal = {arXiv e-prints},
         year = 2025,
        month = sep,
          eid = {arXiv:2509.16033},
        pages = {arXiv:2509.16033},
          doi = {10.48550/arXiv.2509.16033},
archivePrefix = {arXiv},
       eprint = {2509.16033},
 primaryClass = {astro-ph.GA},
       adsurl = {https://ui.adsabs.harvard.edu/abs/2025arXiv250916033H}
}

@ARTICLE{collett2014a,
       author = {{Collett}, Thomas E. and {Auger}, Matthew W.},
        title = "{Cosmological constraints from the double source plane lens SDSSJ0946+1006}",
      journal = {\mnras},
         year = 2014,
        month = sep,
       volume = {443},
       number = {2},
        pages = {969-976},
          doi = {10.1093/mnras/stu1190},
archivePrefix = {arXiv},
       eprint = {1403.5278},
 primaryClass = {astro-ph.CO},
       adsurl = {https://ui.adsabs.harvard.edu/abs/2014MNRAS.443..969C}
}

@ARTICLE{sheu2024,
       author = {{Sheu}, William and {Cikota}, Aleksandar and {Huang}, Xiaosheng and {Glazebrook}, Karl and {Storfer}, Christopher and {Agarwal}, Shrihan and {Schlegel}, David J. and {Suzuki}, Nao and {Barone}, Tania M. and {Bian}, Fuyan and {Jeltema}, Tesla and {Jones}, Tucker and {Kacprzak}, Glenn G. and {O'Donnell}, Jackson H. and {G.~C.}, Keerthi Vasan},
        title = "{The Carousel Lens: A Well-modeled Strong Lens with Multiple Sources Spectroscopically Confirmed by VLT/MUSE}",
      journal = {\apj},
         year = 2024,
        month = sep,
       volume = {973},
       number = {1},
          eid = {3},
        pages = {3},
          doi = {10.3847/1538-4357/ad65d3},
archivePrefix = {arXiv},
       eprint = {2408.10320},
 primaryClass = {astro-ph.GA},
       adsurl = {https://ui.adsabs.harvard.edu/abs/2024ApJ...973....3S}
}

@ARTICLE{agarwal2025,
       author = {{Agarwal}, Shrihan and {Huang}, Xiaosheng and {Sheu}, William and {Storfer}, Christopher J. and {Tamargo-Arizmendi}, Marcos and {Tabares-Tarquinio}, Suchitoto and {Schlegel}, D.~J. and {Aldering}, G. and {Bolton}, A. and {Cikota}, A. and {Dey}, Arjun and {Filipp}, A. and {Jullo}, E. and {Kwon}, K.~J. and {Perlmutter}, S. and {Shu}, Y. and {Sukay}, E. and {Suzuki}, N. and {Aguilar}, J. and {Ahlen}, S. and {BenZvi}, S. and {Brooks}, D. and {Claybaugh}, T. and {Doel}, P. and {Forero-Romero}, J.~E. and {Gazta{\~n}aga}, E. and {Gontcho}, S. Gontcho A and {Gutierrez}, G. and {Honscheid}, K. and {Ishak}, M. and {Juneau}, S. and {Kehoe}, R. and {Kisner}, T. and {Koposov}, S.~E. and {Lambert}, A. and {Landriau}, M. and {Le Guillou}, L. and {de la Macorra}, A. and {Meisner}, A. and {Miquel}, R. and {Moustakas}, J. and {Myers}, A.~D. and {Poppett}, C. and {Prada}, F. and {P{\'e}rez-R{\`a}fols}, I. and {Rossi}, G. and {Sanchez}, E. and {Schubnell}, M. and {Sprayberry}, D. and {Tarl{\'e}}, G. and {Weaver}, B.~A. and {Zou}, H.},
        title = "{DESI Strong Lens Foundry III: Keck Spectroscopy for Strong Lenses Discovered Using Residual Neural Networks}",
      journal = {arXiv e-prints},
         year = 2025,
        month = sep,
          eid = {arXiv:2509.18086},
        pages = {arXiv:2509.18086},
          doi = {10.48550/arXiv.2509.18086},
archivePrefix = {arXiv},
       eprint = {2509.18086},
 primaryClass = {astro-ph.CO},
       adsurl = {https://ui.adsabs.harvard.edu/abs/2025arXiv250918086A}
}

@ARTICLE{bolton2012b,
       author = {{Bolton}, Adam S. and {Brownstein}, Joel R. and {Kochanek}, Christopher S. and {Shu}, Yiping and {Schlegel}, David J. and {Eisenstein}, Daniel J. and {Wake}, David A. and {Connolly}, Natalia and {Maraston}, Claudia and {Arneson}, Ryan A. and {Weaver}, Benjamin A.},
        title = "{The BOSS Emission-Line Lens Survey. II. Investigating Mass-density Profile Evolution in the SLACS+BELLS Strong Gravitational Lens Sample}",
      journal = {\apj},
         year = 2012,
        month = sep,
       volume = {757},
       number = {1},
          eid = {82},
        pages = {82},
          doi = {10.1088/0004-637X/757/1/82},
archivePrefix = {arXiv},
       eprint = {1201.2988},
 primaryClass = {astro-ph.CO},
       adsurl = {https://ui.adsabs.harvard.edu/abs/2012ApJ...757...82B}
}

@ARTICLE{sahu2024,
       author = {{Sahu}, Nandini and {Tran}, Kim-Vy and {Suyu}, Sherry H. and {Shajib}, Anowar J. and {Ertl}, Sebastian and {Kacprzak}, Glenn G. and {Glazebrook}, Karl and {Jones}, Tucker and {G.~C.}, Keerthi Vasan and {Barone}, Tania M. and {Baker}, A. Makai and {Skobe}, Hannah and {Derkenne}, Caro and {Lewis}, Geraint F. and {Sweet}, Sarah M. and {Lopez}, Sebastian},
        title = "{AGEL: Is the Conflict Real? Investigating Galaxy Evolution Models Using Strong Lensing at 0.3 < z < 0.9}",
      journal = {\apj},
         year = 2024,
        month = jul,
       volume = {970},
       number = {1},
          eid = {86},
        pages = {86},
          doi = {10.3847/1538-4357/ad4ce3},
archivePrefix = {arXiv},
       eprint = {2405.15427},
 primaryClass = {astro-ph.GA},
       adsurl = {https://ui.adsabs.harvard.edu/abs/2024ApJ...970...86S}
}

@ARTICLE{silver2025,
       author = {{Silver}, Ethan and {Wang}, R. and {Huang}, Xiaosheng and {Bolton}, Adam S. and {Storfer}, Christopher J. and {Banka}, S.},
        title = "{ML-driven Strong Lens Discoveries: Down to {\ensuremath{\theta}}$_{E}${\ensuremath{\sim}}0.″03 and M$_{halo}$ < {}10$^{11}$M$_{{\ensuremath{\odot}}}$}",
      journal = {\apj},
         year = 2025,
        month = nov,
       volume = {994},
       number = {1},
          eid = {117},
        pages = {117},
          doi = {10.3847/1538-4357/adf3b0},
archivePrefix = {arXiv},
       eprint = {2507.01943},
 primaryClass = {astro-ph.CO},
       adsurl = {https://ui.adsabs.harvard.edu/abs/2025ApJ...994..117S}
}

@ARTICLE{lin2025,
       author = {{Lin}, Emerald and {Toro Bertolla}, Ivonne and {Cikota}, Aleksandar and {Huang}, Xiaosheng and {Storfer}, Christopher J. and {Tamargo-Arizmendi}, Marcos and {Schlegel}, David J. and {Sheu}, William and {Suzuki}, Nao},
        title = "{DESI Strong Lens Foundry IV: Spectroscopic Confirmation of DESI Lens Candidates with VLT/MUSE}",
      journal = {arXiv e-prints},
         year = 2025,
        month = sep,
          eid = {arXiv:2509.18078},
        pages = {arXiv:2509.18078},
          doi = {10.48550/arXiv.2509.18078},
archivePrefix = {arXiv},
       eprint = {2509.18078},
 primaryClass = {astro-ph.CO},
       adsurl = {https://ui.adsabs.harvard.edu/abs/2025arXiv250918078L}
}

@ARTICLE{huang2025b,
       author = {{Huang}, Xiaosheng and {Inchausti}, Jose Carlos and {Storfer}, Christopher J. and {Tabares-Tarquinio}, S. and {Moustakas}, J. and {Sheu}, W. and {Agarwal}, S. and {Tamargo-Arizmendi}, M. and {Schlegel}, D.~J. and {Aguilar}, J. and {Ahlen}, S. and {Aldering}, G. and {Bailey}, S. and {Banka}, S. and {BenZvi}, S. and {Bianchi}, D. and {Bolton}, A. and {Brooks}, D. and {Cikota}, A. and {Claybaugh}, T. and {Dawson}, K.~S. and {de la Macorra}, A. and {Dey}, A. and {Doel}, P. and {Edelstein}, J. and {Forero-Romero}, J.~E. and {Gaztanaga}, E. and {Gontcho}, S. Gontcho A and {Gonzalez-Morales}, A.~X. and {Gu}, A. and {Honscheid}, K. and {Ishak}, M. and {Juneau}, S. and {Kehoe}, R. and {Kisner}, T. and {Koposov}, S.~E. and {Kwon}, K.~J. and {Lambert}, A. and {Landriau}, M. and {Lang}, D. and {Le Guillou}, L. and {Levi}, M.~E. and {Liu}, J. and {Meisner}, A. and {Miquel}, R. and {Myers}, A.~D. and {Perlmutter}, S. and {Palanque-Delabrouille}, N. and {Perez-Rafols}, I. and {Poppett}, C. and {Prada}, F. and {Rossi}, G. and {Rubin}, D. and {Sanchez}, E. and {Schubnell}, M. and {Shu}, Y. and {Silver}, E. and {Sprayberry}, D. and {Suzuki}, N. and {Tarle}, G. and {Weaver}, B.~A. and {Zou}, H.},
        title = "{DESI Strong Lens Foundry II: DESI Spectroscopy for Strong Lens Candidates}",
      journal = {arXiv e-prints},
         year = {2025b},
        month = sep,
          eid = {arXiv:2509.18089},
        pages = {arXiv:2509.18089},
          doi = {10.48550/arXiv.2509.18089},
archivePrefix = {arXiv},
       eprint = {2509.18089},
 primaryClass = {astro-ph.CO},
       adsurl = {https://ui.adsabs.harvard.edu/abs/2025arXiv250918089H}
}

@ARTICLE{urcelay2025,
       author = {{Urcelay}, F. and {Jullo}, E. and {Barrientos}, L.~F. and {Huang}, X. and {Hernandez}, J.},
        title = "{A compact group lens modeled with GIGA-Lens: Enhanced inference for complex systems}",
      journal = {\aap},
         year = 2025,
        month = feb,
       volume = {694},
          eid = {A35},
        pages = {A35},
          doi = {10.1051/0004-6361/202449261},
archivePrefix = {arXiv},
       eprint = {2412.04567},
 primaryClass = {astro-ph.CO},
       adsurl = {https://ui.adsabs.harvard.edu/abs/2025A&A...694A..35U}
}

@ARTICLE{inchausti2025,
       author = {{Inchausti}, Jose Carlos and {Storfer}, Christopher J. and {Huang}, Xiaosheng and {Hsu}, Yuan-Ming and {Kaufmann}, Brandt and {Pasupala}, Chaitanya and {Banka}, S. and {Dey}, A. and {Lang}, D. and {Meisner}, A. and {Moustakas}, J. and {Myers}, A.~D. and {Schlafly}, E.~F. and {Schlegel}, D.~J.},
        title = "{Strong Lens Discoveries in DESI Legacy Imaging Surveys DR10 with Two Deep Learning Architectures}",
      journal = {arXiv e-prints},
         year = 2025,
        month = aug,
          eid = {arXiv:2508.20087},
        pages = {arXiv:2508.20087},
          doi = {10.48550/arXiv.2508.20087},
archivePrefix = {arXiv},
       eprint = {2508.20087},
 primaryClass = {astro-ph.CO},
       adsurl = {https://ui.adsabs.harvard.edu/abs/2025arXiv250820087I}
}

@article{kochanek1991a,
  author =        {{Kochanek}, C.~S.},
  journal =       {\apj},
  month =         jun,
  pages =         {354-368},
  title =         {{The implications of lenses for galaxy structure}},
  volume =        {373},
  year =          {1991},
  doi =           {10.1086/170057},
}

@article{blandford1992a,
  author =        {{Blandford}, R.~D. and {Narayan}, R.},
  journal =       {\araa},
  pages =         {311-358},
  title =         {{Cosmological applications of gravitational lensing}},
  volume =        {30},
  year =          {1992},
  doi =           {10.1146/annurev.astro.30.1.311},
}

@article{broadhurst2000a,
  author =        {{Broadhurst}, Tom and {Huang}, Xiaosheng and
                   {Frye}, Brenda and {Ellis}, Richard},
  journal =       {\apjl},
  month =         may,
  number =        {1},
  pages =         {L15-L18},
  title =         {{A Spectroscopic Redshift for the Cl 0024+16 Multiple
                   Arc System: Implications for the Central Mass
                   Distribution}},
  volume =        {534},
  year =          {2000},
  doi =           {10.1086/312651},
}

@article{koopmans2002a,
  author =        {{Koopmans}, L.~V.~E. and {Treu}, T.},
  journal =       {\apjl},
  month =         mar,
  pages =         {L5-L8},
  title =         {{The Stellar Velocity Dispersion of the Lens Galaxy
                   in MG 2016+112 at z=1.004}},
  volume =        {568},
  year =          {2002},
  doi =           {10.1086/340143},
}

@article{bolton2006a,
  author =        {{Bolton}, A.~S. and {Burles}, S. and
                   {Koopmans}, L.~V.~E. and {Treu}, T. and
                   {Moustakas}, L.~A.},
  journal =       {\apj},
  month =         feb,
  pages =         {703-724},
  title =         {{The Sloan Lens ACS Survey. I. A Large
                   Spectroscopically Selected Sample of Massive
                   Early-Type Lens Galaxies}},
  volume =        {638},
  year =          {2006},
  doi =           {10.1086/498884},
}

@article{koopmans2006a,
  author =        {{Koopmans}, L{\'e}on V.~E. and {Treu}, Tommaso and
                   {Bolton}, Adam S. and {Burles}, Scott and
                   {Moustakas}, Leonidas A.},
  journal =       {\apj},
  month =         oct,
  number =        {2},
  pages =         {599-615},
  title =         {{The Sloan Lens ACS Survey. III. The Structure and
                   Formation of Early-Type Galaxies and Their Evolution
                   since z \raisebox{-0.5ex}\textasciitilde 1}},
  volume =        {649},
  year =          {2006},
  doi =           {10.1086/505696},
}

@article{bradac2008a,
  author =        {Bradač, Maruša and Allen, Steven W. and
                   Treu, Tommaso and Ebeling, Harald and Massey, Richard and
                   Morris, R. Glenn and von der Linden, Anja and
                   Applegate, Douglas},
  journal =       {\apj},
  month =         {Nov},
  number =        {2},
  pages =         {959–967},
  publisher =     {American Astronomical Society},
  title =         {Revealing the Properties of Dark Matter in the
                   Merging Cluster MACS J0025.4−1222},
  volume =        {687},
  year =          {2008},
  doi =           {10.1086/591246},
  issn =          {1538-4357},
  url =           {http://dx.doi.org/10.1086/591246},
}

@article{huang2025a,
      title={DESI Strong Lens Foundry I: HST Observations and Modeling with GIGA-Lens}, 
      author={X. Huang and S. Baltasar and N. Ratier-Werbin and C. Storfer and W. Sheu and S. Agarwal and M. Tamargo-Arizmendi and D. J. Schlegel and J. Aguilar and S. Ahlen and G. Aldering and S. Banka and S. BenZvi and D. Bianchi and A. Bolton and D. Brooks and A. Cikota and T. Claybaugh and A. de la Macorra and A. Dey and P. Doel and J. Edelstein and A. Filipp and J. E. Forero-Romero and E. Gaztanaga and S. Gontcho A Gontcho and A. Gu and G. Gutierrez and K. Honscheid and E. Jullo and S. Juneau and R. Kehoe and D. Kirkby and T. Kisner and A. Kremin and K. J. Kwon and A. Lambert and M. Landriau and D. Lang and L. Le Guillou and J. Liu and A. Meisner and R. Miquel and J. Moustakas and A. D. Myers and S. Perlmutter and I. Perez-Rafols and F. Prada and G. Rossi and D. Rubin and E. Sanchez and M. Schubnell and Y. Shu and E. Silver and D. Sprayberry and N. Suzuki and G. Tarle and B. A. Weaver and H. Zou},
      year={2025a},
      eprint={2502.03455},
      archivePrefix={arXiv},
      primaryClass={astro-ph.CO},
      url={https://arxiv.org/abs/2502.03455}, 
}

@article{vegetti2009a,
  author =        {{Vegetti}, Simona and {Koopmans}, L.~V.~E.},
  journal =       {\mnras},
  month =         {Dec},
  pages =         {1583-1592},
  title =         {{Statistics of mass substructure from strong
                   gravitational lensing: quantifying the mass fraction
                   and mass function}},
  volume =        {400},
  year =          {2009},
  doi =           {10.1111/j.1365-2966.2009.15559.x},
}

@article{huang2009a,
  author =        {{Huang}, X. and {Morokuma}, T. and {Fakhouri}, H.~K. and
                   {Aldering}, G. and {Amanullah}, R. and {Barbary}, K. and
                   {Brodwin}, M. and {Connolly}, N.~V. and
                   {Dawson}, K.~S. and {Doi}, M. and {Faccioli}, L. and
                   {Fadeyev}, V. and {Fruchter}, A.~S. and
                   {Goldhaber}, G. and {Gladders}, M.~D. and
                   {Hennawi}, J.~F. and {Ihara}, Y. and {Jee}, M.~J. and
                   {Kowalski}, M. and {Konishi}, K. and {Lidman}, C. and
                   {Meyers}, J. and {Moustakas}, L.~A. and
                   {Perlmutter}, S. and {Rubin}, D. and
                   {Schlegel}, D.~J. and {Spadafora}, A.~L. and
                   {Suzuki}, N. and {Takanashi}, N. and {Yasuda}, N.},
  journal =       {\apj},
  month =         {Dec},
  pages =         {L12-L16},
  title =         {{Hubble Space Telescope Discovery of a z = 3.9
                   Multiply Imaged Galaxy Behind the Complex Cluster
                   Lens Warps J1415.1+36 at z = 1.026}},
  volume =        {707},
  year =          {2009},
  doi =           {10.1088/0004-637X/707/1/L12},
}

@article{jullo2010a,
  author =        {{Jullo}, E. and {Natarajan}, P. and {Kneib}, J. -P. and
                   {D'Aloisio}, A. and {Limousin}, M. and {Richard}, J. and
                   {Schimd}, C.},
  journal =       {Science},
  month =         aug,
  pages =         {924-927},
  title =         {{Cosmological constraints from strong gravitational
                   lensing in clusters of galaxies.}},
  volume =        {329},
  year =          {2010},
  doi =           {10.1126/science.1185759},
}

@article{tessore2016a,
  author =        {{Tessore}, Nicolas and {Bellagamba}, Fabio and
                   {Metcalf}, R. Benton},
  journal =       {\mnras},
  month =         {Dec},
  pages =         {3115-3128},
  title =         {{LENSED: a code for the forward reconstruction of
                   lenses and sources from strong lensing observations}},
  volume =        {463},
  year =          {2016},
  doi =           {10.1093/mnras/stw2212},
}

@article{jauzac2018a,
  author =        {{Jauzac}, Mathilde and {Harvey}, David and
                   {Massey}, Richard},
  journal =       {\mnras},
  month =         jul,
  number =        {3},
  pages =         {4046-4051},
  title =         {{The shape of galaxy dark matter haloes in massive
                   galaxy clusters: insights from strong gravitational
                   lensing}},
  volume =        {477},
  year =          {2018},
  doi =           {10.1093/mnras/sty909},
}

@article{shajib2019a,
  author =        {{Shajib}, A.~J. and {Birrer}, S. and {Treu}, T. and
                   {Auger}, M.~W. and {Agnello}, A. and {Anguita}, T. and
                   {Buckley-Geer}, E.~J. and {Chan}, J.~H.~H. and
                   {Collett}, T.~E. and {Courbin}, F. and
                   {Fassnacht}, C.~D. and {Frieman}, J. and {Kayo}, I. and
                   {Lemon}, C. and {Lin}, H. and {Marshall}, P.~J. and
                   {McMahon}, R. and {More}, A. and {Morgan}, N.~D. and
                   {Motta}, V. and {Oguri}, M. and {Ostrovski}, F. and
                   {Rusu}, C.~E. and {Schechter}, P.~L. and {Shanks}, T. and
                   {Suyu}, S.~H. and {Meylan}, G. and {Abbott}, T.~M.~C. and
                   {Allam}, S. and {Annis}, J. and {Avila}, S. and
                   {Bertin}, E. and {Brooks}, D. and
                   {Carnero Rosell}, A. and {Carrasco Kind}, M. and
                   {Carretero}, J. and {Cunha}, C.~E. and
                   {da Costa}, L.~N. and {De Vicente}, J. and
                   {Desai}, S. and {Doel}, P. and {Flaugher}, B. and
                   {Fosalba}, P. and {Garc{\'\i}a-Bellido}, J. and
                   {Gerdes}, D.~W. and {Gruen}, D. and {Gruendl}, R.~A. and
                   {Gutierrez}, G. and {Hartley}, W.~G. and
                   {Hollowood}, D.~L. and {Hoyle}, B. and {James}, D.~J. and
                   {Kuehn}, K. and {Kuropatkin}, N. and {Lahav}, O. and
                   {Lima}, M. and {Maia}, M.~A.~G. and {March}, M. and
                   {Marshall}, J.~L. and {Melchior}, P. and
                   {Menanteau}, F. and {Miquel}, R. and {Plazas}, A.~A. and
                   {Sanchez}, E. and {Scarpine}, V. and
                   {Sevilla-Noarbe}, I. and {Smith}, M. and
                   {Soares-Santos}, M. and {Sobreira}, F. and
                   {Suchyta}, E. and {Swanson}, M.~E.~C. and {Tarle}, G. and
                   {Walker}, A.~R.},
  journal =       {\mnras},
  month =         mar,
  number =        {4},
  pages =         {5649-5671},
  title =         {{Is every strong lens model unhappy in its own way?
                   Uniform modelling of a sample of 13 quadruply+ imaged
                   quasars}},
  volume =        {483},
  year =          {2019},
  doi =           {10.1093/mnras/sty3397},
}

@article{meneghetti2020a,
  author =        {{Meneghetti}, Massimo and {Davoli}, Guido and
                   {Bergamini}, Pietro and {Rosati}, Piero and
                   {Natarajan}, Priyamvada and {Giocoli}, Carlo and
                   {Caminha}, Gabriel B. and {Metcalf}, R. Benton and
                   {Rasia}, Elena and {Borgani}, Stefano and
                   {Calura}, Francesco and {Grillo}, Claudio and
                   {Mercurio}, Amata and {Vanzella}, Eros},
  journal =       {Science},
  month =         sep,
  number =        {6509},
  pages =         {1347-1351},
  title =         {{An excess of small-scale gravitational lenses
                   observed in galaxy clusters}},
  volume =        {369},
  year =          {2020},
  doi =           {10.1126/science.aax5164},
}

@article{filipp2023a,
  author =        {{Filipp}, Andreas and {Shu}, Yiping and
                   {Pakmor}, R{\"u}diger and {Suyu}, Sherry H. and
                   {Huang}, Xiaosheng},
  journal =       {\aap},
  month =         sep,
  pages =         {A113},
  title =         {{Simulation-guided galaxy evolution inference: A case
                   study with strong lensing galaxies}},
  volume =        {677},
  year =          {2023},
  doi =           {10.1051/0004-6361/202346594},
  eid =           {A113},
}

@article{vegetti2014a,
  author =        {{Vegetti}, S. and {Koopmans}, L.~V.~E. and
                   {Auger}, M.~W. and {Treu}, T. and {Bolton}, A.~S.},
  journal =       {\mnras},
  month =         aug,
  number =        {3},
  pages =         {2017-2035},
  title =         {{Inference of the cold dark matter substructure mass
                   function at z = 0.2 using strong gravitational
                   lenses}},
  volume =        {442},
  year =          {2014},
  doi =           {10.1093/mnras/stu943},
}

@article{hezaveh2016a,
  author =        {{Hezaveh}, Yashar D. and {Dalal}, Neal and
                   {Marrone}, Daniel P. and {Mao}, Yao-Yuan and
                   {Morningstar}, Warren and {Wen}, Di and
                   {Blandford}, Roger D. and {Carlstrom}, John E. and
                   {Fassnacht}, Christopher D. and {Holder}, Gilbert P. and
                   {Kemball}, Athol and {Marshall}, Philip J. and
                   {Murray}, Norman and {Perreault Levasseur}, Laurence and
                   {Vieira}, Joaquin D. and {Wechsler}, Risa H.},
  journal =       {\apj},
  month =         may,
  number =        {1},
  pages =         {37},
  title =         {{Detection of Lensing Substructure Using ALMA
                   Observations of the Dusty Galaxy SDP.81}},
  volume =        {823},
  year =          {2016},
  doi =           {10.3847/0004-637X/823/1/37},
  eid =           {37},
}

@article{vegetti2018a,
  author =        {{Vegetti}, S. and {Despali}, G. and {Lovell}, M.~R. and
                   {Enzi}, W.},
  journal =       {\mnras},
  month =         dec,
  number =        {3},
  pages =         {3661-3669},
  title =         {{Constraining sterile neutrino cosmologies with
                   strong gravitational lensing observations at redshift
                   z {\ensuremath{\sim}} 0.2}},
  volume =        {481},
  year =          {2018},
  doi =           {10.1093/mnras/sty2393},
}

@article{ritondale2019a,
  author =        {{Ritondale}, E. and {Vegetti}, S. and {Despali}, G. and
                   {Auger}, M.~W. and {Koopmans}, L.~V.~E. and
                   {McKean}, J.~P.},
  journal =       {\mnras},
  month =         {May},
  number =        {2},
  pages =         {2179-2193},
  title =         {{Low-mass halo perturbations in strong gravitational
                   lenses at redshift z ̃ 0.5 are consistent with CDM}},
  volume =        {485},
  year =          {2019},
  doi =           {10.1093/mnras/stz464},
}

@article{diazrivero2020a,
  author =        {{Diaz Rivero}, Ana and {Dvorkin}, Cora},
  journal =       {\prd},
  month =         jan,
  number =        {2},
  pages =         {023515},
  title =         {{Direct detection of dark matter substructure in
                   strong lens images with convolutional neural
                   networks}},
  volume =        {101},
  year =          {2020},
  doi =           {10.1103/PhysRevD.101.023515},
  eid =           {023515},
}

@article{cagan-sengul2022a,
  author =          {{{\c{C}}a{\v{g}}an {\c{S}}eng{\"u}l}, Atin{\c{c}}
  {\c{C}}agan and {Dvorkin}, Cora and {Ostdiek}, Bryan and {Tsang}, Arthur},
  journal =       {\mnras},
  month =         sep,
  number =        {3},
  pages =         {4391-4401},
  title =         {{Substructure detection reanalysed: dark perturber
                   shown to be a line-of-sight halo}},
  volume =        {515},
  year =          {2022},
  doi =           {10.1093/mnras/stac1967},
}

@ARTICLE{cikota2023a,
       author = {{Cikota}, Aleksandar and {Bertolla}, Ivonne Toro and {Huang}, Xiaosheng and {Baltasar}, Saul and {Ratier-Werbin}, Nicolas and {Sheu}, William and {Storfer}, Christopher and {Suzuki}, Nao and {Schlegel}, David J. and {Cartier}, Regis and {Torres}, Simon and {Cikota}, Stefan and {Jullo}, Eric},
        title = "{DESI-253.2534+26.8843: A New Einstein Cross Spectroscopically Confirmed with Very Large Telescope/MUSE and Modeled with GIGA-Lens}",
      journal = {\apjl},
         year = 2023,
        month = aug,
       volume = {953},
       number = {1},
          eid = {L5},
        pages = {L5},
          doi = {10.3847/2041-8213/ace9da},
archivePrefix = {arXiv},
       eprint = {2307.12470},
 primaryClass = {astro-ph.GA},
       adsurl = {https://ui.adsabs.harvard.edu/abs/2023ApJ...953L...5C}
}

@article{nierenberg2023a,
  author =        {{Nierenberg}, A.~M. and {Keeley}, R.~E. and
                   {Sluse}, D. and {Gilman}, D. and {Birrer}, S. and
                   {Treu}, T. and {Abazajian}, K.~N. and {Anguita}, T. and
                   {Benson}, A.~J. and {Bennert}, V.~N. and
                   {Djorgovski}, S.~G. and {Du}, X. and
                   {Fassnacht}, C.~D. and {Hoenig}, S.~F. and
                   {Kusenko}, A. and {Lemon}, C. and {Malkan}, M. and
                   {Motta}, V. and {Moustakas}, L.~A. and {Stern}, D. and
                   {Wechsler}, R.~H.},
  journal =       {arXiv e-prints},
  month =         sep,
  pages =         {arXiv:2309.10101},
  title =         {{JWST lensed quasar dark matter survey I: Description
                   and First Results}},
  year =          {2023},
  doi =           {10.48550/arXiv.2309.10101},
  eid =           {arXiv:2309.10101},
}

@article{abbott2017a,
  author =        {{Abbott}, B.~P. and {Abbott}, R. and {Abbott}, T.~D. and
                   {Acernese}, F. and {Ackley}, K. and {Adams}, C. and
                   {Adams}, T. and {Addesso}, P. and {Adhikari}, R.~X. and
                   {Adya}, V.~B. and {Affeldt}, C.},
  journal =       {\nat},
  month =         nov,
  number =        {7678},
  pages =         {85-88},
  title =         {{A gravitational-wave standard siren measurement of
                   the Hubble constant}},
  volume =        {551},
  year =          {2017},
  doi =           {10.1038/nature24471},
}

@article{abbott2018b,
  author =        {{Abbott}, T.~M.~C. and {Abdalla}, F.~B. and
                   {Annis}, J. and {Bechtol}, K. and {Blazek}, J. and
                   {Benson}, B.~A. and {Bernstein}, R.~A. and
                   {Bernstein}, G.~M. and {Bertin}, E. and
                   {Dark Energy Survey Collaboration} and
                   {South Pole Telescope Collaboration}},
  journal =       {\mnras},
  month =         nov,
  number =        {3},
  pages =         {3879-3888},
  title =         {{Dark Energy Survey Year 1 Results: A Precise H$_{0}$
                   Estimate from DES Y1, BAO, and D/H Data}},
  volume =        {480},
  year =          {2018},
  doi =           {10.1093/mnras/sty1939},
}

@ARTICLE{ahumada2020a,
       author = {{Ahumada}, Romina and {Allende Prieto}, Carlos and {Almeida}, Andr{\'e}s and {Anders}, Friedrich and {Anderson}, Scott F. and {Andrews}, Brett H. and {Anguiano}, Borja and {Arcodia}, Riccardo and {Armengaud}, Eric and {Aubert}, Marie and {Avila}, Santiago and {Avila-Reese}, Vladimir and {Badenes}, Carles and {Balland}, Christophe and {Barger}, Kat and {Barrera-Ballesteros}, Jorge K. and {Basu}, Sarbani and {Bautista}, Julian and {Beaton}, Rachael L. and {Beers}, Timothy C. and {Benavides}, B. Izamar T. and {Bender}, Chad F. and {Bernardi}, Mariangela and {Bershady}, Matthew and {Beutler}, Florian and {Bidin}, Christian Moni and {Bird}, Jonathan and {Bizyaev}, Dmitry and {Blanc}, Guillermo A. and {Blanton}, Michael R. and {Boquien}, M{\'e}d{\'e}ric and {Borissova}, Jura and {Bovy}, Jo and {Brandt}, W.~N. and {Brinkmann}, Jonathan and {Brownstein}, Joel R. and {Bundy}, Kevin and {Bureau}, Martin and {Burgasser}, Adam and {Burtin}, Etienne and {Cano-D{\'\i}az}, Mariana and {Capasso}, Raffaella and {Cappellari}, Michele and {Carrera}, Ricardo and {Chabanier}, Sol{\`e}ne and {Chaplin}, William and {Chapman}, Michael and {Cherinka}, Brian and {Chiappini}, Cristina and {Doohyun Choi}, Peter and {Chojnowski}, S. Drew and {Chung}, Haeun and {Clerc}, Nicolas and {Coffey}, Damien and {Comerford}, Julia M. and {Comparat}, Johan and {da Costa}, Luiz and {Cousinou}, Marie-Claude and {Covey}, Kevin and {Crane}, Jeffrey D. and {Cunha}, Katia and {Ilha}, Gabriele da Silva and {Dai}, Yu Sophia and {Damsted}, Sanna B. and {Darling}, Jeremy and {Davidson}, Jr., James W. and {Davies}, Roger and {Dawson}, Kyle and {De}, Nikhil and {de la Macorra}, Axel and {De Lee}, Nathan and {Queiroz}, Anna B{\'a}rbara de Andrade and {Deconto Machado}, Alice and {de la Torre}, Sylvain and {Dell'Agli}, Flavia and {du Mas des Bourboux}, H{\'e}lion and {Diamond-Stanic}, Aleksandar M. and {Dillon}, Sean and {Donor}, John and {Drory}, Niv and {Duckworth}, Chris and {Dwelly}, Tom and {Ebelke}, Garrett and {Eftekharzadeh}, Sarah and {Davis Eigenbrot}, Arthur and {Elsworth}, Yvonne P. and {Eracleous}, Mike and {Erfanianfar}, Ghazaleh and {Escoffier}, Stephanie and {Fan}, Xiaohui and {Farr}, Emily and {Fern{\'a}ndez-Trincado}, Jos{\'e} G. and {Feuillet}, Diane and {Finoguenov}, Alexis and {Fofie}, Patricia and {Fraser-McKelvie}, Amelia and {Frinchaboy}, Peter M. and {Fromenteau}, Sebastien and {Fu}, Hai and {Galbany}, Llu{\'\i}s and {Garcia}, Rafael A. and {Garc{\'\i}a-Hern{\'a}ndez}, D.~A. and {Garma Oehmichen}, Luis Alberto and {Ge}, Junqiang and {Geimba Maia}, Marcio Antonio and {Geisler}, Doug and {Gelfand}, Joseph and {Goddy}, Julian and {Gonzalez-Perez}, Violeta and {Grabowski}, Kathleen and {Green}, Paul and {Grier}, Catherine J. and {Guo}, Hong and {Guy}, Julien and {Harding}, Paul and {Hasselquist}, Sten and {Hawken}, Adam James and {Hayes}, Christian R. and {Hearty}, Fred and {Hekker}, S. and {Hogg}, David W. and {Holtzman}, Jon A. and {Horta}, Danny and {Hou}, Jiamin and {Hsieh}, Bau-Ching and {Huber}, Daniel and {Hunt}, Jason A.~S. and {Ider Chitham}, J. and {Imig}, Julie and {Jaber}, Mariana and {Jimenez Angel}, Camilo Eduardo and {Johnson}, Jennifer A. and {Jones}, Amy M. and {J{\"o}nsson}, Henrik and {Jullo}, Eric and {Kim}, Yerim and {Kinemuchi}, Karen and {Kirkpatrick}, IV, Charles C. and {Kite}, George W. and {Klaene}, Mark and {Kneib}, Jean-Paul and {Kollmeier}, Juna A. and {Kong}, Hui and {Kounkel}, Marina and {Krishnarao}, Dhanesh and {Lacerna}, Ivan and {Lan}, Ting-Wen and {Lane}, Richard R. and {Law}, David R. and {Le Goff}, Jean-Marc and {Leung}, Henry W. and {Lewis}, Hannah and {Li}, Cheng and {Lian}, Jianhui and {Lin}, Lihwai and {Long}, Dan and {Longa-Pe{\~n}a}, Pen{\'e}lope and {Lundgren}, Britt and {Lyke}, Brad W. and {Mackereth}, J. Ted and {MacLeod}, Chelsea L. and {Majewski}, Steven R. and {Manchado}, Arturo and {Maraston}, Claudia and {Martini}, Paul and {Masseron}, Thomas and {Masters}, Karen L. and {Mathur}, Savita and {McDermid}, Richard M. and {Merloni}, Andrea and {Merrifield}, Michael and {M{\'e}sz{\'a}ros}, Szabolcs and {Miglio}, Andrea and {Minniti}, Dante and {Minsley}, Rebecca and {Miyaji}, Takamitsu and {Mohammad}, Faizan Gohar and {Mosser}, Benoit and {Mueller}, Eva-Maria and {Muna}, Demitri and {Mu{\~n}oz-Guti{\'e}rrez}, Andrea and {Myers}, Adam D. and {Nadathur}, Seshadri and {Nair}, Preethi and {Nandra}, Kirpal and {Correa do Nascimento}, Janaina and {Nevin}, Rebecca Jean and {Newman}, Jeffrey A. and {Nidever}, David L. and {Nitschelm}, Christian and {Noterdaeme}, Pasquier and {O'Connell}, Julia E. and {Olmstead}, Matthew D. and {Oravetz}, Daniel and {Oravetz}, Audrey and {Osorio}, Yeisson and {Pace}, Zachary J. and {Padilla}, Nelson and {Palanque-Delabrouille}, Nathalie and {Palicio}, Pedro A.},
        title = "{The 16th Data Release of the Sloan Digital Sky Surveys: First Release from the APOGEE-2 Southern Survey and Full Release of eBOSS Spectra}",
      journal = {\apjs},
         year = 2020,
        month = jul,
       volume = {249},
       number = {1},
          eid = {3},
        pages = {3},
          doi = {10.3847/1538-4365/ab929e},
archivePrefix = {arXiv},
       eprint = {1912.02905},
 primaryClass = {astro-ph.GA},
       adsurl = {https://ui.adsabs.harvard.edu/abs/2020ApJS..249....3A}
}

@ARTICLE{desidrone2025a,
       author = {{DESI Collaboration} and {Abdul-Karim}, M. and {Adame}, A.~G. and {Aguado}, D. and {Aguilar}, J. and {Ahlen}, S. and {Alam}, S. and {Aldering}, G. and {Alexander}, D.~M. and {Alfarsy}, R. and {Allen}, L. and {Allende Prieto}, C. and {Alves}, O. and {Anand}, A. and {Andrade}, U. and {Armengaud}, E. and {Avila}, S. and {Aviles}, A. and {Awan}, H. and {Bailey}, S. and {Baleato Lizancos}, A. and {Ballester}, O. and {Bault}, A. and {Bautista}, J. and {BenZvi}, S. and {Beraldo e Silva}, L. and {Bermejo-Climent}, J.~R. and {Beutler}, F. and {Bianchi}, D. and {Blake}, C. and {Blum}, R. and {Bolton}, A.~S. and {Bonici}, M. and {Brieden}, S. and {Brodzeller}, A. and {Brooks}, D. and {Buckley-Geer}, E. and {Burtin}, E. and {Canning}, R. and {Carnero Rosell}, A. and {Carr}, A. and {Carrilho}, P. and {Casas}, L. and {Castander}, F.~J. and {Cereskaite}, R. and {Cervantes-Cota}, J.~L. and {Chaussidon}, E. and {Chaves-Montero}, J. and {Chen}, S. and {Chen}, X. and {Claybaugh}, T. and {Cole}, S. and {Cooper}, A.~P. and {Cousinou}, M. -C. and {Cuceu}, A. and {Davis}, T.~M. and {Dawson}, K.~S. and {de Belsunce}, R. and {de la Cruz}, R. and {de la Macorra}, A. and {de Mattia}, A. and {Deiosso}, N. and {Della Costa}, J. and {Demina}, R. and {Demirbozan}, U. and {DeRose}, J. and {Dey}, A. and {Dey}, B. and {Ding}, J. and {Ding}, Z. and {Doel}, P. and {Douglass}, K. and {Dowicz}, M. and {Ebina}, H. and {Edelstein}, J. and {Eisenstein}, D.~J. and {Elbers}, W. and {Emas}, N. and {Escoffier}, S. and {Fagrelius}, P. and {Fan}, X. and {Fanning}, K. and {Fawcett}, V.~A. and {Fern\textbackslash'andez-Garc\textbackslash'ia}, E. and {Ferraro}, S. and {Findlay}, N. and {Font-Ribera}, A. and {Forero-Romero}, J.~E. and {Forero-S\textbackslash'anchez}, D. and {Frenk}, C.~S. and {G\textbackslash''ansicke}, B.~T. and {Galbany}, L. and {Garc\textbackslash'ia-Bellido}, J. and {Garcia-Quintero}, C.},
        title = "{Data Release 1 of the Dark Energy Spectroscopic Instrument}",
      journal = {arXiv e-prints},
         year = 2025,
        month = mar,
          eid = {arXiv:2503.14745},
        pages = {arXiv:2503.14745},
          doi = {10.48550/arXiv.2503.14745},
archivePrefix = {arXiv},
       eprint = {2503.14745},
 primaryClass = {astro-ph.CO},
       adsurl = {https://ui.adsabs.harvard.edu/abs/2025arXiv250314745D}
}

@ARTICLE{limousin2009a,
       author = {{Limousin}, M. and {Cabanac}, R. and {Gavazzi}, R. and {Kneib}, J. -P. and {Motta}, V. and {Richard}, J. and {Thanjavur}, K. and {Foex}, G. and {Pello}, R. and {Crampton}, D. and {Faure}, C. and {Fort}, B. and {Jullo}, E. and {Marshall}, P. and {Mellier}, Y. and {More}, A. and {Soucail}, G. and {Suyu}, S. and {Swinbank}, M. and {Sygnet}, J. -F. and {Tu}, H. and {Valls-Gabaud}, D. and {Verdugo}, T. and {Willis}, J.},
        title = "{A new window of exploration in the mass spectrum: strong lensing by galaxy groups in the SL2S}",
      journal = {\aap},
         year = 2009,
        month = aug,
       volume = {502},
       number = {2},
        pages = {445-456},
          doi = {10.1051/0004-6361/200811473},
archivePrefix = {arXiv},
       eprint = {0812.1033},
 primaryClass = {astro-ph},
       adsurl = {https://ui.adsabs.harvard.edu/abs/2009A&A...502..445L}
}

@article{riess2019a,
  author =        {{Riess}, Adam G. and {Casertano}, Stefano and
                   {Yuan}, Wenlong and {Macri}, Lucas M. and
                   {Scolnic}, Dan},
  journal =       {\apj},
  month =         may,
  number =        {1},
  pages =         {85},
  title =         {{Large Magellanic Cloud Cepheid Standards Provide a
                   1\% Foundation for the Determination of the Hubble
                   Constant and Stronger Evidence for Physics beyond
                   {\ensuremath{\Lambda}}CDM}},
  volume =        {876},
  year =          {2019},
  doi =           {10.3847/1538-4357/ab1422},
  eid =           {85},
}

@ARTICLE{verdugo2011a,
       author = {{Verdugo}, T. and {Motta}, V. and {Mu{\~n}oz}, R.~P. and {Limousin}, M. and {Cabanac}, R. and {Richard}, J.},
        title = "{Gravitational lensing and dynamics in SL2S J02140-0535: probing the mass out to large radius}",
      journal = {\aap},
         year = 2011,
        month = mar,
       volume = {527},
          eid = {A124},
        pages = {A124},
          doi = {10.1051/0004-6361/201014965},
archivePrefix = {arXiv},
       eprint = {1005.1566},
 primaryClass = {astro-ph.CO},
       adsurl = {https://ui.adsabs.harvard.edu/abs/2011A&A...527A.124V}
}

@article{wong2019a,
  author =        {{Wong}, Kenneth C. and {Suyu}, Sherry H. and
                   {Chen}, Geoff C. -F. and {Rusu}, Cristian E. and
                   {Millon}, Martin and {Sluse}, Dominique and
                   {Bonvin}, Vivien and {Fassnacht}, Christopher D. and
                   {Taubenberger}, Stefan and {Auger}, Matthew W. and
                   {Birrer}, Simon and {Chan}, James H.~H. and
                   {Courbin}, Frederic and {Hilbert}, Stefan and
                   {Tihhonova}, Olga and {Treu}, Tommaso and
                   {Agnello}, Adriano and {Ding}, Xuheng and {Jee}, Inh and
                   {Komatsu}, Eiichiro and {Shajib}, Anowar J. and
                   {Sonnenfeld}, Alessandro and {Bland ford}, Roger D. and
                   {Koopmans}, Leon V.~E. and {Marshall}, Philip J. and
                   {Meylan}, Georges},
  journal =       {arXiv e-prints},
  month =         {Jul},
  pages =         {arXiv:1907.04869},
  title =         {{H0LiCOW XIII. A 2.4\% measurement of $H_{0}$ from
                   lensed quasars: $5.3\sigma$ tension between early and
                   late-Universe probes}},
  year =          {2019},
  eid =           {arXiv:1907.04869},
}

@article{freedman2019a,
  author =        {{Freedman}, Wendy L. and {Madore}, Barry F. and
                   {Hatt}, Dylan and {Hoyt}, Taylor J. and
                   {Jang}, In Sung and {Beaton}, Rachael L. and
                   {Burns}, Christopher R. and {Lee}, Myung Gyoon and
                   {Monson}, Andrew J. and {Neeley}, Jillian R. and
                   {Phillips}, M.~M. and {Rich}, Jeffrey A. and
                   {Seibert}, Mark},
  journal =       {\apj},
  month =         sep,
  number =        {1},
  pages =         {34},
  title =         {{The Carnegie-Chicago Hubble Program. VIII. An
                   Independent Determination of the Hubble Constant
                   Based on the Tip of the Red Giant Branch}},
  volume =        {882},
  year =          {2019},
  doi =           {10.3847/1538-4357/ab2f73},
  eid =           {34},
}

@article{freedman2020a,
  author =        {{Freedman}, Wendy L. and {Madore}, Barry F. and
                   {Hoyt}, Taylor and {Jang}, In Sung and
                   {Beaton}, Rachael and {Lee}, Myung Gyoon and
                   {Monson}, Andrew and {Neeley}, Jill and
                   {Rich}, Jeffrey},
  journal =       {\apj},
  month =         mar,
  number =        {1},
  pages =         {57},
  title =         {{Calibration of the Tip of the Red Giant Branch}},
  volume =        {891},
  year =          {2020},
  doi =           {10.3847/1538-4357/ab7339},
  eid =           {57},
}

@article{planck2020a,
  author =        {{Planck Collaboration} and {Aghanim}, N. and
                   {Akrami}, Y. and {Ashdown}, M. and {Aumont}, J. and
                   {Baccigalupi}, C. and {Ballardini}, M.},
  journal =       {\aap},
  month =         sep,
  pages =         {A6},
  title =         {{Planck 2018 results. VI. Cosmological parameters}},
  volume =        {641},
  year =          {2020},
  doi =           {10.1051/0004-6361/201833910},
  eid =           {A6},
}

@article{khetan2020a,
  author =        {{Khetan}, N. and {Izzo}, L. and {Branchesi}, M. and
                   {Wojtak}, R. and {Cantiello}, M. and {Murugeshan}, C. and
                   {Cappellaro}, A. Agnello. E. and {Della Valle}, M. and
                   {Gall}, C. and {Hjorth}, J. and {Benetti}, S. and
                   {Brocato}, E. and {Burke}, J. and {Hiramitsu}, D. and
                   {Howell}, D. Andrew and {Tomasella}, L. and
                   {Valenti}, S.},
  journal =       {arXiv e-prints},
  month =         aug,
  pages =         {arXiv:2008.07754},
  title =         {{A new measurement of the Hubble constant using Type
                   Ia supernovae calibrated with surface brightness
                   fluctuations}},
  year =          {2020},
  eid =           {arXiv:2008.07754},
}

@article{philcox2020a,
  author =        {{Philcox}, Oliver H.~E. and {Sherwin}, Blake D. and
                   {Farren}, Gerrit S. and {Baxter}, Eric J.},
  journal =       {arXiv e-prints},
  month =         aug,
  pages =         {arXiv:2008.08084},
  title =         {{Determining the Hubble Constant without the Sound
                   Horizon: Measurements from Galaxy Surveys}},
  year =          {2020},
  eid =           {arXiv:2008.08084},
}

@article{choi2020a,
  author =        {{Choi}, Steve K. and {Hasselfield}, Matthew and
                   {Ho}, Shuay-Pwu Patty and {Koopman}, Brian and
                   {Lungu}, Marius and {Abitbol}, Maximilian H. and
                   {Addison}, Graeme E. and {Ade}, Peter A.~R. and
                   {Aiola}, Simone and {Alonso}, David and
                   {Amiri}, Mandana and {Amodeo}, Stefania and
                   {Angile}, Elio and {Austermann}, Jason E. and
                   {Baildon}, Taylor and {Battaglia}, Nick and
                   {Beall}, James A. and {Bean}, Rachel and
                   {Becker}, Daniel T. and {Bond}, J Richard and
                   {Bruno}, Sarah Marie and {Calabrese}, Erminia and
                   {Calafut}, Victoria and {Campusano}, Luis E. and
                   {Carrero}, Felipe and {Chesmore}, Grace E. and
                   {Cho}, Hsiao-mei and {Clark}, Susan E. and
                   {Cothard}, Nicholas F. and {Crichton}, Devin and
                   {Crowley}, Kevin T. and {Darwish}, Omar and
                   {Datta}, Rahul and {Denison}, Edward V. and
                   {Devlin}, Mark J. and {Duell}, Cody J. and
                   {Duff}, Shannon M. and {Duivenvoorden}, Adriaan J. and
                   {Dunkley}, Jo and {D{\"u}nner}, Rolando and
                   {Essinger-Hileman}, Thomas and {Fankhanel}, Max and
                   {Ferraro}, Simone and {Fox}, Anna E. and
                   {Fuzia}, Brittany and {Gallardo}, Patricio A. and
                   {Gluscevic}, Vera and {Golec}, Joseph E. and
                   {Grace}, Emily and {Gralla}, Megan and {Guan}, Yilun and
                   {Hall}, Kirsten and {Halpern}, Mark and
                   {Han}, Dongwon and {Hargrave}, Peter and
                   {Henderson}, Shawn and {Hensley}, Brandon and
                   {Hill}, J. Colin and {Hilton}, Gene C. and
                   {Hilton}, Matt and {Hincks}, Adam D. and
                   {Hlo{\v{z}}ek}, Ren{\'e}e and {Hubmayr}, Johannes and
                   {Huffenberger}, Kevin M. and {Hughes}, John P. and
                   {Infante}, Leopoldo and {Irwin}, Kent and
                   {Jackson}, Rebecca and {Klein}, Jeff and
                   {Knowles}, Kenda and {Kosowsky}, Arthur and
                   {Lakey}, Vincent and {Li}, Dale and {Li}, Yaqiong and
                   {Li}, Zack and {Lokken}, Martine and {Louis}, Thibaut and
                   {MacInnis}, Amanda and {Madhavacheril}, Mathew and
                   {Maldonado}, Felipe and {Mallaby-Kay}, Maya and
                   {Marsden}, Danica and {Maurin}, Lo{\"\i}c and
                   {McMahon}, Jeff and {Menanteau}, Felipe and
                   {Moodley}, Kavilan and {Morton}, Tim and
                   {Naess}, Sigurd and {Namikawa}, Toshiya and
                   {Nati}, Federico and {Newburgh}, Laura and
                   {Nibarger}, John P. and {Nicola}, Andrina and
                   {Niemack}, Michael D. and {Nolta}, Michael R. and
                   {Orlowski-Sherer}, John and {Page}, Lyman A. and
                   {Pappas}, Christine G. and {Partridge}, Bruce and
                   {Phakathi}, Phumlani and {Prince}, Heather and
                   {Puddu}, Roberto and {Qu}, Frank J. and
                   {Rivera}, Jesus and {Robertson}, Naomi and
                   {Rojas}, Felipe and {Salatino}, Maria and
                   {Schaan}, Emmanuel and {Schillaci}, Alessandro and
                   {Schmitt}, Benjamin L. and {Sehgal}, Neelima and
                   {Sherwin}, Blake D. and {Sierra}, Carlos and
                   {Sievers}, Jon and {Sifon}, Cristobal and
                   {Sikhosana}, Precious and {Simon}, Sara and
                   {Spergel}, David N. and {Staggs}, Suzanne T. and
                   {Stevens}, Jason and {Storer}, Emilie and
                   {Sunder}, Dhaneshwar D. and {Switzer}, Eric R. and
                   {Thorne}, Ben and {Thornton}, Robert and {Trac}, Hy and
                   {Treu}, Jesse and {Tucker}, Carole and
                   {Vale}, Leila R. and {Van Engelen}, Alexander and
                   {Van Lanen}, Jeff and {Vavagiakis}, Eve M. and
                   {Wagoner}, Kasey and {Wang}, Yuhan and
                   {Ward}, Jonathan T. and {Wollack}, Edward J. and
                   {Xu}, Zhilei and {Zago}, Fernando and
                   {Zhu}, Ningfeng},
  journal =       {arXiv e-prints},
  month =         jul,
  pages =         {arXiv:2007.07289},
  title =         {{The Atacama Cosmology Telescope: A Measurement of
                   the Cosmic Microwave Background Power Spectra at 98
                   and 150 GHz}},
  year =          {2020},
  eid =           {arXiv:2007.07289},
}

@article{dhawan2023a,
  author =        {{Dhawan}, Suhail and {Thorp}, Stephen and
                   {Mandel}, Kaisey S. and {Ward}, Sam M. and
                   {Narayan}, Gautham and {Jha}, Saurabh W. and
                   {Chant}, Thaisen},
  journal =       {\mnras},
  month =         sep,
  number =        {1},
  pages =         {235-244},
  title =         {{A BayeSN distance ladder: H$_{0}$ from a consistent
                   modelling of Type Ia supernovae from the optical to
                   the near-infrared}},
  volume =        {524},
  year =          {2023},
  doi =           {10.1093/mnras/stad1590},
}

@article{verde2019a,
  author =        {{Verde}, Licia and {Treu}, Tommaso and
                   {Riess}, Adam G.},
  journal =       {Nature Astronomy},
  month =         sep,
  pages =         {891-895},
  title =         {{Tensions between the early and late Universe}},
  volume =        {3},
  year =          {2019},
  doi =           {10.1038/s41550-019-0902-0},
}

@article{refsdal1964a,
  author =        {{Refsdal}, S.},
  journal =       {\mnras},
  pages =         {307},
  title =         {{On the possibility of determining Hubble's parameter
                   and the masses of galaxies from the gravitational
                   lens effect}},
  volume =        {128},
  year =          {1964},
  doi =           {10.1093/mnras/128.4.307},
}

@article{treu2010a,
  author =        {{Treu}, Tommaso},
  journal =       {\araa},
  month =         sep,
  pages =         {87-125},
  title =         {{Strong Lensing by Galaxies}},
  volume =        {48},
  year =          {2010},
  doi =           {10.1146/annurev-astro-081309-130924},
}

@article{oguri2010a,
  author =        {{Oguri}, M. and {Marshall}, P.~J.},
  journal =       {\mnras},
  month =         jul,
  pages =         {2579-2593},
  title =         {{Gravitationally lensed quasars and supernovae in
                   future wide-field optical imaging surveys}},
  volume =        {405},
  year =          {2010},
  doi =           {10.1111/j.1365-2966.2010.16639.x},
}

@article{suyu2023a,
  author =        {{Suyu}, Sherry H. and {Goobar}, Ariel and
                   {Collett}, Thomas and {More}, Anupreeta and
                   {Vernardos}, Giorgos},
  journal =       {arXiv e-prints},
  month =         jan,
  pages =         {arXiv:2301.07729},
  title =         {{Strong gravitational lensing and microlensing of
                   supernovae}},
  year =          {2023},
  doi =           {10.48550/arXiv.2301.07729},
  eid =           {arXiv:2301.07729},
}

@article{kelly2023a,
  author =        {{Kelly}, Patrick L. and {Rodney}, Steven and
                   {Treu}, Tommaso and {Birrer}, Simon and
                   {Bonvin}, Vivien and {Dessart}, Luc and
                   {Foley}, Ryan J. and {Filippenko}, Alexei V. and
                   {Gilman}, Daniel and {Jha}, Saurabh and
                   {Hjorth}, Jens and {Mandel}, Kaisey and
                   {Millon}, Martin and {Pierel}, Justin and
                   {Thorp}, Stephen and {Zitrin}, Adi and
                   {Broadhurst}, Tom and {Chen}, Wenlei and
                   {Diego}, Jose M. and {Dressler}, Alan and {Graur}, Or and
                   {Jauzac}, Mathilde and {Malkan}, Matthew A. and
                   {McCully}, Curtis and {Oguri}, Masamune and
                   {Postman}, Marc and {Schmidt}, Kasper Borello and
                   {Sharon}, Keren and {Tucker}, Brad E. and
                   {von der Linden}, Anja and {Wambsganss}, Joachim},
  journal =       {\apj},
  month =         may,
  number =        {2},
  pages =         {93},
  title =         {{The Magnificent Five Images of Supernova Refsdal:
                   Time Delay and Magnification Measurements}},
  volume =        {948},
  year =          {2023},
  doi =           {10.3847/1538-4357/ac4ccb},
  eid =           {93},
}

@ARTICLE{pascale2025a,
       author = {{Pascale}, Massimo and {Frye}, Brenda L. and {Pierel}, Justin D.~R. and {Chen}, Wenlei and {Kelly}, Patrick L. and {Cohen}, Seth H. and {Windhorst}, Rogier A. and {Riess}, Adam G. and {Kamieneski}, Patrick S. and {Diego}, Jos{\'e} M. and {Meena}, Ashish K. and {Cha}, Sangjun and {Oguri}, Masamune and {Zitrin}, Adi and {Jee}, M. James and {Foo}, Nicholas and {Leimbach}, Reagen and {Koekemoer}, Anton M. and {Conselice}, C.~J. and {Dai}, Liang and {Goobar}, Ariel and {Siebert}, Matthew R. and {Strolger}, Lou and {Willner}, S.~P.},
        title = "{SN H0pe: The First Measurement of H$_{0}$ from a Multiply Imaged Type Ia Supernova, Discovered by JWST}",
      journal = {\apj},
         year = 2025,
        month = jan,
       volume = {979},
       number = {1},
          eid = {13},
        pages = {13},
          doi = {10.3847/1538-4357/ad9928},
archivePrefix = {arXiv},
       eprint = {2403.18902},
 primaryClass = {astro-ph.CO},
       adsurl = {https://ui.adsabs.harvard.edu/abs/2025ApJ...979...13P}
}

@article{linder2011a,
  author =        {{Linder}, Eric V.},
  journal =       {\prd},
  month =         dec,
  number =        {12},
  pages =         {123529},
  title =         {{Lensing time delays and cosmological
                   complementarity}},
  volume =        {84},
  year =          {2011},
  doi =           {10.1103/PhysRevD.84.123529},
  eid =           {123529},
}

@article{treu2016a,
  author =        {{Treu}, Tommaso and {Marshall}, Philip J.},
  journal =       {Astronomy and Astrophysics Review},
  month =         {Jul},
  pages =         {11},
  title =         {{Time delay cosmography}},
  volume =        {24},
  year =          {2016},
  doi =           {10.1007/s00159-016-0096-8},
  eid =           {11},
}

@article{pierel2021a,
  author =        {{Pierel}, J.~D.~R. and {Rodney}, S. and
                   {Vernardos}, G. and {Oguri}, M. and {Kessler}, R. and
                   {Anguita}, T.},
  journal =       {\apj},
  month =         feb,
  number =        {2},
  pages =         {190},
  title =         {{Projected Cosmological Constraints from Strongly
                   Lensed Supernovae with the Roman Space Telescope}},
  volume =        {908},
  year =          {2021},
  doi =           {10.3847/1538-4357/abd8d3},
  eid =           {190},
}

@article{sharma2022a,
  author =        {{Sharma}, Divij and {Linder}, Eric V.},
  journal =       {arXiv e-prints},
  month =         apr,
  pages =         {arXiv:2204.03020},
  title =         {{Double Source Lensing Probing High Redshift
                   Cosmology}},
  year =          {2022},
  eid =           {arXiv:2204.03020},
}

@article{sharma2023a,
  author =        {{Sharma}, Divij and {Collett}, Thomas E. and
                   {Linder}, Eric V.},
  journal =       {\jcap},
  month =         apr,
  number =        {4},
  pages =         {001},
  title =         {{Testing cosmology with double source lensing}},
  volume =        {2023},
  year =          {2023},
  doi =           {10.1088/1475-7516/2023/04/001},
  eid =           {001},
}

@article{collett2018a,
  author =        {{Collett}, Thomas E. and {Oldham}, Lindsay J. and
                   {Smith}, Russell J. and {Auger}, Matthew W. and
                   {Westfall}, Kyle B. and {Bacon}, David and
                   {Nichol}, Robert C. and {Masters}, Karen L. and
                   {Koyama}, Kazuya and {van den Bosch}, Remco},
  journal =       {Science},
  month =         jun,
  number =        {6395},
  pages =         {1342-1346},
  title =         {{A precise extragalactic test of General Relativity}},
  volume =        {360},
  year =          {2018},
  doi =           {10.1126/science.aao2469},
}

@article{dey2019a,
  author =        {{Dey}, Arjun and {Schlegel}, David J. and
                   {Lang}, Dustin and {Blum}, Robert and
                   {Burleigh}, Kaylan and {Fan}, Xiaohui and
                   {Findlay}, Joseph R. and {Finkbeiner}, Doug and
                   {Herrera}, David and {Juneau}, St{\'e}phanie and
                   {Landriau}, Martin and {Levi}, Michael and
                   {McGreer}, Ian and {Meisner}, Aaron and
                   {Myers}, Adam D. and {Moustakas}, John and
                   {Nugent}, Peter and {Patej}, Anna and
                   {Schlafly}, Edward F. and {Walker}, Alistair R. and
                   {Valdes}, Francisco and {Weaver}, Benjamin A. and
                   {Y{\`e}che}, Christophe and {Zou}, Hu and {Zhou}, Xu and
                   {Abareshi}, Behzad and {Abbott}, T.~M.~C. and
                   {Abolfathi}, Bela and {Aguilera}, C. and
                   {Alam}, Shadab and {Allen}, Lori and {Alvarez}, A. and
                   {Annis}, James and {Ansarinejad}, Behzad and
                   {Aubert}, Marie and {Beechert}, Jacqueline and
                   {Bell}, Eric F. and {BenZvi}, Segev Y. and
                   {Beutler}, Florian and {Bielby}, Richard M. and
                   {Bolton}, Adam S. and {Brice{\~n}o}, C{\'e}sar and
                   {Buckley-Geer}, Elizabeth J. and {Butler}, Karen and
                   {Calamida}, Annalisa and {Carlberg}, Raymond G. and
                   {Carter}, Paul and {Casas}, Ricard and
                   {Castander}, Francisco J. and {Choi}, Yumi and
                   {Comparat}, Johan and {Cukanovaite}, Elena and
                   {Delubac}, Timoth{\'e}e and {DeVries}, Kaitlin and
                   {Dey}, Sharmila and {Dhungana}, Govinda and
                   {Dickinson}, Mark and {Ding}, Zhejie and
                   {Donaldson}, John B. and {Duan}, Yutong and
                   {Duckworth}, Christopher J. and
                   {Eftekharzadeh}, Sarah and {Eisenstein}, Daniel J. and
                   {Etourneau}, Thomas and {Fagrelius}, Parker A. and
                   {Farihi}, Jay and {Fitzpatrick}, Mike and
                   {Font-Ribera}, Andreu and {Fulmer}, Leah and
                   {G{\"a}nsicke}, Boris T. and {Gaztanaga}, Enrique and
                   {George}, Koshy and {Gerdes}, David W. and
                   {Gontcho}, Satya Gontcho A. and {Gorgoni}, Claudio and
                   {Green}, Gregory and {Guy}, Julien and
                   {Harmer}, Diane and {Hernand ez}, M. and
                   {Honscheid}, Klaus and {(Wendy Huang}, Lijuan and
                   {James}, David J. and {Jannuzi}, Buell T. and
                   {Jiang}, Linhua and {Joyce}, Richard and
                   {Karcher}, Armin and {Karkar}, Sonia and
                   {Kehoe}, Robert and {Jean-Paul} and {Kneib} and
                   {Kueter-Young}, Andrea and {Lan}, Ting-Wen and
                   {Lauer}, Tod R. and {Le Guillou}, Laurent and
                   {Le Van Suu}, Auguste and {Lee}, Jae Hyeon and
                   {Lesser}, Michael and {Perreault Levasseur}, Laurence and
                   {Li}, Ting S. and {Mann}, Justin L. and
                   {Marshall}, Robert and
                   {Mart{\'\i}nez-V{\'a}zquez}, C.~E. and
                   {Martini}, Paul and {du Mas des Bourboux}, H{\'e}lion and
                   {McManus}, Sean and {Meier}, Tobias Gabriel and
                   {M{\'e}nard}, Brice and {Metcalfe}, Nigel and
                   {Mu{\~n}oz-Guti{\'e}rrez}, Andrea and {Najita}, Joan and
                   {Napier}, Kevin and {Narayan}, Gautham and
                   {Newman}, Jeffrey A. and {Nie}, Jundan and
                   {Nord}, Brian and {Norman}, Dara J. and
                   {Olsen}, Knut A.~G. and {Paat}, Anthony and
                   {Palanque-Delabrouille}, Nathalie and {Peng}, Xiyan and
                   {Poppett}, Claire L. and {Poremba}, Megan R. and
                   {Prakash}, Abhishek and {Rabinowitz}, David and
                   {Raichoor}, Anand and {Rezaie}, Mehdi and
                   {Robertson}, A.~N. and {Roe}, Natalie A. and
                   {Ross}, Ashley J. and {Ross}, Nicholas P. and
                   {Rudnick}, Gregory and {Safonova}, Sasha and
                   {Saha}, Abhijit and {S{\'a}nchez}, F. Javier and
                   {Savary}, Elodie and {Schweiker}, Heidi and
                   {Scott}, Adam and {Seo}, Hee-Jong and
                   {Shan}, Huanyuan and {Silva}, David R. and
                   {Slepian}, Zachary and {Soto}, Christian and
                   {Sprayberry}, David and {Staten}, Ryan and
                   {Stillman}, Coley M. and {Stupak}, Robert J. and
                   {Summers}, David L. and {Sien Tie}, Suk and
                   {Tirado}, H. and {Vargas-Maga{\~n}a}, Mariana and
                   {Vivas}, A. Katherina and {Wechsler}, Risa H. and
                   {Williams}, Doug and {Yang}, Jinyi and {Yang}, Qian and
                   {Yapici}, Tolga and {Zaritsky}, Dennis and
                   {Zenteno}, A. and {Zhang}, Kai and {Zhang}, Tianmeng and
                   {Zhou}, Rongpu and {Zhou}, Zhimin},
  journal =       {\aj},
  month =         {May},
  number =        {5},
  pages =         {168},
  title =         {{Overview of the DESI Legacy Imaging Surveys}},
  volume =        {157},
  year =          {2019},
  doi =           {10.3847/1538-3881/ab089d},
  eid =           {168},
}

@article{des2005a,
  author =        {{The Dark Energy Survey Collaboration}},
  journal =       {arXiv e-prints},
  month =         {Oct},
  pages =         {astro-ph/0510346},
  title =         {{The Dark Energy Survey}},
  year =          {2005},
  eid =           {astro-ph/0510346},
}

@article{huang2020a,
  author =        {X. Huang and C. Storfer and V. Ravi and A. Pilon and
                   M. Domingo and D. J. Schlegel and S. Bailey and
                   A. Dey and R. R. Gupta and D. Herrera and S. Juneau and
                   M. Landriau and D. Lang and A. Meisner and
                   J. Moustakas and A. D. Myers and E. F. Schlafly and
                   F. Valdes and B. A. Weaver and J. Yang and
                   C. Y{\`{e}}che},
  journal =       {The Astrophysical Journal},
  month =         {may},
  number =        {1},
  pages =         {78},
  publisher =     {American Astronomical Society},
  title =         {Finding Strong Gravitational Lenses in the {DESI}
                   {DECam} Legacy Survey},
  volume =        {894},
  year =          {2020},
  doi =           {10.3847/1538-4357/ab7ffb},
  url =           {https://doi.org/10.3847%2F1538-4357%2Fab7ffb},
}

@article{huang2021a,
  author =        {{Huang}, X. and {Storfer}, C. and {Gu}, A. and
                   {Ravi}, V. and {Pilon}, A. and {Sheu}, W. and
                   {Venguswamy}, R. and {Banka}, S. and {Dey}, A. and
                   {Landriau}, M. and {Lang}, D. and {Meisner}, A. and
                   {Moustakas}, J. and {Myers}, A.~D. and {Sajith}, R. and
                   {Schlafly}, E.~F. and {Schlegel}, D.~J.},
  journal =       {\apj},
  month =         mar,
  number =        {1},
  pages =         {27},
  title =         {{Discovering New Strong Gravitational Lenses in the
                   DESI Legacy Imaging Surveys}},
  volume =        {909},
  year =          {2021},
  doi =           {10.3847/1538-4357/abd62b},
  eid =           {27},
}

@article{storfer2024a,
  author =        {{Storfer}, C. and {Huang}, X. and {Gu}, A. and
                   {Sheu}, W. and {Banka}, S. and {Dey}, A. and
                   {Inchausti Reyes}, J. and {Jain}, A. and
                   {Kwon}, K.~J. and {Lang}, D. and {Lee}, V. and
                   {Meisner}, A. and {Moustakas}, J. and {Myers}, A.~D. and
                   {Tabares-Tarquinio}, S. and {Schlafly}, E.~F. and
                   {Schlegel}, D.~J.},
  journal =       {\apjs},
  month =         sep,
  number =        {1},
  pages =         {16},
  title =         {{New Strong Gravitational Lenses from the DESI Legacy
                   Imaging Surveys Data Release 9}},
  volume =        {274},
  year =          {2024},
  doi =           {10.3847/1538-4365/ad527e},
  eid =           {16},
}

@article{dawes2023a,
  author =        {C. Dawes and C. Storfer and X. Huang and G. Aldering and
                   Aleksandar Cikota and Arjun Dey and D. J. Schlegel},
  journal =       {The Astrophysical Journal Supplement Series},
  month =         {dec},
  number =        {2},
  pages =         {61},
  publisher =     {The American Astronomical Society},
  title =         {Finding Multiply Lensed and Binary Quasars in the
                   DESI Legacy Imaging Surveys},
  volume =        {269},
  year =          {2023},
  doi =           {10.3847/1538-4365/ad015a},
  url =           {https://dx.doi.org/10.3847/1538-4365/ad015a},
}

@article{gu2022a,
  author =        {{Gu}, A. and {Huang}, X. and {Sheu}, W. and
                   {Aldering}, G. and {Bolton}, A.~S. and {Boone}, K. and
                   {Dey}, A. and {Filipp}, A. and {Jullo}, E. and
                   {Perlmutter}, S. and {Rubin}, D. and
                   {Schlafly}, E.~F. and {Schlegel}, D.~J. and {Shu}, Y. and
                   {Suyu}, S.~H.},
  journal =       {\apj},
  month =         aug,
  number =        {1},
  pages =         {49},
  title =         {{GIGA-Lens: Fast Bayesian Inference for Strong
                   Gravitational Lens Modeling}},
  volume =        {935},
  year =          {2022},
  doi =           {10.3847/1538-4357/ac6de4},
  eid =           {49},
}

@article{gelman1992a,
  author =        {Andrew Gelman and Donald B. Rubin},
  journal =       {Statistical Science},
  number =        {4},
  pages =         {457 -- 472},
  publisher =     {Institute of Mathematical Statistics},
  title =         {{Inference from Iterative Simulation Using Multiple
                   Sequences}},
  volume =        {7},
  year =          {1992},
  doi =           {10.1214/ss/1177011136},
  url =           {https://doi.org/10.1214/ss/1177011136},
}

@book{gelman2014a,
  author =        {{Gelman}, Andrew and {Carlin}, John B. and
                   {Stern}, Hal S. and {Dunson}, David B. and
                   {Vehtari}, Aki and {Rubin}, Donald B.},
  title =         {{Bayesian Data Analysis}},
  year =          {2014},
}

@article{birrer2015a,
  author =        {{Birrer}, Simon and {Amara}, Adam and
                   {Refregier}, Alexandre},
  journal =       {\apj},
  month =         nov,
  number =        {2},
  pages =         {102},
  title =         {{Gravitational Lens Modeling with Basis Sets}},
  volume =        {813},
  year =          {2015},
  doi =           {10.1088/0004-637X/813/2/102},
  eid =           {102},
}

@article{tamm2006a,
  author =        {{Tamm}, A. and {Tenjes}, P.},
  journal =       {\aap},
  month =         apr,
  number =        {1},
  pages =         {67-78},
  title =         {{Surface photometry and structure of high redshift
                   disk galaxies in the HDF-S NICMOS field}},
  volume =        {449},
  year =          {2006},
  doi =           {10.1051/0004-6361:20054065},
}

@ARTICLE{shajib2020b,
       author = {{Shajib}, Anowar J. and {Treu}, Tommaso and {Birrer}, Simon and
         {Sonnenfeld}, Alessandro},
        title = "{Massive elliptical galaxies at $z \sim 0.2$ are well described by stars and a Navarro-Frenk-White dark matter halo}",
      journal = {arXiv e-prints},
         year = 2020,
        month = aug,
          eid = {arXiv:2008.11724},
        pages = {arXiv:2008.11724},
archivePrefix = {arXiv},
       eprint = {2008.11724},
 primaryClass = {astro-ph.GA},
       adsurl = {https://ui.adsabs.harvard.edu/abs/2020arXiv200811724S}
}

@article{mandel2022a,
  author =        {{Mandel}, Kaisey S. and {Thorp}, Stephen and
                   {Narayan}, Gautham and {Friedman}, Andrew S. and
                   {Avelino}, Arturo},
  journal =       {\mnras},
  month =         mar,
  number =        {3},
  pages =         {3939-3966},
  title =         {{A hierarchical Bayesian SED model for Type Ia
                   supernovae in the optical to near-infrared}},
  volume =        {510},
  year =          {2022},
  doi =           {10.1093/mnras/stab3496},
}

@misc{TensorFlow,
  author =        {Mart\'{\i}n~Abadi and Ashish~Agarwal and Paul~Barham and
                   Eugene~Brevdo and Zhifeng~Chen and Craig~Citro and
                   Greg~S.~Corrado and Andy~Davis and Jeffrey~Dean and
                   Matthieu~Devin and Sanjay~Ghemawat and Ian~Goodfellow and
                   Andrew~Harp and Geoffrey~Irving and Michael~Isard and
                   Yangqing Jia and Rafal~Jozefowicz and Lukasz~Kaiser and
                   Manjunath~Kudlur and Josh~Levenberg and
                   Dandelion~Man\'{e} and Rajat~Monga and Sherry~Moore and
                   Derek~Murray and Chris~Olah and Mike~Schuster and
                   Jonathon~Shlens and Benoit~Steiner and Ilya~Sutskever and
                   Kunal~Talwar and Paul~Tucker and Vincent~Vanhoucke and
                   Vijay~Vasudevan and Fernanda~Vi\'{e}gas and
                   Oriol~Vinyals and Pete~Warden and Martin~Wattenberg and
                   Martin~Wicke and Yuan~Yu and Xiaoqiang~Zheng},
  note =          {Software available from tensorflow.org},
  title =         {{TensorFlow}: Large-Scale Machine Learning on
                   Heterogeneous Systems},
  year =          {2015},
  url =           {https://www.tensorflow.org/},
}

@misc{dillon2017a,
  author =        {Joshua V. Dillon and Ian Langmore and Dustin Tran and
                   Eugene Brevdo and Srinivas Vasudevan and Dave Moore and
                   Brian Patton and Alex Alemi and Matt Hoffman and
                   Rif A. Saurous},
  title =         {TensorFlow Distributions},
  year =          {2017},
}

@misc{bradbury2018a,
  author =        {James Bradbury and Roy Frostig and Peter Hawkins and
                   Matthew James Johnson and Chris Leary and
                   Dougal Maclaurin and George Necula and Adam Paszke and
                   Jake Vander{P}las and Skye Wanderman-{M}ilne and
                   Qiao Zhang},
  title =         {{JAX}: composable transformations of
                   {P}ython+{N}um{P}y programs},
  year =          {2018},
  url =           {http://github.com/google/jax},
}

@misc{optax2020,
  author =        {Matteo Hessel and David Budden and Fabio Viola and
                   Mihaela Rosca and Eren Sezener and Tom Hennigan},
  title =         {Optax: composable gradient transformation and
                   optimisation, in JAX!},
  year =          {2020},
  url =           {http://github.com/deepmind/optax},
}

@article{birrer2018a,
  author =        {{Birrer}, Simon and {Amara}, Adam},
  journal =       {Physics of the Dark Universe},
  month =         dec,
  pages =         {189-201},
  title =         {{lenstronomy: Multi-purpose gravitational lens
                   modelling software package}},
  volume =        {22},
  year =          {2018},
  doi =           {10.1016/j.dark.2018.11.002},
}

@article{hunter2007a,
  author =        {Hunter, J. D.},
  journal =       {Computing in Science \& Engineering},
  number =        {3},
  pages =         {90--95},
  publisher =     {IEEE COMPUTER SOC},
  title =         {Matplotlib: A 2D graphics environment},
  volume =        {9},
  year =          {2007},
  doi =           {10.1109/MCSE.2007.55},
}

@misc{bradley2023a,
  author =        {Larry Bradley and Brigitta Sip{\H o}cz and
                   Thomas Robitaille and Erik Tollerud and
                   Z\`e Vin{\'{\i}}cius and Christoph Deil and
                   Kyle Barbary and Tom J Wilson and Ivo Busko and
                   Axel Donath and Hans Moritz G{\"u}nther and
                   Mihai Cara and P. L. Lim and Sebastian Me{\ss}linger and
                   Simon Conseil and Azalee Bostroem and
                   Michael Droettboom and E. M. Bray and
                   Lars Andersen Bratholm and Geert Barentsen and
                   Matt Craig and Shivangee Rathi and Sergio Pascual and
                   Gabriel Perren and Iskren Y. Georgiev and
                   Miguel de Val-Borro and Wolfgang Kerzendorf and
                   Yoonsoo P. Bach and Bruno Quint and
                   Harrison Souchereau},
  month =         may,
  publisher =     {Zenodo},
  title =         {astropy/photutils: 1.8.0},
  year =          {2023},
  doi =           {10.5281/zenodo.7946442},
  url =           {https://doi.org/10.5281/zenodo.7946442},
}

@article{waskom2021a,
  author =        {Michael L. Waskom},
  journal =       {Journal of Open Source Software},
  number =        {60},
  pages =         {3021},
  publisher =     {The Open Journal},
  title =         {seaborn: statistical data visualization},
  volume =        {6},
  year =          {2021},
  doi =           {10.21105/joss.03021},
  url =           {https://doi.org/10.21105/joss.03021},
}

@article{foreman2016a,
  author =        {Daniel Foreman-Mackey},
  journal =       {The Journal of Open Source Software},
  month =         {jun},
  number =        {2},
  pages =         {24},
  publisher =     {The Open Journal},
  title =         {corner.py: Scatterplot matrices in Python},
  volume =        {1},
  year =          {2016},
  doi =           {10.21105/joss.00024},
  url =           {https://doi.org/10.21105/joss.00024},
}

@article{harris2020a,
  author =        {Charles R. Harris and K. Jarrod Millman and
                   St{\'{e}}fan J. van der Walt and Ralf Gommers and
                   Pauli Virtanen and David Cournapeau and Eric Wieser and
                   Julian Taylor and Sebastian Berg and
                   Nathaniel J. Smith and Robert Kern and Matti Picus and
                   Stephan Hoyer and Marten H. van Kerkwijk and
                   Matthew Brett and Allan Haldane and
                   Jaime Fern{\'{a}}ndez del R{\'{i}}o and Mark Wiebe and
                   Pearu Peterson and Pierre G{\'{e}}rard-Marchant and
                   Kevin Sheppard and Tyler Reddy and Warren Weckesser and
                   Hameer Abbasi and Christoph Gohlke and
                   Travis E. Oliphant},
  journal =       {\nat},
  month =         sep,
  number =        {7825},
  pages =         {357--362},
  publisher =     {Springer Science and Business Media {LLC}},
  title =         {Array programming with {NumPy}},
  volume =        {585},
  year =          {2020},
  doi =           {10.1038/s41586-020-2649-2},
  url =           {https://doi.org/10.1038/s41586-020-2649-2},
}

@article{brooks1998a,
  author =        {Stephen P. Brooks and Andrew Gelman},
  journal =       {Journal of Computational and Graphical Statistics},
  number =        {4},
  pages =         {434--455},
  publisher =     {ASA Website},
  title =         {General Methods for Monitoring Convergence of
                   Iterative Simulations},
  volume =        {7},
  year =          {1998},
  doi =           {10.1080/10618600.1998.10474787},
  url =           {https://www.tandfonline.com/doi/abs/10.1080/
                  10618600.1998.10474787},
}

\appendix

\section{Effective PSF generation}\label{sec:psf_section}
This section describes the empirical point spread function (PSF) used for lens modeling.
The PSF for each system is generated from stars 
extracted from the respective fits file containing the strong lensing system. 
We show the image containing  \desionefivefour as an example in figure \ref{fig:psf_cutout}. 
With the aim of selecting several stars in the \hst image, we specify  an appropriate pixel value threshold. 
Pixels belonging to other kinds of objects (e.g., galaxies) may be 
included in the list of brightest pixels. 
We therefore compare this list
with the pixel positions in the image, and only keep the pixels that belong to stars. 
Moreover, we avoid very bright, saturated stars.


\begin{minipage}{\linewidth}
\makebox[\linewidth]{
\includegraphics[keepaspectratio=true,scale=0.63]{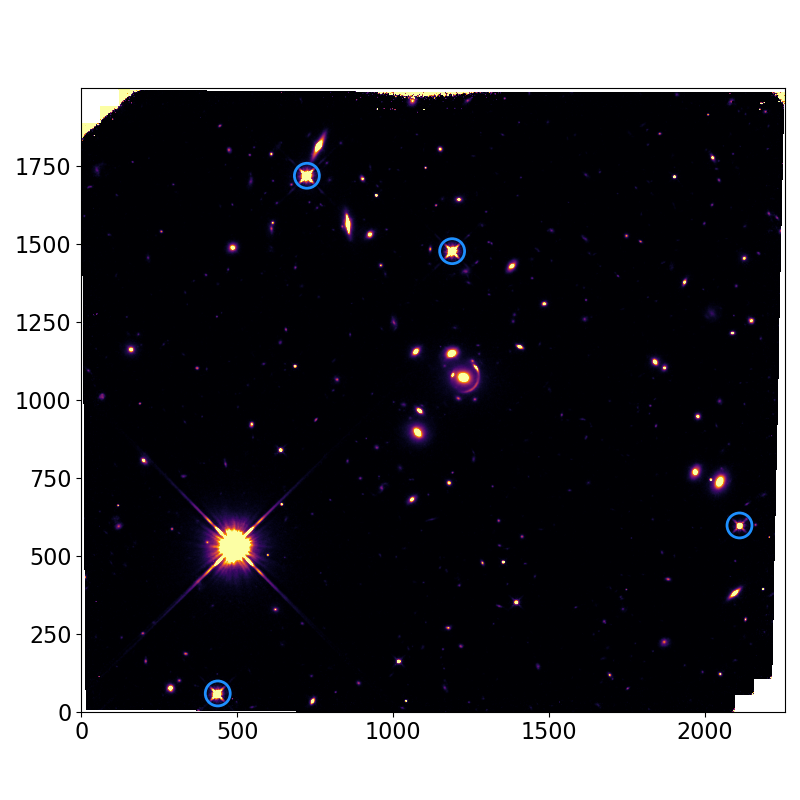}}
\captionof{figure}{\hst image containing \desionefivefour (near the center of the image). Circled in blue are the four stars used to generate the empirical PSF.}
\label{fig:psf_cutout}
\end{minipage}

Finally, the empirical PSF is generated from these cutouts using the \texttt{EPSFBuilder} function of \texttt{photutils}\footnote{\url{https://photutils.readthedocs.io/en/stable/}}. 
As an example, Figure \ref{fig:psf_result} shows an empirical PSF for \desionefivefour.



\begin{minipage}{\linewidth}
\makebox[\linewidth]{
\includegraphics[keepaspectratio=true,scale=0.75]{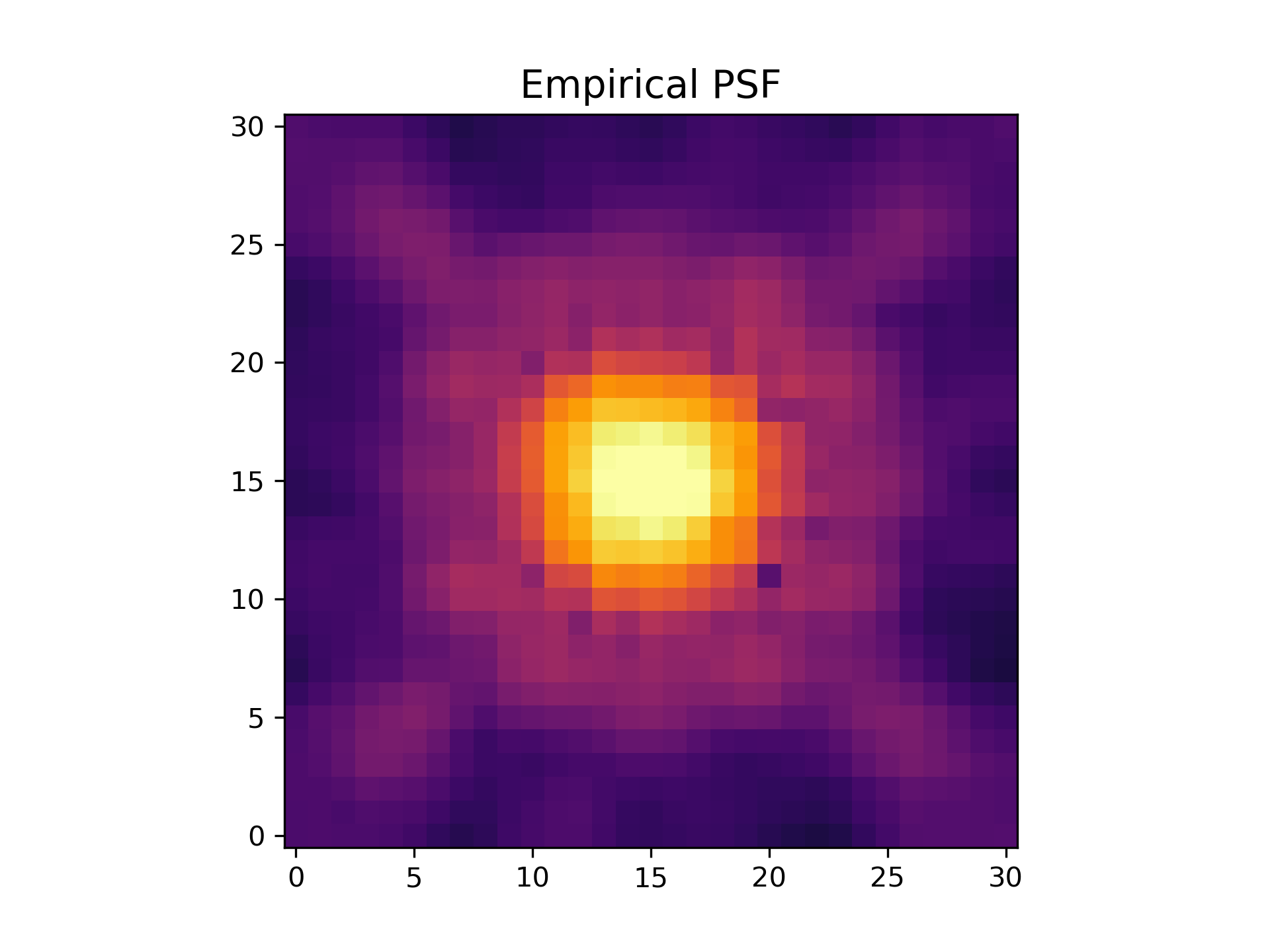}}
\captionof{figure}{Empirical PSF for \desionefivefour, with 31 pixels on the side.}
\label{fig:psf_result}
\end{minipage}

\section{Lens Modeling Results}\label{sec:lens-models-app}
Below we present the modeling results of \desionefivefour (\S~\ref{sec:desi154}),
\desionesixfive (modeled in Paper ~I, briefly summarized in \S~\ref{sec:desi165}),
\desizeroninefour (\S~\ref{sec:desi094}),
\desitwofiveseven (\S~\ref{sec:desi257}), 
and \desitwofoursix (\S~\ref{sec:desi246}),
ordered by increasing lens redshift.

\subsection{\desionefivefour}\label{sec:desi154}
\href{https://www.legacysurvey.org/viewer/?ra=154.6972&dec=-01.3590&layer=ls-dr10-grz&pixscale=0.26&size=101&zoom=16}{\desionefivefour} was discovered in H20, with a ResNet probability of 1.000, and a human grade of A.
The lens redshift, 
$\zd = 0.3884$, is from SDSS~DR16 \citep{ahumada2020a} and
DESI obtained its source redshift, \zs = 1.4304 with a strong [\ion{O}{2}] detection.\footnote{There are actually two DESI spectra targeting the brightest knot of the main arc, yielding two slightly different redshifts at 1.4303 and 1.4305. This is within the redshift uncertainty of DESI (Paper~II). 
In this work, we use the average of  these two values, $\zs = 1.4304$.}

Figure \ref{fig:system154} shows  \desionefivefour.
The Legacy Surveys image for this system features an elliptical galaxy as the lens. 
To its right, there is a long lensed arc with a bright ``knot'' near the top (denoted with the letter ``A'' in Figure~\ref{fig:system154}). 
To the left of the lens is the counter image of the same lensed source, referred to by the letter ``B''.  

The cutout for this system has a size of $120 \times 120$ pixels. Besides the central elliptical galaxy and the lensed source, the cutout also contains other objects. 
There is a small environmental galaxy to the upper left of the main lensed arc (arrow in Figure~\ref{fig:system154}). 
We include it as part of the lens model, and we will refer to it as the ``nearby galaxy'' (we only model its light).
There are a few other even dimmer objects. 
The light and mass of these objects are not included in the lens model. 
They are masked out during the modeling process (circles in Figure \ref{fig:system154}). 
Finally, there is a bright object at about 5.5\twopr from the lensing galaxy, to its upper left (also circled in Figure~\ref{fig:system154}).
DESI DR1 \citep{desidrone2025a} reveals it to be a galaxy at $z = 0.7640$. 
It is far enough away that we do not model its mass and simply mask out its light.

\begin{minipage}{\linewidth}
\makebox[\linewidth]{
\includegraphics[keepaspectratio=true,scale=0.90]{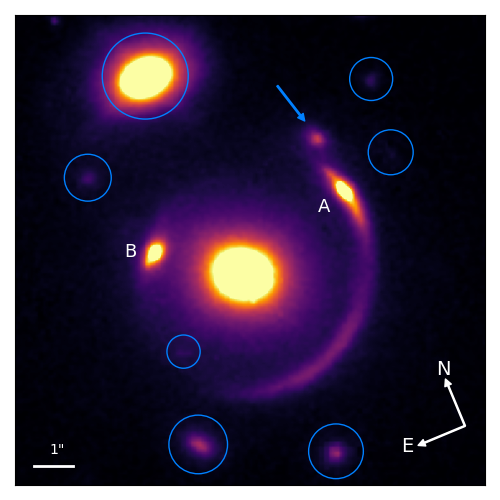}}
\captionof{figure}{\hst image of \desionefivefour.
The brightest parts of the lensed images are labeled as A and B.
Fainted objects circled in blue are masked and not modeled.
We model the light of the small environmental galaxy (arrow) north of image A, and then fix its parameters (see text).}
\label{fig:system154}
\end{minipage}

This is a challenging system to model, for three reasons. Firstly, its size. With an Einstein radius of nearly $\tE \sim 3''$, it stands at the limit between galaxy and group scale strong lensing \citep[e.g.,][]{limousin2009a}.
For group/cluster lenses, the modeling is usually done only using the images positions and not at the pixel level \citep[e.g.,][]{verdugo2011a}.
In addition, given that the area enclosed by \tE is $\sim 9\times$ larger than typical galaxy-scale strong lens with a $\tE \sim 1''$, it is not a surprise that there are more environmental galaxies near the arcs.
Each of these needs to be properly treated.
Secondly, the lens is very bright.
This is expected given that it is a massive lensing galaxy with a large \tE.
It is so luminous that faint diffraction spikes are visible out to $\sim 1.5''$ from the center of the lensing galaxy.
Thus, we need to use a large PSF size of $59 \times 59$ pixels ($\sim 4\times$ larger than for the other systems in this work) to properly model its light at large radii.
Thirdly, the lensed images are also very bright. 
In fact, their peak flux values are comparable to those of the lens; this is unusual.
Typically, the lensed arcs are considerably fainter than the lens, which is certainly the case for the other systems in this work.
But for this system, the signal to noise ratio (SNR) of the lensed images is very high. 
In fact, image B has an SNR around 250, and image A around 225,  calculated by averaging flux value of the top 5\% pixels. 
For comparison, the SNR's of the top 5\% brightest pixels of the four images for \desitwothreefour (\S~\ref{sec:desi234}) range between 63.1 and 84.3.
It is well known that source structure can be difficult to model well.
With a very high SNR for the ``knots'' in the lensed images in this system, 
this poses a significant challenge to lens modeling.

Our model consists of four elliptical \ser's for the source light. 
For lens light, we mask the center of the lens and model it with a single elliptical \ser. 
Finally, for the nearby galaxy, we do not model its mass and its light is modeled by a single elliptical \ser profile. 
Initially, the six light parameters of this galaxy were modeled together with the rest of the model's parameters.
Once a good residual was achieved in that area, 
their values were fixed. This helps reduce the dimensionality of the model parameter space.

This system is modeled with linear inversion to solve for the light intensities. The final model after sampling from HMC is shown in Figure \ref{fig:model154}. The final reduced $\chi^{2}$ is $0.786$. The best-fit mass parameters for the lens appear in Table \ref{tab:best-fit-mass-params154}, and their posterior distribution in Figure \ref{fig:cornermass154}.
The best-fit values for the light parameters appear in Table~\ref{tab:light_params_154}.

\begin{minipage}{\linewidth}
\makebox[\linewidth]{
\includegraphics[keepaspectratio=true,scale=0.53]{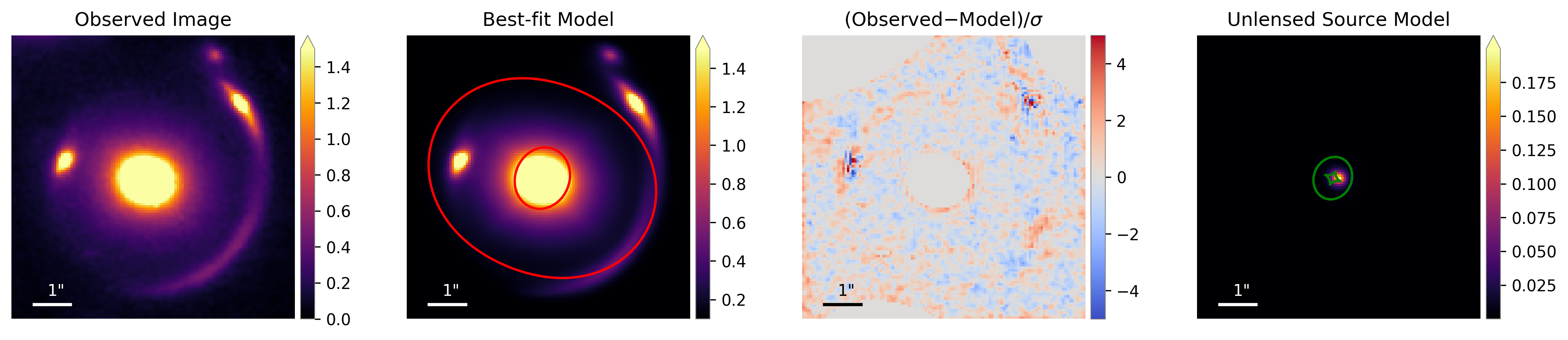}}
\captionof{figure}{
Our best-fit model. 
From left to right, we show:
the observed \emph{Hubble} image (the orientation is the same as in Figure \ref{fig:system154}), our best-fit model with critical curves. Note that since $\gamma<2 $,
there are two critical curves and two caustics), 
the reduced residual,
and the unlensed source with caustics.
}
\label{fig:model154}
\end{minipage}

\begin{deluxetable}{lccccccc}[H]
\tabletypesize{\scriptsize}
\tablecaption{Best-fit mass parameters for \desionefivefour.
\label{tab:best-fit-mass-params154}}
\renewcommand{\arraystretch}{1.4}
\tablehead{
    \colhead{$\theta_E$} &
    \colhead{$\gamma$} &
    \colhead{$\epsilon_1$} &
    \colhead{$\epsilon_2$} &
    \colhead{$x$} &
    \colhead{$y$} &
    \colhead{$\gamma_{ext, 1}$} &
    \colhead{$\gamma_{ext, 2}$} 
}
\startdata
$2.8982_{-0.0011}^{+0.0010}$& 
$1.411\pm{0.016}$&	
$0.0196_{-0.0020}^{+0.0022}$&
$-0.0322\pm{0.0017}$& 
$-0.1649_{-0.0035}^{+0.0033}$& 
$-0.0287\pm{0.0019}$&	
$-0.0076_{-0.0022}^{+0.0024}$&	
$-0.0027\pm{0.0015}$ \\
\enddata
\end{deluxetable}

\vspace{-1.9cm}

\begin{deluxetable}{l|c|cccc}[H]
\tablecaption{Best-fit light parameters for \desionefivefour}
\label{tab:light_params_154}
\renewcommand{\arraystretch}{1.4}
\setlength{\tabcolsep}{8pt}
\tablehead{
    \colhead{Parameter} & 
    \colhead{Lens light} & 
    \multicolumn{4}{c}{Source light} \\
    \colhead{} & 
    \colhead{} & 
    \colhead{Comp 1} & \colhead{Comp 2} & \colhead{Comp 3} & \colhead{Comp 4}
}
\startdata
$\bm{R_e}$ & $2.954_{-0.062}^{+0.064}$ & $0.0680\pm{0.0035}$ & $0.0533_{-0.0069}^{+0.0073}$ & $0.0110_{-0.0004}^{+0.0005}$ & $0.0520_{-0.0020}^{+0.0021}$ \\[2pt]
$\bm{n}$ & $7.334\pm{0.093}$ & $2.57_{-0.13}^{+0.15}$ & $7.14_{-0.48}^{+0.27}$ & $0.709_{-0.045}^{+0.049}$ & $0.344_{-0.029}^{+0.030}$ \\[2pt]
$\bm{\epsilon_{1}}$ & 
$0.0881\pm{0.0008}$ & $-0.099_{-0.012}^{+0.011}$ & $0.010^{+0.024}_{-0.023}$ & $0.039\pm{0.022}$ & $0.375_{-0.015}^{+0.014}$ \\[2pt]
$\bm{\epsilon_{2}}$ & 
$-0.0519\pm{0.0008}$ & 
$-0.073\pm{0.017}$ & $-0.2564^{+0.027}_{-0.029}$ & $-0.246\pm{0.017}$ &
$0.402\pm{0.014}$ \\[2pt]
$\bm{x}$ & 
$-0.1814\pm{0.0009}$ & $0.0178^{+0.0080}_{-0.0083}$ & $-0.0307_{-0.0068}^{+0.0062}$ & $-0.0116_{-0.0074}^{+0.0070}$ & $-0.0228_{-0.0074}^{+0.0071}$ \\[2pt]
$\bm{y}$ & $-0.0893\pm{0.0009}$ & $-0.0553_{-0.0020}^{+0.0021}$ & $0.0171\pm{0.0025}$ & 
$-0.0236\pm{0.0020}$ & $-0.0460_{-0.0022}^{+0.0021}$ \\[2pt]
\enddata
\end{deluxetable}

\begin{figure}[H]
    \centering
    \includegraphics[width=0.85\linewidth]{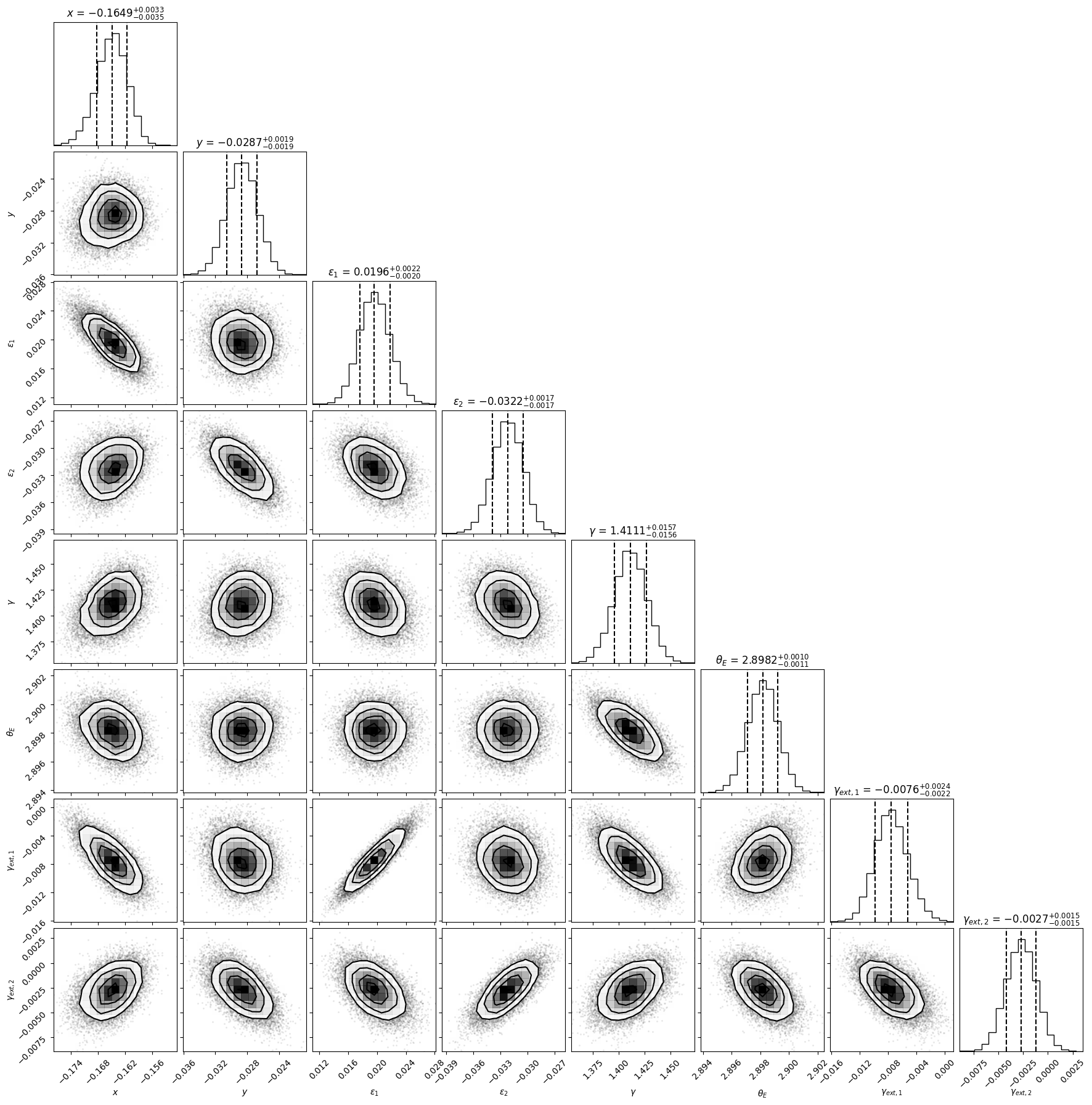}
    \caption{Corner plot for the eight mass parameters for \desionefivefour.}
    \label{fig:cornermass154}
\end{figure}

The \rhat values for all parameters satisfy the convergence criterion, $\rhat<1.1$. In fact, the mean $\rhat$ is $1.0011$ and the maximum $1.0047$. 
 
From the HMC results, we estimate the magnification and the mass enclosed inside the critical curve. For the former, we use three different methods as described in the Appendix of Paper~I. In particular, we computed the magnification for lensed images A (including the main lensed arc) and B separately (see Figure \ref{fig:system154}). The results appear in Table \ref{tab:magnifications_154}. We find the projected enclosed mass within the \tE to be  $M (< \tE) =1.8203\pm_{0.0014}^{0.0013}\times 10^{12}M_{\odot}$. 

\begin{deluxetable}{lccc}[H]
\tabletypesize{\footnotesize}
\renewcommand{\arraystretch}{1.4}
\tablecaption{Magnification calculations for \desionefivefour using three approaches, for images A and B, and the total magnification, respectively.}
\label{tab:magnifications_154}
\tablehead{
    \colhead{Method} &
    \colhead{$A$} &
    \colhead{$B$} &
    \colhead{Total}
}
\startdata
$1^{\text{st}}$ & $88.88\pm^{7.82}_{7.87}$ & $19.66\pm^{1.59}_{1.86}$ & $108.55\pm^{9.42}_{9.55}$\\
$2^{\text{nd}}$ & $82.19\pm^{7.27}_{7.87}$ & $21.83\pm^{1.67}_{1.98}$ & $104.02\pm^{8.74}_{9.95}$\\
$3^{\text{rd}}$ & $80.85\pm^{6.15}_{7.75}$ & $19.97\pm^{1.56}_{2.00}$ & $100.82\pm^{7.57}_{9.87}$\\
\enddata
\end{deluxetable}

\vspace{-1cm}
\subsection{\desionesixfive }\label{sec:desi165}
\desionesixfive was discovered in H21 and modeled in Paper~I.
Based on color and photo-$z$, 
this galaxy appears to be at the center, 
and the brightest member, of a small group.
The lens and source redshifts are $z_d = 0.4834$ and $z_s = 1.6748$.
The Legacy Surveys image of \desionesixfive shows an elliptical galaxy surrounded by what appears to be an almost complete Einstein ring, with 
the \hst image revealing that it is a quadruply lensed system. 
Similar to five other systems in this work (with the exception of \desionefivefour), we used a $27 \times 27$ pixel empirical PSF. 
Our mass model comprises an elliptical power law (EPL) for the main lens with external shear.
We model the lens light with two elliptical \ser profiles,
and the source light with an elliptical S\'{e}rsic profile and a shapelets with $n_{max} = 6$ \citep{birrer2015a}. 
We determined the Einstein radius to be $\theta_E = \thetaEfit$, 
the slope of the power-law mass profile, $\gamma = \gammafit$, 
and the total mass within the critical curve, $M(< \tE) = 1.7390_{-0.0043}^{+0.0046} \times 10^{12} \, M_{\odot}$. 
All \rhat values are below 1.06. 
It is one of three systems in which even an environmental galaxy is included in the full forward modeling.
For more details, we refer the reader to Paper~I.
\subsection{\desizeroninefour}\label{sec:desi094}
\href{https://www.legacysurvey.org/viewer/?ra=94.5639&dec=+50.3059&layer=ls-dr10-grz&pixscale=0.262&zoom=16}{\desizeroninefour}
was discovered in H21, with a ResNet probability of 1.000, and a human grade of A.
The spectroscopic lens redshift, \zd = 0.552 is from Lick and DESI and source redshift, \zs = 3.3332, from Keck NIRES (both were reported in Paper~III).

Figure~\ref{fig:system094} shows \desizeroninefour in a cutout with 120 pixels on the side. 
The image consists of a foreground elliptical galaxy with a large lensed arc and a counter-arc on the opposite side of the lens. There is also a small environmental galaxy towards the lower left of the main arc (arrow in Figure~\ref{fig:system094}).
As with \desionefivefour, we will refer to this object as the nearby galaxy.
We also identify and mask out several faint objects in the cutout image (circles in Figure~\ref{fig:system094}). 

\begin{minipage}[H]{\linewidth}
\makebox[\linewidth]{
\includegraphics[keepaspectratio=true,scale=0.70]{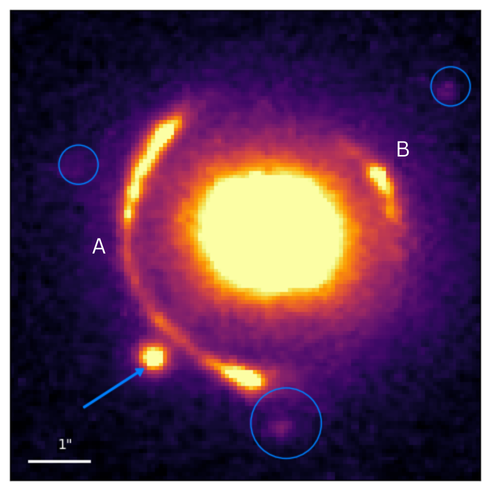}}
\captionof{figure}{\hst image of \desizeroninefour.
Faint objects circled in blue are masked out and not modeled.
We model the light of the small environmental galaxy (arrow) and then fix its parameters.}
\label{fig:system094}
\end{minipage}

 This system is challenging to model due to several factors. First, its Einstein radius of around $\tE \approx 2.3''$ is relatively large.
Furthermore, we find that a single \ser profile is inadequate for modeling the source. 
Lastly, the light of the bright environmental galaxy (arrow in Figure~\ref{fig:system094}) overlaps slightly with the lensed image.

The lens light is modeled with two \ser profiles, and the source light with three \ser's.
We model the light of the nearby galaxy separately, ahead of the main modeling process, using a \ser profile. We then fix its parameters.

For the HMC step, the \rhat values for all parameters are below 1.1; in fact, the average \rhat is 1.02.
The best-fit mass parameter for the lens appear in Table~\ref{tab:VBbest-fit-mass-params}, and their sampled posterior distribution in Figure~\ref{fig:cornermass094}. 
We also show the best-fit values for the light parameters in Table~\ref{tab:light_params_094}. 
We achieve a reduced $\chi^2$ of 1.049 (Figure~\ref{fig:bestfit094}).

\vspace{-.15cm}
\begin{deluxetable*}{lccccccc}[h]
\tabletypesize{\scriptsize}
\tablecaption{Best-fit mass parameters for \desizeroninefour.
\label{tab:VBbest-fit-mass-params}}
\renewcommand{\arraystretch}{1.4}
\tablehead{
    \colhead{$\theta_E$} &
    \colhead{$\gamma$} &
    \colhead{$\epsilon_1$} &
    \colhead{$\epsilon_2$} &
    \colhead{$x$} &
    \colhead{$y$} &
    \colhead{$\gamma_{ext, 1}$} &
    \colhead{$\gamma_{ext, 2}$} 
}
\startdata
$2.2866\pm0.0004$& 
$2.540_{-0.040}^{+0.043}$& 
$0.0633_{-0.0060}^{+0.0068}$&
$-0.0980_{-0.0089}^{+0.0073}$& 
$0.4234\pm0.0015$& 
$0.1215\pm0.0010$&	
$-0.0274\pm0.0019$&	
$-0.0648_{-0.0012}^{+0.0013}$\\
\enddata
\end{deluxetable*}
\vspace{-2.2cm}

\begin{deluxetable}{c|cc|ccc}[H]
\tablecaption{Best-fit light parameters for \desizeroninefour}
\label{tab:light_params_094}
\renewcommand{\arraystretch}{1.4}
\setlength{\tabcolsep}{8pt}
\tablehead{
  \colhead{Parameter} & 
  \multicolumn{2}{c}{Lens light} &
  \multicolumn{3}{c}{Source light} \\
  \colhead{} & 
  \colhead{Comp 1} & \colhead{Comp 2} & 
  \colhead{Comp 1} & \colhead{Comp 2} & \colhead{Comp 3}
}
\startdata
$\bm{R_e}$          & $2.032_{-0.028}^{+0.038}$ & $1.413_{-0.095}^{+0.055}$ & $0.658_{-0.061}^{+0.068}$ & $0.544_{-0.047}^{+0.051}$ & $0.623_{-0.050}^{+0.055}$ \\[2pt]
$\bm{n}$            & $0.531_{-0.025}^{+0.079}$ & $7.43_{-0.14}^{+0.12}$ & $9.24_{-0.79}^{+0.54}$ & $9.75_{-0.34}^{+0.18}$ & $2.40\pm{0.10}$ \\[2pt]
$\bm{\epsilon_{1}}$ & $0.0090_{-0.0076}^{+0.0099}$ & $0.1060_{-0.0011}^{+0.0012}$ & $-0.141\pm{0.018}$ & $-0.257_{-0.015}^{+0.016}$ & $-0.288_{-0.023}^{+0.024}$ \\[2pt]
$\bm{\epsilon_{2}}$ & $0.084_{-0.011}^{+0.010}$ & $-0.0553_{-0.0014}^{+0.0012}$ & $0.2974_{-0.0042}^{+0.0019}$  & $0.083\pm{0.014}$ & $0.170_{-0.018}^{+0.017}$ \\[2pt]
$\bm{x}$            & 
$0.571\pm{0.012}$ & $0.4700_{-0.0008}^{+0.0007}$ & $0.2595_{-0.0048}^{+0.0047}$ & $0.3210_{-0.0033}^{+0.0031}$ & $0.3577_{-0.0026}^{+0.0023}$ \\[2pt]
$\bm{y}$            & $0.002_{-0.016}^{+0.025}$ & $0.1846_{-0.0006}^{+0.0007}$ & $0.1595_{-0.0012}^{+0.0013}$ & $0.0235_{-0.0037}^{+0.0036}$ & $0.0955_{-0.0019}^{+0.0017}$ \\[2pt]
\enddata
\end{deluxetable}
\vspace{-1cm}
\begin{figure}[H]
    \centering
    \includegraphics[keepaspectratio=true,scale=0.52]{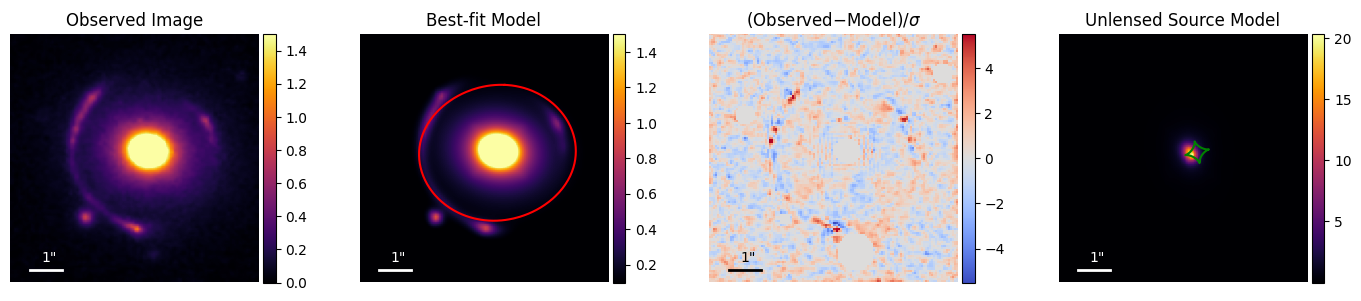}
    \caption{Best Fit Model \desizeroninefour. From left to right, we show: the observed Hubble image, our best-fit model with critical curve, the reduced residual, and the unlensed source with caustic.}
    \label{fig:bestfit094} 
\end{figure}  



Magnifications are computed for lensed images A (including the main lensed arc) and B separately (Figure~\ref{fig:system094}). The results appear in Table~\ref{tab:magnifications_094}. 
Finally, we found the projected mass within \tE to be
$M(< \tE) = 
1.1231 \pm 0.0023 \times 10^{12}  M_{\odot}.$

\begin{deluxetable}{lccc}[H]
\tabletypesize{\footnotesize}
\renewcommand{\arraystretch}{1.4}
\tablecaption{Magnification calculations for \desizeroninefour using three approaches, for images A and B, and the total magnification, respectively.}
\label{tab:magnifications_094}
\tablehead{
    \colhead{Method} &
    \colhead{$A$} &
    \colhead{$B$} &
    \colhead{Total}
}
\startdata
$1^{\text{st}}$ & $2.58^{+0.11}_{-0.13}$ & $9.30^{+0.92}_{-1.21}$ & $11.88^{+1.03}_{-1.34}$ \\
$2^{\text{nd}}$ & $2.34^{+0.16}_{-0.20}$ & $12.86^{+1.01}_{-1.25}$ & $15.20^{+1.17}_{-1.45}$ \\
$3^{\text{rd}}$ & $2.98^{+0.09}_{-0.14}$ & $11.68^{+1.10}_{-1.15}$ & 
$14.66\pm{1.19}$ \\
\enddata
\end{deluxetable}

\begin{figure}[H]
    \centering
    \includegraphics[keepaspectratio=true,width=0.85\linewidth]{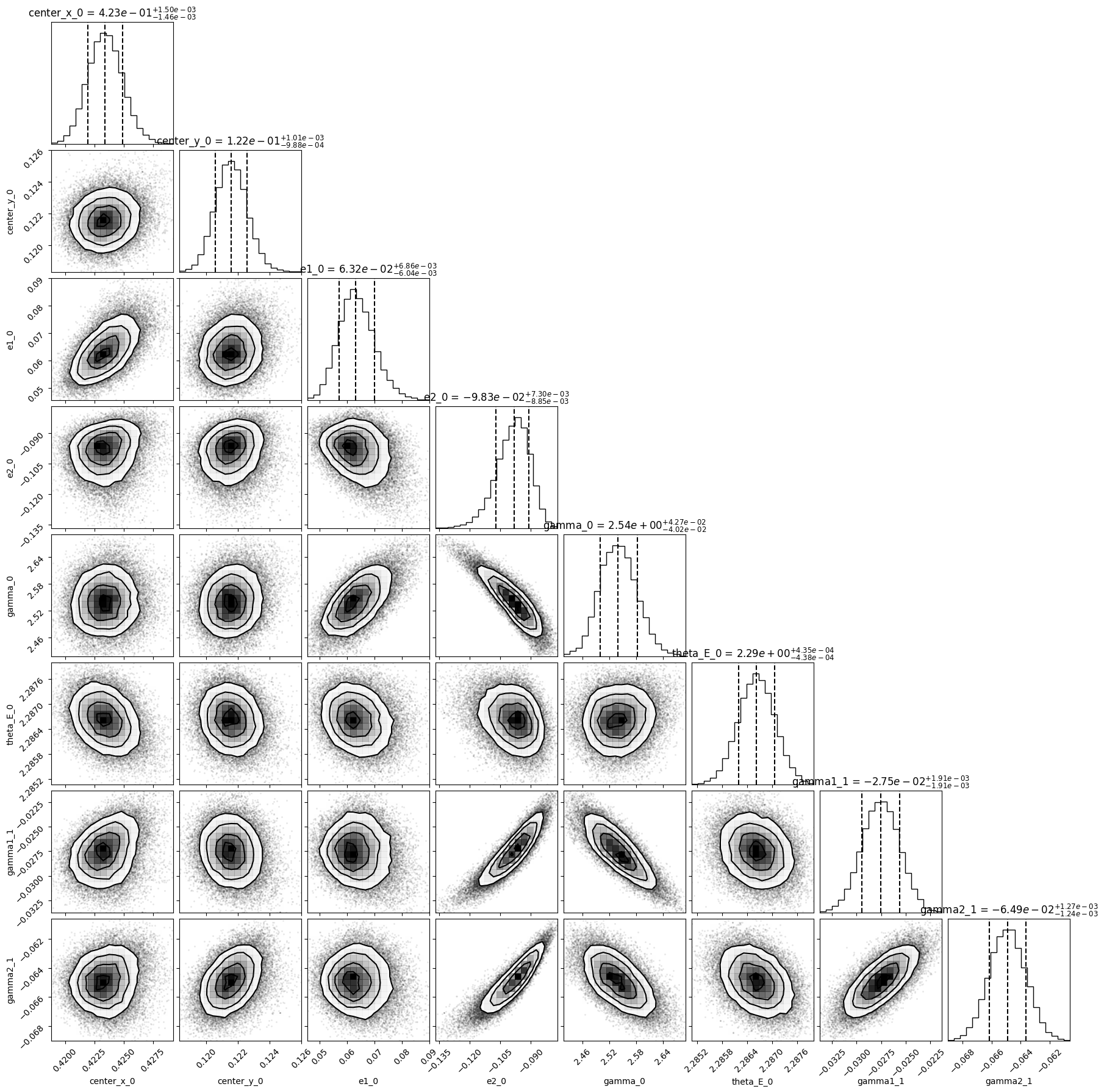}
    \caption{Sampled posterior distribution of mass parameters for system \desizeroninefour.}
    \label{fig:cornermass094}
\end{figure}

\vspace{-0.4 cm}
\subsection{\desitwofiveseven}\label{sec:desi257}
\href{https://www.legacysurvey.org/viewer/?ra=257.434789999999&dec=31.9046&layer=ls-dr8&pixscale=0.26&size=101&zoom=16}{\desitwofiveseven} was discovered in H21, with a numerical grade of 3.0 (out of 4), corresponding to a B~grade.
The lens and source redshifts of $z_d = 0.7464$ and
$z_s = 2.1200$ are obtained from DESI and Keck NIRES, respectively (Paper~III).
The Legacy Surveys image of this system shows an elliptical galaxy surrounded by an arc-counterarc pair, with an Einstein radius of approximately $\sim 2\twopr$.
The \hst image resolves the main arc to be triple images, labeled as A (encompassing two merging images) and B, with the counter image labeled as C (Figure~\ref{fig:desi257mask}).

There is a very faint object south of the lensing system (small blue circle in Figure~\ref{fig:desi257mask}). This object is masked during the modeling process. 
We also mask the brightest part of a diffraction spike (from a bright star, indicated by the big blue circle in Figure~\ref{fig:desi257mask}) that extends to the vicinity of image C.
The mask has a brightness threshold of $2\times$ the background noise. 
We use a $33 \times 33$ pixel empirical PSF, generated using the procedure outlined in Section~\ref{sec:psf_section}.

We model the lens light with two elliptical \ser profiles. The source is modeled using an elliptical \ser profile and a shapelets component with \( n_{max} = 3 \) \citep{birrer2015a}.

\begin{figure}[H]
    \centering
    \includegraphics[width=0.7\linewidth]{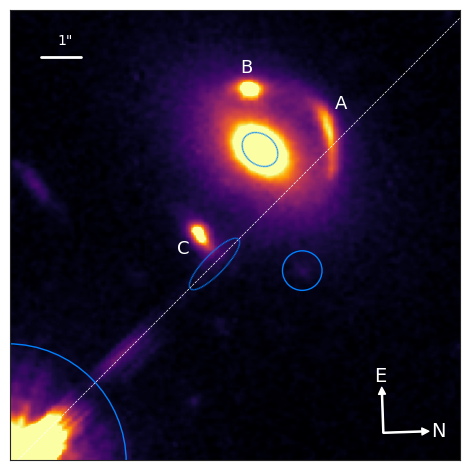}
    \captionof{figure}{
    \desitwofiveseven.
    The three lensed images are labeled counterclockwise as A, B and C. To model this system, we mask out the light from a faint object in the bottom right corner (circle) and the brightest part of its diffraction spike (blue ellipse aligned with the dashed white line). 
}
    \label{fig:desi257mask}
\end{figure}

This system is modeled using linear inversion to solve for the light intensities. 
Our best-fit model, with a reduced $\chi^2$ of 0.813, is shown in Figure~\ref{fig:desi257bestfitmodel}.
The best-fit mass and light parameters after HMC are shown in Tables~\ref{tab:desi257best-fit-mass-params} and \ref{tab:desi257best-fit-light-params} respectively.

\begin{figure}[H]
    \centering
    \includegraphics[keepaspectratio=true,scale=0.605]{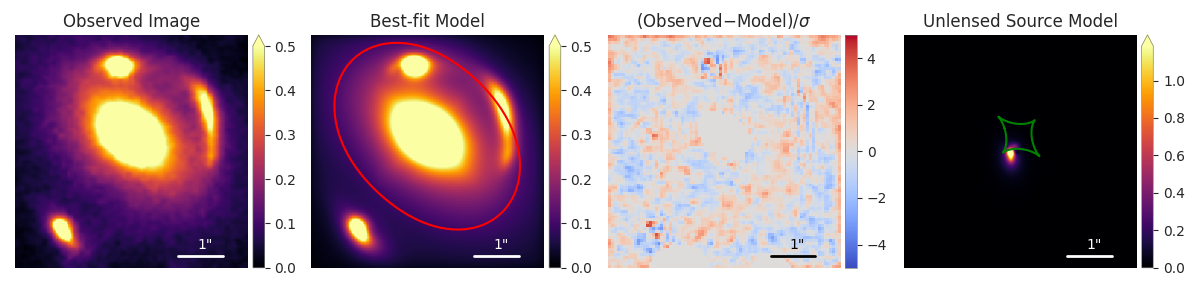}
    \caption{Best Fit Model for \desitwofiveseven. From left to right, we show: the observed Hubble image, our best-fit model with critical curve, the reduced residual, and the unlensed source with caustic.}
    \label{fig:desi257bestfitmodel}
\end{figure}

\begin{deluxetable}{lccccccc}[H]
\tabletypesize{\scriptsize}
\tablecaption{Best-fit mass parameters for \desitwofiveseven.
\label{tab:desi257best-fit-mass-params}}
\renewcommand{\arraystretch}{1.4}
\tablehead{
    \colhead{$\theta_E$} &
    \colhead{$\gamma$} &
    \colhead{$\epsilon_1$} &
    \colhead{$\epsilon_2$} &
    \colhead{$x$} &
    \colhead{$y$} &
    \colhead{$\gamma_{ext, 1}$} &
    \colhead{$\gamma_{ext, 2}$} 
}
\startdata
$1.9899\pm0.0024$ & 
$2.066\pm0.023$ &	
$-0.0162\pm0.0028$ &
$-0.1408_{-0.0066}^{+0.0064}$ & 
$-0.0111_{-0.0030}^{+0.0029}$ &
$0.3386_{-0.0038}^{+0.0039}$ &	
$-0.0148\pm0.0015$ &	
$0.0520\pm0.0043$ \\
\enddata
\end{deluxetable}
\vspace{-1.5cm}

\begin{deluxetable}{l|rr|rr}[H]
\tablecaption{Best-fit light parameters for \desitwofiveseven.}
\label{tab:desi257best-fit-light-params}
\renewcommand{\arraystretch}{1.4}
\setlength{\tabcolsep}{8pt}
\tablehead{
    \colhead{Parameter} & \multicolumn{2}{c}{Lens light} & \multicolumn{2}{c}{Source light} \\
    \colhead{} & \colhead{Comp 1} & \colhead{Comp 2} & \colhead{Comp 1} & \colhead{Comp 2}
}
\startdata
$\bm{R_e}$ &
$1.973_{-0.027}^{+0.028}$ & $0.241_{-0.011}^{+0.012}$ & $0.2146_{-0.0075}^{+0.0080}$ & - \\[4pt]
$\bm{n}$ &                
$1.207_{-0.056}^{+0.060}$ & $2.78\pm0.12$ & $2.096_{-0.093}^{+0.099}$ & - \\[4pt]
$\bm{\epsilon_{1}}$ &     
$0.0243_{-0.0026}^{+0.0025}$ & $0.0711\pm0.0024$ & $-0.2242\pm0.0077$ & - \\[4pt]
$\bm{\epsilon_{2}}$ &  
$-0.2345_{-0.0022}^{+0.0023}$ & $-0.2387\pm0.0022$ & $-0.0469\pm0.0098$ & - \\[4pt]
$\bm{x}$ &
$0.0422_{-0.0044}^{+0.0047}$ & $-0.0132\pm0.0005$ & $-0.2156\pm0.0047$ & $-0.1861_{-0.0043}^{+0.0045}$ \\[4pt]
$\bm{y}$ &
$0.3445_{-0.0038}^{+0.0036}$ & $0.3817\pm0.0004$ & $-0.1539_{-0.0098}^{+0.0095}$ & $0.0393_{-0.0068}^{+0.0065}$ \\[4pt] 
$\bm{\beta_{shp}}$ & - & - & - & $0.0495\pm0.0011$  \\[4pt]
\enddata
\end{deluxetable}

\vspace{-0.9cm}
We successfully sampled this system with $\hat{R}$ values well below 1.01 for all parameters, achieving an average $\hat{R}$ of 1.0002, with the maximum being 1.0012.


The mass parameter corner plot is shown in Fig.~\ref{fig:masscornerplot257} and 
the magnification results are presented in Table \ref{tab:magnifications_257}. 
The total mass within the critical curve is $M(< \tE) = 1.5140^{+0.0074}_{-0.0078} \times 10^{12} M_{\odot}$. 

\vspace{-2cm}

\begin{deluxetable*}{lcccc}[h!]
\tablecaption{Magnification estimates for \desitwofiveseven using three methods. Values are shown for images A through C. \label{tab:magnifications_257}}
\tablehead{
    \colhead{Method} & 
    \colhead{$A$} & 
    \colhead{$B$} & 
    \colhead{$C$} & 
    \colhead{Total}
}
\startdata
$1^{\text{st}}$ & $22.7 \pm 2.7$ & $3.21 \pm 0.26$ & $2.71 \pm 0.22$ & $28.6 \pm 2.7$ \\
$2^{\text{nd}}$ & $18.9 \pm 1.7$ & $3.09 \pm 0.28$ & $2.57 \pm 0.21$ & $24.6 \pm 1.7$ \\
$3^{\text{rd}}$ & $20.4 \pm 2.3$ & $3.16 \pm 0.29$ & $2.64 \pm 0.19$ & $26.2 \pm 2.3$ \\
\enddata
\end{deluxetable*}
\vspace{-2cm}

\begin{figure}[H]
    \centering
    \includegraphics[width=0.85\linewidth]{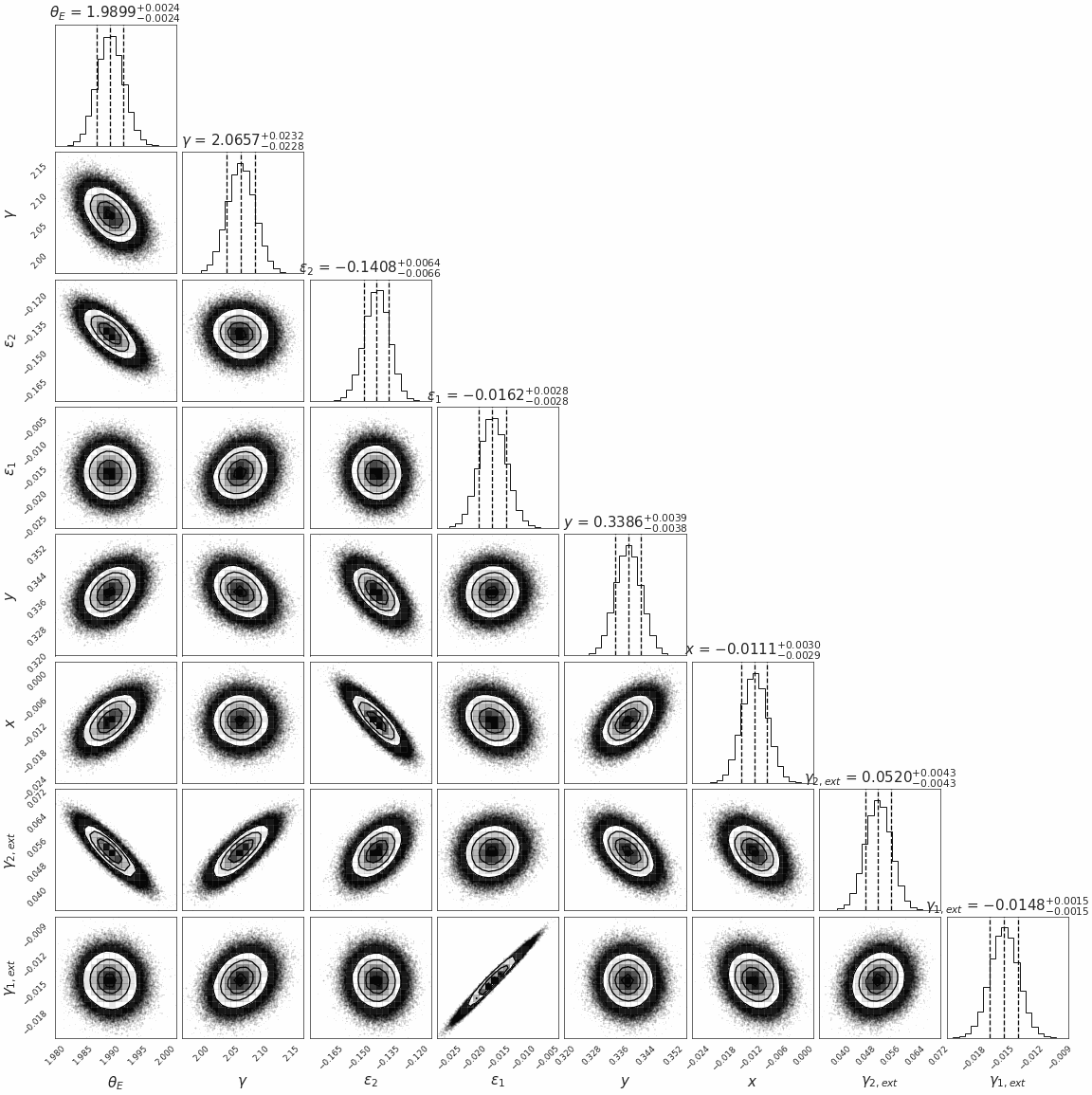}
    \caption{Corner plot for 8 mass parameters for \desitwofiveseven.}
        \label{fig:masscornerplot257}
\end{figure}

\newpage
This system has no environmental galaxies bright enough to require modeling. 
The full posterior for all 29 lens and source parameters is shown in Figure~\ref{fig:257full-corner}.

\begin{figure}[H]
    \centering
   \includegraphics[width=\linewidth]{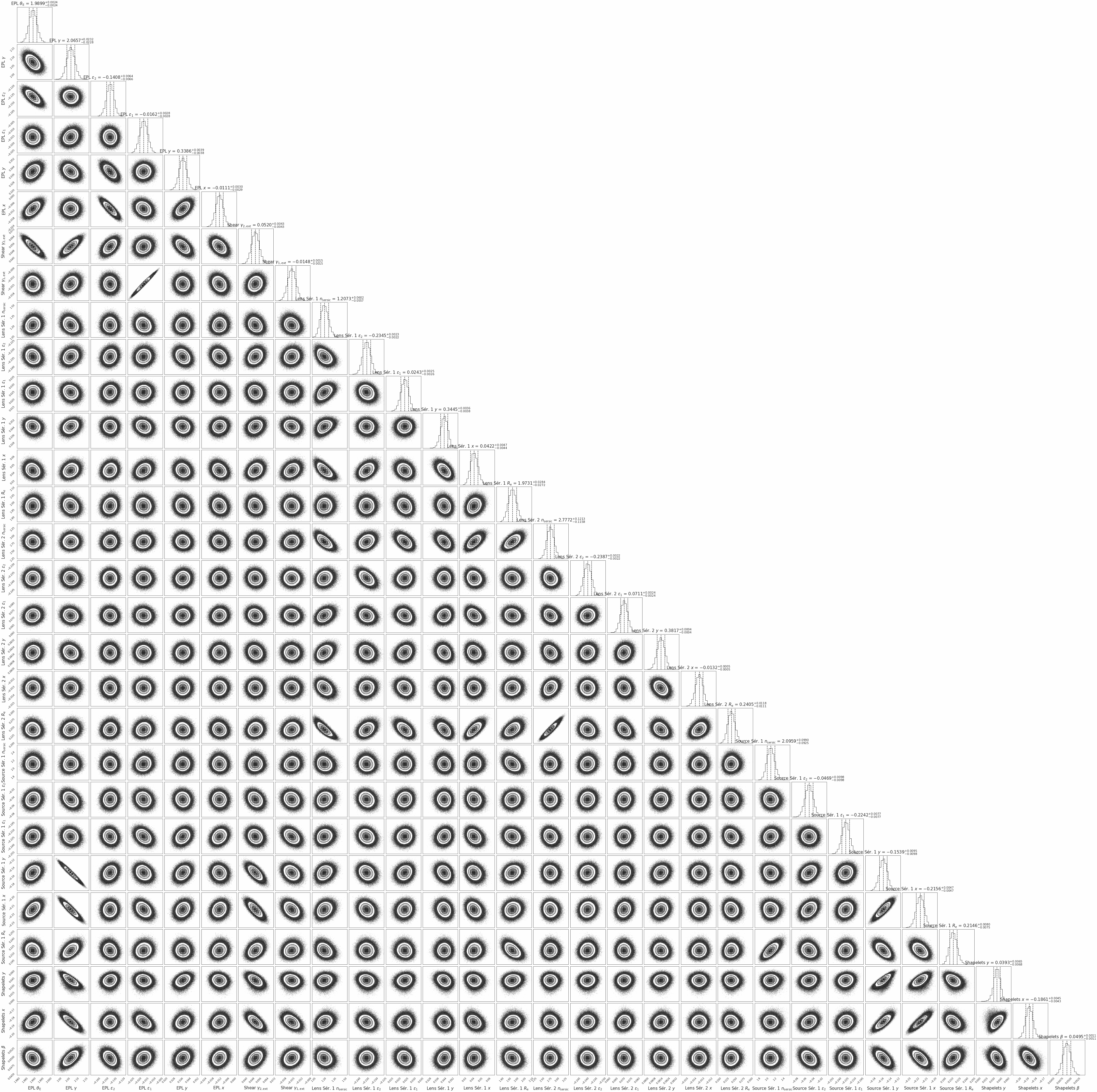}
    \caption{Corner plot of all 29 model parameters for \desitwofiveseven.}
    \label{fig:257full-corner}
\end{figure}

\newpage
\subsection{\desitwofoursix}\label{sec:desi246}
\href{https://www.legacysurvey.org/viewer?ra=246.0062&dec=1.4836&layer=ls-dr10-grz&zoom=16}{\desitwofoursix} was discovered in H21, with a ResNet probability of 1.00, and a human score of 3.5 (out of 4.0), corresponding to an A grade.
The lens redshift, \zd = 1.092, is obtained from VLT/MUSE, based on the 4000 \ang break and Ca~H\&K \citep[][Paper VI]{lin2025}.
This is one of the highest lens redshift galaxy-scale strong lensing systems \citep[][\zd = 1.004]{koopmans2002a}.
This is certainly the highest lens redshift among the confirmed systems in the DESI Strong Lens Foundry project thus far. 
The source redshift, \zs = 2.368, was first determined from Keck NIRES spectra (Paper~III).
In Paper~IV, we further identify multiple weak absorption lines
for different parts of the arc and thus show that they belong to the same source.

The lensing galaxy appears to be at the center and is the brightest member of a small galaxy group (Figure~\ref{fig:246mask}). This system has an Einstein radius of around $\theta_E \approx 2.7''$.
We use a $31 \times 31$ pixel empirical PSF, generated using the procedure outlined in Section~\ref{sec:psf_section}.

We model the lens light with an elliptical \ser profile. We mask out a region around its center with a flux threshold of $0.40$, as the central region of an elliptical galaxy is not always well described by a \ser profile \citep[see, e.g., Paper~I;][]{shajib2020b}. We model the source using five elliptical \ser profiles. This system has the most complex source structure (Figure \ref{fig:246mask}) in this work. 

\begin{figure}[H]
    \centering
    \includegraphics[width=0.58\linewidth]{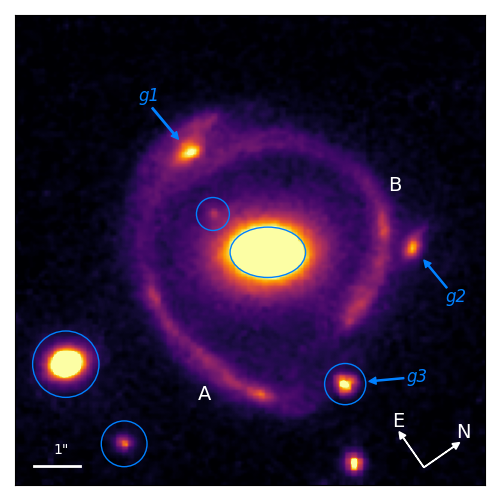}
    \caption{\desitwofoursix. The two lensed arcs are labeled as A and B. To model this system, we mask out the light from the central region of the lensing galaxy (oval).
    We further mask out one faint galaxy to the upper left of the lensing galaxy and two galaxies near the lower left corner (all three are marked with circles). For $g1$, $g2$, and $g3$  (arrows), we model their light and fix their parameters in the subsequent main modeling process.
    $g3$ is also also circled because we mask out its central region.}
    \label{fig:246mask}
\end{figure}

We now briefly discuss the objects $g1$, $g2$, and $g3$ (Figure~\ref{fig:246mask}).
From MUSE data (Paper~VI), 
for $g1$, it is possible that we have detected the \ion{O}{2} doublet (\lamlam~3726, 3729~\ang). If this is a real detection, $g1$ is at a redshift of 1.07.
For $g2$, we determine its redshift to be 1.441 based on a convincingly detected \ion{O}{2} doublet.
For $g3$, the spectrum does not have clear enough emission or absorption features for us to securely determine its redshift.
However, the MUSE image has a slightly better seeing than the Legacy Surveys, and as a result, $g3$ is deblended from the lensed arc, and it clearly has a different color from the lensed arcs, and thus is not part of them.
In addition, based on the surface brightness, $g1$ does not appear to be part of the lensed arcs either. 
In this work, we model the light of $g1$ with two elliptical \ser's. For each of $g2$ and $g3$,  we use an elliptical \ser profile.  
Then, we freeze these 28 parameters. 
We further mask out the central region of $g3$.
It is possible that modeling the mass of $g1$ and $g2$ will further improve the model. We leave this for further work. 

In addition, we mask a small and dim object (small blue circle just to the east of the main lens in Figure~\ref{fig:246mask}; note the orientation of the compass).
 We also mask two objects near the bottom left corner.
We determine the brighter of the two, 
from the MUSE datacube, to be a group member, based on the 4000 \ang break and Ca H\&K absorption (Paper VI).

For the modeling of \desitwofoursix, we do not use linear inversion to solve for the light intensities for this system. Instead, we explicitly fit for the light intensities ($I_e$). 
For the \ser index ($n_{s}$) of the first 
source light component, 
we initially used a lower bound of 0.5 (Gu22).
However, in the modeling process, we find that this parameter is consistently pushed towards its lower limit.
When we lower the lower bound to $0.2$, this behavior stops.
We find the best-fit value for $n_s$ of the first source component to be $0.234^{+0.016}_{-0.15}$. We note that \citet{tamm2006a} reported a similarly low $n_s=0.3\pm0.05$ for a galaxy. 
Moreover, this is just one component of the source light, which as we show below has a complex structure.

To achieve convergence in a reasonable amount of time, 
we freeze the 28 parameters describing the environmental galaxies $g1$, $g2$, and $g3$. It is important to note that these belong to neither the lens nor the source.
We do not observe significant correlation between the mass parameters of the main lens (the main goal of the lens modeling effort) and these fixed parameters.

In modeling this system, we have significantly advanced the frontier of galaxy-scale strong lens modeling: this is a system with a large \tE, very complex source structure, and a large number of environmental galaxies in the cutout, which, given the large cutout, is not unexpected.
In our model, we simultaneously optimize and sample 50 parameters! 
To the best of our knowledge, for a galaxy-scale strong lens, this is unprecedented.

We achieve very good residuals (Figure \ref{fig:246model}), with a reduced $\chi^2=1.1299$.
The central region of {\it g3} is masked when displaying the residuals and computing $\chi^2$, but not during the modeling process.
We present the best-fit mass parameters in Table~\ref{tab:246best-fit-mass-params}, and light parameters in \ref{tab:246best-fit-light-params}. We show the five \ser components for the source in Figure~\ref{fig:246sources}.

We successfully sampled this system with $\hat{R}$ values well below 1.1 for all parameters, achieving an average $\hat{R}$ of 1.003, with the maximum being 1.037. 

\begin{figure}[H]
    \centering  \includegraphics[keepaspectratio=true,scale=0.60]{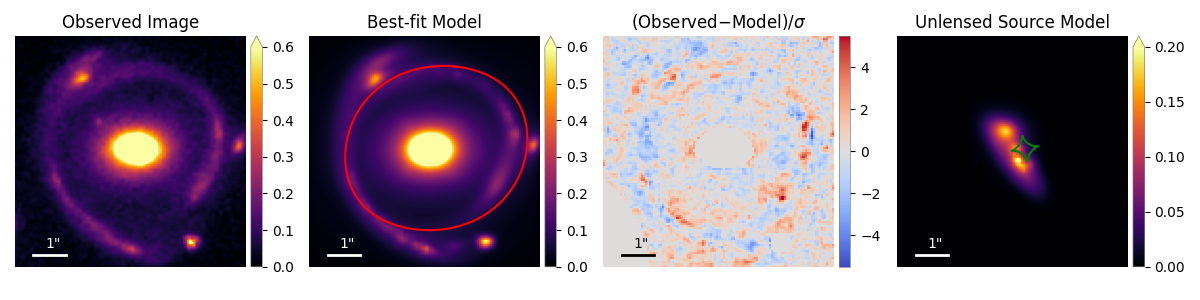}
    \caption{Best Fit Model for DESI-246.0062+01.4836. From left to right, we show: the observed Hubble image, our best-fit model with critical curve, the reduced residual, and the unlensed source with caustic.}
    \label{fig:246model}
\end{figure}

Notably, each \ser component contributes to both arcs. 
This consistency reinforces the reliability of our results, as the five components align coherently, indicating that they belong to the same object.

\begin{figure}[H]
    \centering
    \includegraphics[trim={0cm 1.8cm 0cm 1cm},clip,scale=0.407]{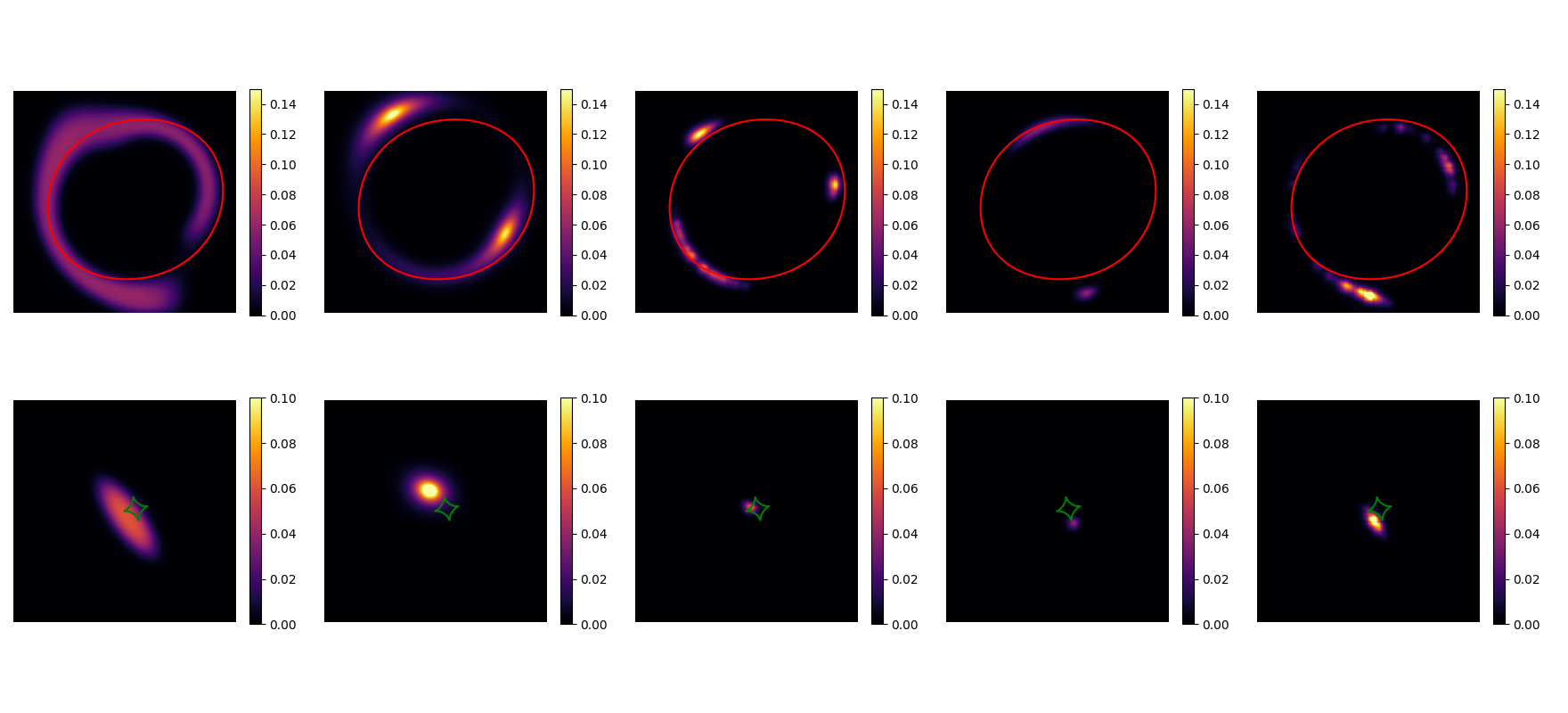}
    \caption{Source light \ser components in the lensed (top row) and unlensed (bottom row) plane for \desitwofoursix. We label them as component 1 to 5 from left to right.}
    \label{fig:246sources}
\end{figure}

\begin{deluxetable}{lccccccc}[H]
\tabletypesize{\scriptsize}
\tablecaption{Best-fit mass parameters for DESI-246.0062+01.4836.
\label{tab:246best-fit-mass-params}}
\renewcommand{\arraystretch}{1.4}
\tablehead{
    \colhead{$\theta_E$} &
    \colhead{$\gamma$} &
    \colhead{$\epsilon_1$} &
    \colhead{$\epsilon_2$} &
    \colhead{$x$} &
    \colhead{$y$} &
    \colhead{$\gamma_{ext, 1}$} &
    \colhead{$\gamma_{ext, 2}$} 
}
\startdata
$2.6982\pm0.0007$ & 
$2.616\pm0.017$&	
$0.0537_{-0.0027}^{+0.0029}$ &
$0.1789_{-0.0069}^{+0.0070}$ & 
$0.3664\pm0.0007$ &
$0.1028\pm0.0006$ &	
$-0.0521\pm0.0006$ &	
$0.0307_{-0.0005}^{+0.0004}$ \\
\enddata
\end{deluxetable}
\vspace{-1.8cm}
\begin{deluxetable}{l|r|rrrrr}[H]
\tablecaption{Best-fit light parameters for DESI-246.0062+01.4836.
\label{tab:246best-fit-light-params}}
\renewcommand{\arraystretch}{1.4}
\setlength{\tabcolsep}{8pt}
\tablehead{
    \colhead{Parameter} & \colhead{Lens light} &
    \multicolumn{5}{c}{Source light}
    \\
      \colhead{} & 
  \colhead{} & \colhead{Comp 1} & 
  \colhead{Comp 2} & \colhead{Comp 3} & \colhead{Comp 4} & \colhead{Comp 5} 
}

\startdata
$\bm{R_e}$ &
$2.158_{-0.090}^{+0.096}$ &
$0.5005_{-0.012}^{+0.010}$ &
$0.3464_{-0.0072}^{+0.0076}$ &
$0.0345_{-0.0009}^{+0.0010}$ &
$0.0671_{-0.0043}^{+0.0048}$ &
$0.0356\pm0.0008$ \\[4pt]
$\bm{n}$ &               
$6.97_{-0.31}^{+0.32}$ &
$0.234_{-0.015}^{+0.016}$ &
$1.046_{-0.046}^{+0.048}$ &
$0.527_{-0.050}^{+0.056}$ &
$0.522_{-0.017}^{+0.037}$ &
$0.644_{-0.054}^{+0.062}$ \\[4pt]
$\bm{\epsilon_{1}}$ &    
$0.1749\pm0.0017$ &
$-0.1645\pm+0.0085$ &
$0.114\pm+0.011$ &
$0.5149_{-0.0088}^{+0.0087}$ &
$-0.148_{-0.051}^{+0.054}$ &
$-0.4395\pm0.0038$ \\[4pt]
$\bm{\epsilon_{2}}$ &    
$0.0085\pm0.0016$ &
$-0.4660_{-0.0064}^{+0.0065}$ &
$-0.101\pm0.015$ &
$-0.5955_{-0.0033}^{+0.0062}$ &
$0.0239\pm0.047$ &
$-0.8516_{-0.0029}^{+0.0032}$ \\[4pt]
$\bm{x}$ &
$0.1885\pm0.0019$ &
$0.094\pm0.010$ &
$-0.1945\pm0.0086$ &
$0.1222\pm0.0031$ &
$0.5136_{-0.0039}^{+0.0041}$ &
$0.2042_{-0.0026}^{+0.0025}$ \\[4pt]
$\bm{y}$ &
$0.063\pm0.0013$ &
$-0.227_{-0.017}^{+0.016}$ &
$0.6497_{-0.0060}^{+0.0059}$ &
$0.1110\pm0.0017$ &
$-0.3990_{-0.0095}^{+0.0094}$ &
$-0.3530_{-0.0066}^{+0.0069}$ \\[4pt]
$\bm{I_e}$ &
$10.50_{-0.83}^{+0.87}$ &
$11.59_{-0.26}^{+0.27}$ &
$11.93\pm0.52$ &
$143\pm7.8$ &
$23.0_{-2.1}^{+2.3}$ &
$462_{-27}^{+29}$ \\
\enddata
\end{deluxetable}

\vspace{-1.3cm}




The corner plot for the mass parameters is displayed in Figure~\ref{fig:246cornermass}.
The best-fit Einstein radius is $\theta_E = 2.6982''$$\pm0.0007$, and the mass profile slope, $\gamma = 2.616\pm0.017$. The projected total mass within the critical curve is $M (<\tE) = 4.0735^{+0.0109}_{-0.0089} \times 10^{12} M_{\odot}$. 
The magnification results for each of the three methods are shown in Table \ref{tab:magnifications_246}.

\begin{deluxetable*}{lccc}[h]
\tablecaption{Magnification estimates for \desitwofoursix using three methods. Values are shown for components A and B. \label{tab:magnifications_246}}
\tablehead{
    \colhead{Method} & 
    \colhead{$A$} & 
    \colhead{$B$} & 
    \colhead{Total}
}
\startdata
$1^{\text{st}}$ & $7.60\pm1.79$ & $2.87\pm0.54$ & $10.1\pm1.9$ \\
$2^{\text{nd}}$ & $5.50\pm1.63$ & $2.11\pm0.54$ & $7.6\pm1.7$ \\
$3^{\text{rd}}$ & $6.2\pm1.4$   & $2.51\pm0.52$ & $8.7\pm1.5$ \\
\enddata
\end{deluxetable*}

\begin{figure}[H]
    \centering
    \includegraphics[width=0.85\linewidth]{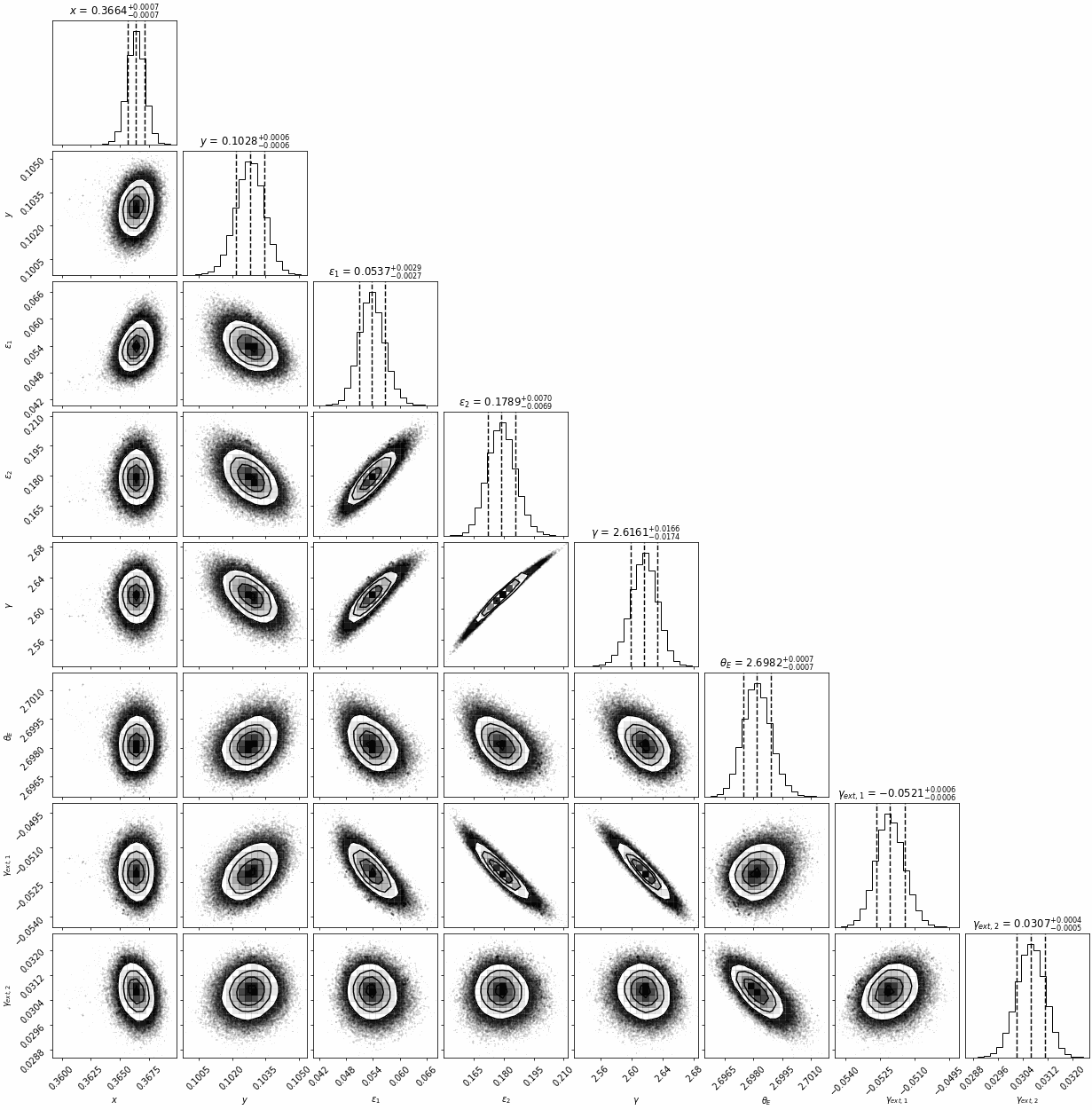}
    \caption{Corner plot for the eight mass parameters for \desitwofoursix.}
    \label{fig:246cornermass}
\end{figure}


\section{Convergence Metrics}\label{sec:append-converg}










\subsection{\rhat Definitional Differences}

Following \citet[][BG98]{brooks1998a}, Equation~(1.1)

\begin{equation}\label{eq:bg98rhat}
    \hat{R}_{BG98} = \frac{\hat{V}} {W}
\end{equation}

with 

\begin{equation}\label{eq:vhat}
    \hat{V} = \hat{\sigma}_+^2 + \frac{B}{mn}
\end{equation}

and

\begin{equation}\label{eq:sigplus}
    \hat{\sigma}_+^2 = \frac{n-1}{n} W + \frac{B}{n}
\end{equation}

\noindent
where $m$ is the number of chains and $n$ is the number of samples.

The definition of $\hat{\sigma}_+^2$ (Equation~\ref{eq:sigplus}) is the same as in \citet[][G14]{gelman2014a}, Equation~(11.3), written as $\widehat{\mathrm{var}}^+$.
The definitions of $W$ and $B$ (not given here) are also the same in BG98 and G14.

In G14, \rhat is defined as the following, in their Equation~(11.4),

\begin{equation}\label{eq:g14rhat}
    \hat{R}_{G14} = \sqrt{\frac{\widehat{\mathrm{var}}^+} {W}}
\end{equation}

Comparing Equations~\ref{eq:bg98rhat} and \ref{eq:g14rhat}, which are the definitions of \rhat in BG98 and G14, respectively, there are two differences.
First, BG98 deals with variances (as explicitly pointed out in their Section~1.2), thus no square root; whereas G14 deals with the actual ``scale" (the ``scale'' in the Potential \emph{Scale} Reduction Factor), and thus the square root is applied.
Second, in BG98, there is an extra term in the numerator of the \rhat definition: $\frac{B}{mn}$ (the second term in Equation~\ref{eq:vhat}). 
The product of $mn$ tends to be very large, and this represents a small correction.
Therefore, the main difference results from whether the square root is applied.


The \tf documentation states that using the BG98 definition,  
$\rhat < 1.2$ (without the square root) indicates approximate convergence\footnote{See \url{https://www.tensorflow.org/probability/api_docs/python/tfp/mcmc/potential_scale_reduction}} (BG98, p.\ 444\footnote{Somewhat confusingly, in Section 3.2 of BG98, when introducing the threshold of 1.2, they \emph{put \rhat under the square root}. However, that does not change the fact that the definitions of \rhat in BG98 and G14 differ by the square root and that \tf implemented the BG98 version without the square root.}).
\citet[][G14]{gelman2014a} provides the current definition of \rhat, which is approximately the square root of the definition that BG98 (and \tf) used.\footnote{See also the Stan Reference Manual Section 15.3.1, \url{https://mc-stan.org/docs/2_18/reference-manual/notation-for-samples-chains-and-draws.html}.}
and the recommended threshold $\rhat < 1.1$ for convergence (G14, p.\ 287; consistent with BG98).
 This is the definition we use in this paper.

 Finally, in Paper~I, we used the G14 definition for \rhat. However, in \citet{cikota2023a}, we used the BG98 definition (i.e., the \rhat value as reported by \tf), without taking its square root --- this was before we realized the definitional difference discussed in this section.
 Therefore, using the G14 definition of \rhat, all parameters in \citet{cikota2023a} are below the threshold of 1.1.

\end{document}